\documentclass[review,3p]{elsarticle}
\usepackage{graphicx}
\usepackage{amsthm}
\usepackage{amssymb,color}
\usepackage{bm,amsmath,longtable,booktabs, array,rotfloat}
\usepackage{subfigure}

\theoremstyle{remark}

\journal{Elseiver Science}

\begin{document}

\begin{frontmatter}



\title{A modified lattice Bhatnagar-Gross-Krook model for convection heat transfer in porous media}

%
\author[label1,label2]{Liang Wang}
\address[label1]{State Key Laboratory of Coal Combustion, Huazhong University of Science and Technology, Wuhan, 430074, P.R. China}
\author[label2]{Jianchun Mi}
\address[label2]{State Key Laboratory of Turbulence and Complex Systems, Peking University, Beijing, 100871, P.R. China}
\author[label1]{Zhaoli Guo\corref{cor1}}
\ead{zlguo@hust.edu.cn}
\cortext[cor1]{Corresponding author.}

\begin{abstract}
The lattice Bhatnagar-Gross-Krook (LBGK) model has become the most popular one in the lattice Boltzmann method for simulating the convection heat transfer in porous media. However, the LBGK model generally suffers from numerical instability at low fluid viscosities and effective thermal diffusivities. In this paper, a modified LBGK model is developed for incompressible thermal flows in porous media at the representative elementary volume scale, in which the shear rate and temperature gradient are incorporated into the equilibrium distribution functions. With two additional parameters, the relaxation times in the collision process can be fixed at a proper value invariable to the viscosity and the effective thermal diffusivity. In addition, by constructing a modified equilibrium distribution function and a source term in the evolution equation of temperature field, the present model can recover the macroscopic equations correctly through the Chapman-Enskog analysis, which is another key point different from previous LBGK models. Several benchmark problems are simulated to validate the present model with the proposed local computing scheme for the shear rate and temperature gradient, and the numerical results agree well with analytical solutions and/or those well-documented data in previous studies. It is also shown that the present model and the computational schemes for the gradient operators have a second-order accuracy in space, and better numerical stability of the present modified LBGK model than previous LBGK models is demonstrated.
\end{abstract}


\end{frontmatter}


\section{Introduction}
\label{Sec:Intro}
Fluid flow and convection heat transfer in porous media have long been a subject of research due to its comprehensive relevance in engineering and scientific applications, such as geothermal energy extraction, heat exchangers and electronic cooling instruments, chemical catalytic reactors, and
pollution transport and mineral processing. Over the past several decades, considerable investigations and applications have been devoted to the convection heat transfer in porous media by many researchers using various traditional numerical methods (eg. finite difference, element and volume methods). A comprehensive review on this subject has been made by Cheng \cite{ChengP78}, Nield and Bejan \cite{Bejan06} and Vafai \cite{Vafai05}.

As a powerful computational tool based on the kinetic theory, the lattice Boltzmann method (LBM) has been successfully applied to simulating complex fluid flows and modeling the physics in fluids \cite{ChenS98,Succi01,GuoS13}. Some attractive advantages of LBM over the traditional numerical methods have been explained in references \cite{Succi08} and \cite{Mohammad11}. Thanks to the mesoscopic and kinetic nature, the LBM has been widely applied to the fluid flow and thermal problems in porous media \cite{Guo02,Guo05,Seta06} after its emergence \cite{Succi89}. Generally speaking, applications of the LBM to porous flows in the literature center around two scales: the pore scale and the representative elementary volume (REV) scale. At the pore scale \cite{Succi89,Martys96,Pan06,Kang07,Parmigiani11,Landry14}, the standard lattice Boltzmann equation (LBE) is used to simulate fluid flows in pores, and the local information of flow can be directly obtained. Therefore, the LBM at the pore scale can be severed as the most straightforward way to investigate the macroscopic relations and reveal the microscopic mechanism of porous flows. However, the detailed geometric information of the pores is needed for this approach, and thus the computational domain size cannot be too large in view of the limited computer resources. An alternative approach is to investigate the averaged quantities at the REV scale. In the LBM at the REV scale, an additional term based on some semi-empirical models is incorporated into the standard LBE to account for the presence of a porous medium \cite{Guo02,Guo05,Seta06}. Evidence from the literature has demonstrated the REV approach to be simple and computationally efficient for modeling flows and transport problems in porous media \cite{Spaid97,Dardis98,Kang02,Zarghami14}. Guo and Zhao \cite{Guo02} proposed a lattice Boltzmann (LB) model for simulating fluid flow in porous media, in which the porosity is introduced in the equilibrium distribution function (EDF), and a forcing term is included in the LBE to account for the linear and nonlinear drag forces of the porous media. Further, Guo and Zhao \cite{Guo05} extended the isothermal LBM to thermal flows in porous media by adding a temperature distribution function for the evolution of temperature field. Subsequently, Seta \cite{Seta06} confirmed the reliability and computational efficiency of the LBM in simulating natural convection in porous media. Shokouhmand et al. \cite{Shokouhmand09} conducted simulations on laminar flow and convective heat transfer in conduits partially and fully filled with porous media. Rong et al. \cite{Rong10} proposed a LB model particularly for axisymmetric thermal flows in porous media. Abrach et al. \cite{Abrach13} employed the LBM for the heat and mass transfer during drying of deformable saturated porous media. Recently, Gao et al. \cite{Gaoetal14} developed a thermal LB model for simulating the non-equilibrium natural convection problems in porous media.

All the mentioned-above LB models for porous flow and heat transfer problems at the REV scale are based on the Bhatnagar-Gross-Krook (BGK) collision model \cite{BGK54}. Among these models, both the viscosity of fluid and the effective thermal diffusivity of solid matrix are directly determined by the relaxation times. The most well-known criticism on these BGK-based models is the numerical instability for moderately low viscosity and/or effective thermal diffusivity. One alternative option to resolve the shortcomings of the BGK model is to employ the multiple-relaxation-time (MRT) collision model \cite{Higuera89,Hume92,Lalleme&Luo00}. Very recently, such effort has been made by Liu et al. \cite{Liu14} who constructed a MRT-LB model within the double-distribution-function (DDF) framework. They showed that numerical stability of the BGK model is well improved by the MRT model when simulating convection heat transfer in porous media. On the other hand, compared with the MRT model, the BGK counterpart has become the most used form of the LBM owing to its extreme simplicity and computational efficiency. Therefore, it is of much significance for us to develop a new lattice BGK (LBGK) model which can enhance numerical stability at low viscosities and thermal diffusivities.

Within the framework of LBGK model, Inamuro \cite{Inamuro02} proposed a lattice kinetic scheme (LKS) for simulating fluid flows with heat transfer in the absence of porous medium. In this method, an additional term relating with the shear rate (and temperature) is incorporated into the EDF to represent the fluid viscosity (and the thermal diffusivity). The relaxation time is fixed at unity while the fluid  viscosity (and the thermal diffusivity) is determined by another parameter independent of the relaxation time. Thus, better numerical stability than the standard LBGK model can be achieved by the LKS at relatively small viscosity (and thermal diffusivity). Later, this LKS was extended to develop LBGK models for two-phase fluid flows \cite{Inamuro06,Nishiyama13} and non-Newtonian fluid flows \cite{Yoshino07}. However, in these LKS schemes, the mass conservation is not satisfied unless the fluid density is constant \cite{WangL14}, and the gradient operators of velocity and temperature occurred in the EDF are calculated by a finite-difference scheme. This non-local treatment not only spoils the computational efficiency, but also brings about some difficulties to the implementation of complex boundary conditions with a local lattice scheme. Although Peng et al. \cite{PengY04} computed the stress tensor locally in viscous thermal flows, the problem of mass nonconservation still remains. Recently, Wang et al. \cite{WangL14} improved the original LKS to develop a mass-conserving and localized LBGK model for non-Newtonian fluid flows. However, as far as we know, no works have been reported on extending the LKS to improve the numerical stability of LBGK models for incompressible thermal flows in fluid-saturated porous media.

In this work, we propose a modified LBGK model for simulating convection heat transfer in porous media at the REV scale. By introducing the shear rate and temperature gradient into the EDFs as the original LKS, the dimensionless relaxation times can be fixed at proper values invariant to the fluid viscosity and thermal diffusivity which are determined by two additional parameters. In addition, with a modified EDF and a source term in the evolution equation for the temperature field, the macroscopic equations can be correctly recovered from the present LBGK model through the Chapman-Enskog analysis. In the following, the modified LBGK model is first presented, and the Chapman-Enskog analysis is then given to recover the macroscopic equations. Subsequently, a local scheme, instead of the non-local finite-difference schemes, is presented for computing the shear rate and temperature gradient. Finally, some benchmark numerical tests are carried out to validate the present model. The numerical results show that the present modified LBGK model and the local scheme for the gradient operators are both second-order accurate in space. It is also confirmed that the present LBGK model are much more stable compared with the standard LBGK model.

\section{Macroscopic equations}\label{Sec:VAM}
For a homogeneous, isotropic and fluid-saturated porous medium, the local thermal equilibrium assumption between the fluid and solid is invoked \cite{Bejan06}. The viscous fluid flow in porous media is assumed to be incompressible and the Boussinesq approximation is valid. Neglecting viscous heat dissipation and compression work done by the pressure, the macroscopic governing equations for the convection heat transfer in porous media at the REV scale can be written as \cite{GuoS13,Guo05,Liu14,ChengP90}
\begin{subequations}\label{GoverneEq}
 \begin{equation}\label{MsEq1}
 \nabla\cdot\bm{u}=0,
 \end{equation}
 \begin{equation}\label{MMeq2}
   \frac{\partial\bm{u}}{\partial t}+(\bm{u}\cdot \nabla)\left(\frac{\bm{u}}{\varepsilon}\right)=-\frac{1}{\rho_0}\nabla(\varepsilon p)+\nu_e\nabla^2\bm{u}+\bm{F},
 \end{equation}
 \begin{equation}\label{EnEq3}
   \sigma\frac{\partial T}{\partial t}+\bm{u}\cdot\nabla T=\nabla\cdot(\alpha_e\nabla T)+Q,
 \end{equation}
\end{subequations}
where $\rho_0$ is the mean fluid density, $\bm{u}$, $p$ and $T$ are the volume-averaged fluid velocity, pressure and temperature, respectively; $\varepsilon$ is the porosity of the porous medium, $\nu_e$ is the effective kinematic viscosity, and $Q$ is the internal heat source term; $\sigma=[\varepsilon\rho_fc_{pf}+(1-\varepsilon)\rho_sc_{ps}]/(\rho_fc_{pf})$ is the thermal capacity ratio between the solid and fluid phases, with $\rho_f(\rho_s)$ and $c_{pf}(c_{ps})$ being the density and specific heat of fluid (solid) phase, respectively; $\alpha_e$ is the effective thermal diffusivity, and relates with the effective conductivity ($k_e$) as $\alpha_e=k_e/(\rho_fc_{pf})$. $\bm{F}$ represents the total body force stemming from the presence of a porous medium and other external force, and are expressed as
\begin{equation}\label{ForceEq}
  \bm{F}=-\frac{\varepsilon\nu}{K}\bm{u}-\frac{\varepsilon F_{\varepsilon}}{\sqrt{K}}|\bm{u}|\bm{u}+\varepsilon\bm{G},
\end{equation}
where $\nu$ is the kinematic viscosity (not necessarily identical to $\nu_e$), $K$ is the permeability, $F_\varepsilon$ is the geometric function, and $\bm{G}$ denotes the external body force. With the Boussinesq approximation, the body force $\bm{G}$ encompassing the buoyancy force and other external forces is described by
\begin{equation}
  \bm{G}=g\beta(T-T_0)\bm{j}+\bm{a},
\end{equation}
where $g$ is the gravity acceleration, $\beta$ is the coefficient of thermal expansion, $T_0$ is the reference temperature, $\bm{j}$ is the unit vector in a direction opposite to gravity, and $\bm{a}$ is the acceleration induced by other external force. Based on Ergun's empirical relation, $F_\varepsilon$ and $K$ can be written as
\begin{equation}
   F_\varepsilon=\frac{1.75}{\sqrt{150\varepsilon^3}},\qquad K=\frac{\varepsilon^3d_p^2}{150(1-\varepsilon)^2},
\end{equation}
where $d_p$ is the diameter of filling solid particle.

The convection heat transfer problems governed by Eq. \eqref{GoverneEq} can be characterized by several dimensionless parameters: the Darcy number $Da$, the viscosity ratio $J_e$, the Reynolds number $Re$ (for mixed convection flow), the Rayleigh number $Ra$ (for natural convection), the internal Rayleigh number $Ra_I$ (for thermal convection flows with internal heat source), the Prandtl number $Pr$, which are defined as follows:
\begin{align}
   Da=\frac{K}{L^2},\quad J_e=\frac{\nu_e}{\nu},\quad Re=\frac{LU}{\nu},\quad Ra=\frac{g\beta\Delta TL^3}{\nu\alpha_e}
   \quad Ra_I=\frac{g\beta QL^5}{\nu\alpha_e^2}, \quad Pr=\frac{\nu}{\alpha_e},
\end{align}
where $L$ is the characteristic length, $U$ is the characteristic velocity and $\Delta T$ is the characteristic temperature difference.

\section{Modified LBGK model for thermal flows in porous media}
In this section, we follow the DDF method to develop a thermal LB model for fluid flow and convection heat transfer in porous media at the REV scale. The model is based on the BGK collision operator, and is constructed with the idea of the LKS. In this model, the flow field is modeled by a LBE of the density distribution function, and the temperature field is modeled by another evolution equation of the temperature distribution function. For the shear rate and temperature gradient, a local computational scheme is presented especially.

\subsection{Lattice Boltzmann equation for the flow field}
For the pressure and velocity fields governed by Eqs. \eqref{MsEq1} and \eqref{MMeq2}, the evolution equation of the modified LBGK model is given by
\begin{equation}\label{LBEeq}
  f_i(\bm{x}+\bm{c}_i\delta_t, t+\delta_t)-f_i(\bm{x}, t)=-\frac{1}{\tau_f}\left[f_i(\bm{x}, t)-f_i^{(eq)}(\bm{x}, t)\right]+\delta_tF_i(\bm{x}, t),
\end{equation}
where $f_i(\bm{x}, t)$ is the density distribution function for the particle with velocity $\bm{c}_i$ at time $t$ and position $\bm{x}$, $\delta_t$ is the time increment, $\tau_f$ is the dimensionless relaxation time, $f_i^{(eq)}(\bm{x}, t)$ is the equilibrium distribution function, and $F_i$ is the external force term.

In this work, the original LKS for fluid flows in plain media is extended to incompressible fluid flows in porous media, in which the above EDF is defined as \cite{Liu14,Inamuro02,WangL14,GuoS00}
\begin{equation}\label{EQFIN}
  f_i^{(eq)}(\bm{x}, t)=
  \begin{cases}
     \rho_0-(1-\omega_0)\frac{\varepsilon p}{c_s^2}+\rho_0s_0(\bm{u})+\rho_0r_0(\bm{u}),\quad i=0\\
     \omega_i\frac{\varepsilon p}{c_s^2}+\rho_0s_i(\bm{u})+\rho_0r_i(\bm{u}),\qquad\qquad\qquad\; i\neq0
  \end{cases}
\end{equation}
where $\omega_i$ is the weight coefficient, $c_s$ is the speed of sound, and
\begin{align}\label{EQTSHIN}
  s_i(\bm{u})=\omega_i\left[\frac{\bm{c}_i\cdot\bm{u}}{c_s^2}+\frac{\bm{u}\bm{u}:(\bm{c}_i\bm{c}_i-c_s^2\bm{I})}{2\varepsilon c_s^4}\right],\qquad r_i(\bm{u})=\omega_i\frac{A\delta_t\bm{S}:(\bm{c}_i\bm{c}_i-c_s^2\bm{I})}{2c_s^2}.
\end{align}
Here, $\bm{I}$ denotes the identity tensor, and the shear rate $\bm{S}$ ($\bm{S}=\nabla\bm{u}+\left(\nabla\bm{u}\right)^T$) together with an additional parameter $A$ are included in the EDF. For the forcing term, the appropriate form of $F_i$ to achieve correct hydrodynamic equations is taken as \cite{Guo02,Guo05,Seta06}:
\begin{equation}\label{FtermF}
  F_i=\omega_i\rho_0\left(1-\frac{1}{2\tau_f}\right)\left[\frac{\bm{c}_i\cdot\bm{F}}{c_s^2}
  +\frac{\left(\bm{u}\bm{F}+\bm{F}\bm{u}\right):(\bm{c}_i\bm{c}_i-c_s^2\bm{I})}{2\varepsilon c_s^4}\right]
\end{equation}

The pressure and fluid velocity are defined as
\begin{equation}
  p=\frac{c_s^2}{\varepsilon(1-\omega_0)}\left[\sum_{i\neq0}f_i+\tau_f\delta_tF_0+\rho_0s_0(\bm{u})\right],\quad \bm{u}=\frac{1}{\rho_0}\sum_i\bm{c}_if_i+\frac{\delta_t}{2}\bm{F}
\end{equation}
Due to the nonlinear relation of $\bm{F}$ with $\bm{u}$ in Eq. \eqref{ForceEq}, the fluid velocity $\bm{u}$ is calculated by a temporal velocity $\bm{v}$ and is given by
\begin{equation}
   \bm{u}=\frac{\bm{v}}{c_0+\sqrt{c_0^2+c_1|\bm{v}|}},\qquad \bm{v}=\frac{1}{\rho_0}\sum_i\bm{c}_if_i+\frac{\delta_t}{2}\varepsilon\bm{G},
\end{equation}
where the two parameters $c_0$ and $c_1$ are respectively given by
\begin{equation}
   c_0=\frac{1}{2}\left(1+\varepsilon\frac{\delta_t}{2}\frac{\nu}{K}\right),\qquad c_1=\varepsilon\frac{\delta_t}{2}\frac{F_\varepsilon}{\sqrt{K}}.
\end{equation}

Note that implementation of the above LBGK model depends on the underlying lattice. For convenience of study, the flow problems considered in this paper are limited to be two dimensional. But it is straightforward for us to extend the present model to three dimensional case. For the D2Q9 model, the discrete nine velocities are given by $\bm{c}_0=\bm{0}$, $\bm{c}_i=c\left(\mathrm{cos}((i-1)\pi/2),~\mathrm{sin}((i-1)\pi/2)\right)$ for $i=1-4$, and $\bm{c}_i=\sqrt{2}c\left(\mathrm{cos}((i-5)\pi/2+\pi/4),~\mathrm{sin}((i-5)\pi/2+\pi/4)\right)$ for $i=5-8$. Here, $c=\delta_x/\delta_t$ is the lattice speed with $\delta_x$ denoting the lattice spacing. Accordingly, the sound speed $c_s=c/\sqrt{3}$, and the weight coefficients are given by $\omega_0=4/9$, $\omega_i=1/9$ for $i=1-4$, and $\omega_i=1/36$ for $i=5-8$.

Through the Taylor expansion and the Chapman-Enskog analysis, the macroscopic equations \eqref{MsEq1} and \eqref{MMeq2} can be derived from the LBE \eqref{LBEeq} (The details are presented in \ref{Appen:MRTMod}), and the effective kinematic viscosity is determined by $\nu_e=c_s^2(\tau_f-A-1/2)\delta_t$. We would like to point out that when $A$ in the EDF \eqref{EQTSHIN} is set to be zero, the present LBE \eqref{LBEeq} is simplified to the standard LBGK model in which the viscosity $\nu_e$ is directly determined by $\tau_f$, i.e., $\nu_e=c_s^2(\tau_f-1/2)\delta_t$.

\subsection{Lattice Boltzmann equation for the temperature field}
The governing equation \eqref{EnEq3} is a convection-diffusion equation (CDE) with a heat source term, in which the velocity $\bm{u}$ obeys the incompressible generalized Navier-Stokes equations \eqref{MsEq1} and \eqref{MMeq2}. For most of the existing LBGK models for CDE \eqref{EnEq3}, several unwanted deviation terms are ignored with the additional condition, $t_0\gg L/c_s$ ($t_0$ and $L$ are the characteristic time and length, respectively) in the Chapman-Enskog analysis to derive Eq. \eqref{EnEq3} (cf. Ref. \cite{Guo05} and references therein). In order to eliminate this imperfectness, as well as inspired by the idea of Chai and Zhao \cite{Chai13}, a modified EDF and a source term should be constructed in the evolution equation of the temperature field. Additionally, it is noted that that the temperature gradient is included in the LKS. Thus, the evolution equation of the present modified LBGK model for CDE \eqref{EnEq3} is written as follows
\begin{equation}\label{LBEHeq}
  g_i(\bm{x}+\bm{c}_i\delta_t, t+\delta_t)-g_i(\bm{x}, t)=-\frac{1}{\tau_T}\left[g_i(\bm{x}, t)-g_i^{(eq)}(\bm{x}, t)\right]+\delta_tP_i(\bm{x}, t)+\delta_tQ_i(\bm{x}, t),
\end{equation}
where $\tau_T$ is the dimensionless relaxation time, $g_i$ is the temperature distribution function, and $g_i^{(eq)}$ is the temperature EDF and defined by
\begin{equation}\label{TeEDFEQ}
  g_i^{(eq)}=\omega_i T \left[\sigma+\frac{\bm{c}_i\cdot\bm{u}}{c_s^2}+\frac{\bm{u}\bm{u}:(\bm{c}_i\bm{c}_i-c_s^2\bm{I})}{2\varepsilon c_s^4}\right]+\varpi_iT\frac{\varepsilon p}{c_s^2\rho_0}+\sigma\omega_i B \delta_t\bm{c}_i\cdot\nabla T,
\end{equation}
where the coefficient $\varpi_i$ is given by $\varpi_i=\omega_i~(i\neq0)$,~$\varpi_0=-\sum_{i\neq0}\varpi_i$. The two source terms $P_i$ and $Q_i$ are taken as
\begin{align}
  P_i&=\omega_i\left(1-\frac{1}{2\tau_T}\right)\left[\frac{\bm{c}_i\cdot\left(T\bm{F}+\varepsilon p\nabla T/\rho_0\right)}{c_s^2}+\left(\frac{1}{\varepsilon}-\frac{1}{\sigma}\right)\frac{\bm{u}\bm{u}:(\bm{c}_i\nabla T)}{c_s^2}\right], \label{Eq:Tforce}\\
  Q_i&=\omega_i\left(1-\frac{1}{2\tau_T}\right)\left(1+\frac{\bm{c}_i\cdot\bm{u}}{\sigma c_s^2}\right)Q.
\end{align}
It is noted that the discrete source term $Q_i$ corresponds to the source term $Q$ resided in Eq. \eqref{EnEq3}.

The LBE \eqref{LBEHeq} is implemented separately with two substeps, i.e., the collision step and streaming step:
\begin{align}
   &\text{Collision}:\quad g_i^{*}(\bm{x},t)=g_i(\bm{x},t)-\frac{1}{\tau_T}\left[g_i(\bm{x}, t)-g_i^{(eq)}(\bm{x}, t)\right]+\delta_tP_i(\bm{x}, t)+\delta_tQ_i(\bm{x}, t),\label{TEqColli}\\
   &\text{Streaming}:\quad g_i(\bm{x}+\bm{c}_i\delta_t, t+\delta_t)=g_i^{*}(\bm{x},t),
  \end{align}
where $g_i^{*}(\bm{x},t)$ denotes the postcollision distribution function. At each time step when the two substeps are completed, the macroscopic temperature $T$ are computed by
\begin{equation}
  \sigma T=\sum_ig_i+\frac{\delta_t}{2}Q.
\end{equation}
In the LBE \eqref{LBEHeq} for the temperature filed, both the EDF $g_i^{(eq)}$ and the source term $P_i$ depend on $\varepsilon$ and $\sigma$ with which the influence of porous media is introduced. It should be noted that although a linear EDF is widely employed in previous LBGK models for Eq. \eqref{EnEq3} \cite{Guo05,Gaoetal14,Liu14,Gao11}, large errors will be encountered in solving the CDE \eqref{EnEq3} by these models when the resulted numerical diffusion coefficient from the Chapman-Enskog analysis is proportional to the velocity square \cite{Chai13,Chopard09}. Additionally, when $\sigma=\varepsilon=1$, $B=0$ and without the forcing term $Q_i$, the present LBGK model will reduce to that proposed by Chai and Zhao \cite{Chai13} for the CDE without porous media, which can thus be considered as a special case of the present model.

By the Chapman-Enskog analysis on the LBE \eqref{LBEHeq}, the temperature equation \eqref{EnEq3} can be recovered exactly. In what follows, the detailed mathematic derivations will be shown. To this end, the Chapman-Enskog expansions are first applied to the distribution function $g_i$, the derivatives of time and space, and the internal heat source term $Q$:
\begin{subequations} \label{ChapExp}
  \begin{align}
   g_i&=g_i^{(0)}+\lambda g_i^{(1)}+\lambda^2g_i^{(2)}+\cdots,\\
   \partial_t&=\lambda\partial_{t_1}+\lambda^2\partial_{t_2}, \qquad \nabla=\lambda\nabla_1,\\
   Q&=\lambda Q^{(1)},
  \end{align}
\end{subequations}
where $\lambda$ is a small expansion parameter having the magnitude of the Knudsen number. Since the spatial derivative of temperature is contained in the EDF $g_i^{(eq)}$ and the force term $P_i$, the following multiscaling expansions are also introduced \cite{WangL14}:
\begin{equation}\label{ChapLKS}
 g_i^{(eq)}=g_i^{e(0)}+\lambda g_i^{e(1)}, \qquad P_i=\lambda P_i^{(1)},
\end{equation}
where $g_i^{e(0)}$, $g_i^{e(1)}$, and $P_i^{(1)}$ can be explicitly written as
\begin{subequations}
  \begin{align}
    g_i^{e(0)}&=\omega_i T \left[\sigma+\frac{\bm{c}_i\cdot\bm{u}}{c_s^2}+\frac{\bm{u}\bm{u}:(\bm{c}_i\bm{c}_i-c_s^2\bm{I})}{2\varepsilon c_s^4}\right]+\varpi_iT\frac{\varepsilon p}{c_s^2\rho_0}, \\
    g_i^{e(1)}&=\sigma\omega_i B \delta_t\bm{c}_i\cdot\nabla_1 T,\\
    P_i^{(1)}&=\omega_i\left(1-\frac{1}{2\tau_T}\right)\left[\frac{\bm{c}_i\cdot\left(T\bm{F}+\varepsilon p\nabla_1 T/\rho_0\right)}{c_s^2}+\left(\frac{1}{\varepsilon}-\frac{1}{\sigma}\right)\frac{\bm{u}\bm{u}:(\bm{c}_i\nabla_1 T)}{c_s^2}\right].
  \end{align}
\end{subequations}
After some computations, we can get the following velocity moments
\begin{subequations}\label{MeOtEq}
  \begin{align}
    \sum_ig_i^{e(0)}&=\sigma T,\quad \sum_i\bm{c}_ig_i^{e(0)}=T\bm{u},\quad \sum_i\bm{c}_i\bm{c}_ig_i^{e(0)}=\frac{T\bm{u}\bm{u}}{\varepsilon}+c_s^2\sigma T\bm{I}+\frac{\varepsilon p T}{\rho_0}\bm{I},\quad \sum_i\bm{c}_i\bm{c}_i\bm{c}_ig_i^{e(0)}=c_s^2 T\Delta\cdot\bm{u}, \\
    \sum_ig_i^{e(1)}&=0,\quad \sum_i\bm{c}_ig_i^{e(1)}=\sigma c_s^2 B\delta_t\nabla_1T, \quad \sum_i\bm{c}_i\bm{c}_ig_i^{e(1)}=\bm{0},\\
    \sum_iP_i^{(1)}&=0, \quad  \sum_i\bm{c}_iP_i^{(1)}=\left(1-\frac{1}{2\tau_T}\right)\left[T\bm{F}+\frac{\varepsilon p\nabla_1 T}{\rho_0}+\left(\frac{1}{\varepsilon}-\frac{1}{\sigma}\right)\bm{u}\bm{u}\cdot\nabla_1T\right], \quad \sum_i\bm{c}_i\bm{c}_iP_i^{(1)}=\bm{0},\\
    \sum_iQ_i^{(1)}&=\left(1-\frac{1}{2\tau_T}\right)Q^{(1)}, \quad  \sum_i\bm{c}_iQ_i^{(1)}=\left(1-\frac{1}{2\tau_T}\right)\frac{\bm{u}}{\sigma}Q^{(1)},
  \end{align}
\end{subequations}
where $\Delta\cdot\bm{u}=u_\alpha\delta_{\beta\gamma}+u_\beta\delta_{\alpha\gamma}+u_\gamma\delta_{\alpha\beta}$, in which $\delta_{\alpha\beta}$ denotes the Kronecker delta with two indices, and $Q_i^{(1)}=\lambda Q_i$.

The Taylor series expansion applied to Eq. \eqref{LBEHeq} up to second order in $\delta_t$ leads to
\begin{equation}\label{EqTay}
  D_ig_i+\frac{\delta_t}{2}D_i^2g_i=-\frac{1}{\tau_T\delta_t}\left[g_i-g_i^{(eq)}\right]+\delta_tP_i+\delta_tQ_i,
\end{equation}
where $D_i=\partial_t+\bm{c}_i\cdot\nabla$. Substituting Eqs. \eqref{ChapExp} and \eqref{ChapLKS} into Eq. \eqref{EqTay}, one can obtain the consecutive orders of Eq. \eqref{EqTay} in terms of $\lambda$:
\begin{subequations}
  \begin{align}
     \lambda^0:& \quad g_i^{(0)}=g_i^{e(0)},\label{EqTemO0}\\
     \lambda^1:& \quad D_{1i}g_i^{(0)}=-\frac{1}{\tau_T\delta_t}\left[g_i^{(1)}-g_i^{e(1)}\right]+P_i^{(1)}+Q_i^{(1)}, \label{EqTemO1}\\
     \lambda^2:& \quad \partial_{t_2}g_i^{(0)}+D_{1i}g_i^{(1)}+\frac{\delta_t}{2}D^2_{1i}g_i^{(0)}=-\frac{1}{\tau_T\delta_t}g_i^{(2)}\label{EqTemO2},
  \end{align}
\end{subequations}
where $D_{1i}=\partial_{t_1}+\bm{c}_i\cdot\nabla_1$. With Eq. \eqref{EqTemO1}, Eq. \eqref{EqTemO2} can be rewritten as
\begin{equation}\label{Eqt2}
 \partial_{t_2}g_i^{(0)}+\left(1-\frac{1}{2\tau_T}\right)D_{1i}g_i^{(1)}+\frac{1}{2\tau_T}D_{1i}g_i^{e(1)}
  +\frac{\delta_t}{2}D_{1i}\left(P_i^{(1)}+Q_i^{(1)}\right)=-\frac{1}{\tau_T\delta_t}g_i^{(2)}.
\end{equation}
Note that $\sigma T=\sum_ig_i^{(eq)}=\sum_ig_i+\frac{\delta_t}{2}Q$. Based on Eqs. \eqref{ChapLKS}-\eqref{MeOtEq}, it is easy from Eq. \eqref{EqTemO0} to obtain that
\begin{equation}
  \sum_ig_i^{(0)}=\sigma T,\qquad \sum_ig_i^{(1)}=-\frac{\delta_t}{2}Q_i^{(1)}, \qquad \sum_ig_i^{(k)}=0~(k>1)~.
\end{equation}
Along with these derived equations, by taking summations of Eqs. \eqref{EqTemO1} and \eqref{Eqt2} over $i$, the macroscopic equations at the $t_1=\lambda t$ and $t_2=\lambda^2 t$ time scales can be obtained
\begin{subequations}
   \begin{equation}\label{TemEqt1}
     \partial_{t_1}(\sigma T)+\nabla_1\cdot(T\bm{u})=Q^{(1)},
   \end{equation}
   \begin{equation}\label{TemEqt2}
     \begin{split}
     \partial_{t_2}(\sigma T)+\left(1-\frac{1}{2\tau_T}\right)&\nabla_1\cdot\left(\sum_i\bm{c}_ig_i^{(1)}\right)+\frac{1}{2\tau_T}\nabla_1\cdot(\sigma c_s^2 B\delta_t\nabla_1T)\\&\quad +\left(1-\frac{1}{2\tau_T}\right)\frac{\delta_t}{2}\nabla_1\cdot\biggl[T\bm{F}^{(1)}+\frac{\varepsilon p \nabla_1T}{\rho_0}+\left(\frac{1}{\varepsilon}-\frac{1}{\sigma}\right)\bm{u}\bm{u}\cdot\nabla_1 T+\frac{\bm{u}}{\sigma}Q^{(1)}\biggr]=0.
     \end{split}
   \end{equation}
\end{subequations}

From Eq. \eqref{EqTemO1}, $g_i^{(1)}$ can be draw out to further compute $\sum_i\bm{c}_ig_i^{(1)}$. After some tedious but standard algebraic manipulations, we can get
\begin{equation}\label{TfluxEq}
  \begin{split}
   \sum_i\bm{c}_ig_i^{(1)}=-\tau_T\delta_t\biggl[\partial_{t_1}(T\bm{u})+\nabla_1\cdot\left(\frac{T\bm{u}\bm{u}}{\varepsilon}\right)
   +\nabla_1\left(\frac{\varepsilon pT}{\rho_0}\right)-\left(1-\frac{1}{2\tau_T}\right)\left(T\bm{F}^{(1)}+\frac{\varepsilon p \nabla_1T}{\rho_0}\right)\\
  -\left(1-\frac{1}{2\tau_T}\right)\left(\frac{1}{\varepsilon}-\frac{1}{\sigma}\right)\bm{u}\bm{u}\cdot\nabla_1 T-\left(1-\frac{1}{2\tau_T}\right)\frac{\bm{u}}{\sigma}Q^{(1)}\biggr]+ c_s^2\sigma\left(B-\tau_T\right)\delta_t\nabla_1T.
  \end{split}
\end{equation}
With the aid of the incompressible Navier-Stokes equations at $t_1$ time scale (see \ref{Appen:MRTMod})
\begin{subequations}
\begin{equation}
   \nabla_1\cdot\bm{u}=0,
\end{equation}
  \begin{equation}
   \partial_{t_1}\bm{u}+\nabla_1\cdot\left(\frac{\bm{u}\bm{u}}{\varepsilon}\right)=-\frac{1}{\rho_0}\nabla_1(\varepsilon p)+\bm{F}^{(1)},
  \end{equation}
\end{subequations}
and Eq. \eqref{TemEqt1}, the first two terms in the square bracket on the right hand side of Eq. \eqref{TfluxEq} can be evaluated. After some standard algebra, it can give us
\begin{equation}\label{EqTuPl}
  \begin{split}
  \partial_{t_1}(T\bm{u})+\nabla_1\cdot\left(\frac{T\bm{u}\bm{u}}{\varepsilon}\right)&=
  \partial_{t_1}(T\bm{u})+\nabla_1\cdot\left(\frac{T\bm{u}\bm{u}}{\sigma}\right)+\left(\frac{1}{\varepsilon}-\frac{1}{\sigma}\right)
  \nabla_1\cdot\left(T\bm{u}\bm{u}\right)\\
  &=T\left[\partial_{t_1}\bm{u}+\nabla_1\cdot\left(\frac{\bm{u}\bm{u}}{\varepsilon}\right)\right]
  +\frac{\bm{u}}{\sigma}\bigl[\partial_{t_1}(\sigma T)+\nabla_1\cdot(T\bm{u})\bigr]+\left(\frac{1}{\varepsilon}-\frac{1}{\sigma}\right)
  \bm{u}\bm{u}\cdot\nabla_1 T\\
  &=T\bm{F}^{(1)}-\frac{ T\nabla_1(\varepsilon p)}{\rho_0}+\frac{\bm{u}}{\sigma}Q^{(1)}+\left(\frac{1}{\varepsilon}-\frac{1}{\sigma}\right)
  \bm{u}\bm{u}\cdot\nabla_1 T.
  \end{split}
\end{equation}
Further, application of Eq. \eqref{EqTuPl} to Eq. \eqref{TfluxEq} yields
\begin{equation}\label{TfFintemEq}
   \sum_i\bm{c}_ig_i^{(1)}=-\frac{\delta_t}{2}\left(T\bm{F}^{(1)}+\frac{\varepsilon p \nabla_1T}{\rho_0}\right)-\frac{\delta_t}{2}\left(\frac{1}{\varepsilon}-\frac{1}{\sigma}\right)\bm{u}\bm{u}\cdot\nabla_1 T-\frac{\delta_t}{2}\frac{\bm{u}}{\sigma}Q^{(1)}+c_s^2\sigma\left(B-\tau_T\right)\delta_t\nabla_1 T.
\end{equation}
As a consequence, Eq. \eqref{TemEqt2} can be rewritten as
\begin{equation}\label{TemTerEqt2}
  \partial_{t_2}(\sigma T)+\nabla_1\cdot\left[\sigma c_s^2\left(B-\tau_T+\frac{1}{2}\right)\delta_t\nabla_1T\right]=0.
\end{equation}

Combining the derived equations at $t_1$ and $t_2$ time scales (Eqs. \eqref{TemEqt1} and \eqref{TemTerEqt2}), the final CDE can be recovered
\begin{equation}\label{FinalCDE}
  \partial_t\left(\sigma T\right)+\nabla\cdot\left(T\bm{u}\right)=\nabla\cdot\left(\alpha_e\nabla T\right)+Q,
\end{equation}
where $\alpha_e$ is the effective thermal diffusivity and determined as
\begin{equation}\label{Eq:Therdiff}
   \alpha_e=\sigma c_s^2\left(\tau_T- B-\frac{1}{2}\right)\delta_t.
\end{equation}
Under incompressible flows condition, i.e., $\nabla\cdot\bm{u}=0$, it should be seen that the derived CDE \eqref{FinalCDE} is just Eq. \eqref{EnEq3}.

From the presented Chapman-Enskog analysis, it is clearly seen that the macroscopic equations Eq. \eqref{GoverneEq} for convection heat transfer in porous media can be exactly recovered from the present modified LBGK model. Now, some comments on the present modified LBGK model are in order. First, the fluid viscosity and the thermal diffusivity are not determined only by the relaxation times. Thus, as compared with the non-modified LBGK model, the present LBGK model can achieve better numerical stability and boarder range of Prandtl number values in the simulations. Second, the EDF $g_i^{eq}$(Eq. \eqref{TeEDFEQ}) in the present LBGK model is nonlinear in terms of $\bm{u}$, and thus avoids the numerical diffusion coefficient resulted in the recovered CDE. Third, the temperature gradient $\nabla T$ appearing in $g_i^{(eq)}$ and $P_i$ can be computed by a local scheme instead of a finite difference scheme (as will be presented in the forthcoming section). Thus, the locality of collision process (Eq. \eqref{TEqColli}) of the present model is preserved. Finally, as $\varepsilon\rightarrow1$, one can see that the macroscopic equation Eq. \eqref{GoverneEq} will be that for free convection heat transfer without porous media. Accordingly, the present LBGK model is actually reduced to that for thermal convective flows in the absence of porous media.

\subsection{Local scheme for the shear rate and the temperature gradient}
In the framework of LBM, it is direct to use the traditionally nonlocal finite-difference schemes to compute spatial gradients. However, the shear rate and the temperature gradient can be locally calculated by the nonequilibrium part of the distribution function \cite{Chai13,Boyd06,YongL12} without the influence of porous media. In this work, we will provide a local computational scheme for these two gradient terms in the presence of porous media, which has not been reported in the literature yet.

First, the local computing scheme for the shear rate is derived. Reviewing Eq. \eqref{Eq:Shear} in \ref{Appen:MRTMod}:
\begin{equation*}
  \sum_i\bm{c}_i\bm{c}_if_i^{(1)}=c_s^2\rho_0(A-\tau_f)\delta_t\bm{S}_1-
  \frac{\delta_t}{2}\rho_0\left(\frac{\bm{u}\bm{F}^{(1)}}{\varepsilon}+\frac{\bm{F}^{(1)}\bm{u}}{\varepsilon}\right),
\end{equation*}
and multiplying $\xi$ on its both sides with the relations $\xi f_i^{(1)}=f_i-f_i^{(0)}+O(\xi^2)$ and $f_i^{(0)}=f_i^{e(0)}$, we can obtain that
\begin{equation}\label{ShearEqCom}
 \bm{S}=\frac{\sum_i\bm{c}_i\bm{c}_i\left[f_i-f_i^{e(0)}\right]+\frac{\delta_t}{2}\rho_0\left(\frac{\bm{uF}}{\varepsilon}
 +\frac{\bm{Fu}}{\varepsilon}\right)}{c_s^2\rho_0\left(A-\tau_f\right)\delta_t},
\end{equation}
where $f_i^{e(0)}$ is given by Eq. \eqref{EQFIN}, and its velocity moments can be found in Eq. \eqref{EqZEqu}.

Next, we turn to the computational scheme for the temperature gradient $\nabla T$. Similar to the shear rate, multiplying Eq. \eqref{TfFintemEq} with $\lambda$ on both sides and recognizing that $\lambda g_i^{(1)}=g_i-g_i^{(0)}+O(\lambda^2)$ and $g_i^{(0)}=g_i^{e(0)}$, we can get
\begin{equation}\label{SonaTtem}
  \sum_i\bm{c}_i\left[g_i-g_i^{e(0)}\right]+\frac{\delta_t}{2}\left(T\bm{F}+\frac{\bm{u}}{\sigma}Q\right)=-\frac{\varepsilon p\delta_t}{2\rho_0}\nabla T-\frac{\delta_t}{2}\left(\frac{1}{\varepsilon}-\frac{1}{\sigma}\right)\bm{u}\bm{u}\cdot\nabla T+c_s^2\sigma \left(B-\tau_T\right)\delta_t \nabla T.
\end{equation}
Due to the tensor $\bm{uu}$ contained in the above equation, we cannot directly obtain $\nabla T$. Fortunately, based on the equality $\nabla T=\bm{I}\cdot \nabla T$, Eq. \eqref{SonaTtem} can be put into the following tensor form
\begin{equation}\label{SonaT}
  \left[-\frac{\varepsilon p\delta_t}{2\rho_0}\bm{I}-\frac{\delta_t}{2}\left(\frac{1}{\varepsilon}-\frac{1}{\sigma}\right)\bm{u}\bm{u}+c_s^2\sigma \left(B-\tau_T\right)\delta_t\bm{I}\right]\cdot\nabla T= \sum_i\bm{c}_i\left[g_i-g_i^{e(0)}\right]+\frac{\delta_t}{2}\left(T\bm{F}+\frac{\bm{u}}{\sigma}Q\right).
\end{equation}
Now we can obtain $\nabla T$ by solving Eq. \eqref{SonaT}. Mathematically, this can be done by solving the following equation in the matrix form
\begin{equation}\label{TeEqCom}
  \langle M \rangle \langle\nabla T\rangle=\langle N \rangle,
\end{equation}
where $\langle M \rangle$, $\langle\nabla T\rangle$ and $\langle N \rangle$ are three matrixes, which are respectively formed by the elements of tensors $-\frac{\varepsilon p\delta_t}{2\rho_0}\bm{I}-\frac{\delta_t}{2}\left(\frac{1}{\varepsilon}-\frac{1}{\sigma}\right)\bm{u}\bm{u}+c_s^2\sigma \left(B-\tau_T\right)\delta_t\bm{I}$, $\nabla T$ and $\sum_i\bm{c}_i\left[g_i-g_i^{e(0)}\right]+\frac{\delta_t}{2}\left(T\bm{F}+\frac{\bm{u}}{\sigma}Q\right)$
\begin{align}
  \langle M \rangle_{\alpha\beta}&=-\frac{\varepsilon p\delta_t}{2\rho_0}\delta_{\alpha\beta}-\frac{\delta_t}{2}\left(\frac{1}{\varepsilon}-\frac{1}{\sigma}\right)u_\alpha u_\beta
  +c_s^2\sigma \left(B-\tau_T\right)\delta_t\delta_{\alpha\beta}, \label{MatrM}\\
  \langle \nabla T\rangle_\beta&=\partial_\beta T,\label{MatrTe}\\
  \langle N \rangle_\alpha&=\sum_i\bm{c}_{i\alpha}\left[g_i-g_i^{e(0)}\right]+\frac{\delta_t}{2}\left(TF_\alpha
  +\frac{u_\alpha}{\sigma}Q\right). \label{MatrN}
\end{align}
By some simple computations, the determinant of the matrix $\langle M \rangle$ can be obtained
\begin{equation}\label{Eq:detMatr}
  det(\langle M \rangle)=\left[c_s^2\sigma \left(B-\tau_T\right)\delta_t-\frac{\varepsilon p\delta_t}{2\rho_0}\right]\left[c_s^2\sigma \left(B-\tau_T\right)\delta_t-\frac{\varepsilon p\delta_t}{2\rho_0}-\frac{\delta_t}{2}\left(\frac{1}{\varepsilon}-\frac{1}{\sigma}\right)\mid\bm{u}\mid^2\right].
\end{equation}
From Eq. \eqref{Eq:Therdiff}, one can see that $B<\tau_T$. Furthermore, in the simulations of this paper, $\sigma$ is set to be $1$, while $\varepsilon$ is varied within $0<\varepsilon<1$. Hence, $1/\varepsilon-1/\sigma>0$. With these inequalities, we can conclude from Eq. \eqref{Eq:detMatr} that $det(\langle M \rangle)>0$, that is, $\langle M \rangle$ is an invertible linear matrix. Consequently, $\langle \nabla T\rangle$ can be determined by $\langle \nabla T\rangle=\langle M \rangle^{-1}\langle N \rangle$.

It is clear that the above formulae to compute the shear rate and the temperature gradient are local schemes while no finite difference schemes are employed. This ensures that the collision process of the present LBGK model can be locally performed. A few remarks regarding the local schemes for computing the shear rate and the temperature gradient are given here. First of all, the local computing schemes are still available for the case without porous media only by taking the porosity $\varepsilon=1$. Second, as $A$ and $B$ are equal to zero, the shear rate $\bm{S}$ disappears from the present model, however, the temperature gradient $\nabla T$ still needs to be computed since it remains in the source term $P_i$. Third, when $\varepsilon=\sigma$ or $\bm{u}\cdot\nabla T=0$ (this corresponds to the pure diffusion system), the second term on the right hand side of Eq. \eqref{SonaTtem} equals to zero. As such, the temperature gradient can be directly obtained, and is given by
\begin{equation}
  \nabla T=\frac{\sum_i\bm{c}_i\left[g_i-g_i^{e(0)}\right]
  +\frac{\delta_t}{2}\left(T\bm{F}+\frac{\bm{u}}{\sigma}Q\right)}{\left[c_s^2\sigma\left(B-\tau_T\right)-\frac{\varepsilon p}{2\rho_0}\right]\delta_t}.
\end{equation}

\subsection{Boundary conditions\label{Sec:NEES}}
In this work, to specify physical boundary conditions for the distribution functions, we will extend the nonequilibrium extrapolation scheme (NEES) \cite{GuoNEES} to the present LBGK model due to its simplicity, second-order accuracy, capability for different boundary conditions, and good robustness.

The basic idea of NEES is to decompose the distribution function at a boundary node $\bm{x}_b$ into the equilibrium part and the nonequilibrium part. Specifically, for the density distribution function $f_i$, the decomposition at $\bm{x}_b$ is described as
\begin{equation}
  f_i(\bm{x}_b,t)=f_i^{(eq)}(\bm{x}_b,t)+f_i^{(neq)}(\bm{x}_b,t).
\end{equation}
The physical boundary conditions are embodied through the equilibrium part, while the nonequilibrium part is approximated by certain extrapolation schemes. With the same idea in the present LBGK model, the NEES is first generalized for the velocity boundary condition where $\bm{u}(\bm{x}_b,t)$ is known but $p(\bm{x}_b,t)$ is unknown. From Eqs. \eqref{EQFIN} and \eqref{EQTSHIN}, one can see that the pressure $p(\bm{x}_b,t)$ and the shear rate $\bm{S}(\bm{x}_b,t)$ need to be evaluated to determine the equilibrium part. For this purpose, we use the pressure $p(\bm{x}_f,t)$ and the shear rate $\bm{S}(\bm{x}_f,t)$ at $\bm{x}_f$ to respectively replace $p(\bm{x}_b,t)$ and $\bm{S}(\bm{x}_b,t)$, and then the EDF $f_i^{(eq)}(\bm{x}_b,t)$ at $\bm{x}_b$ is approximated as
\begin{equation}\label{FNEESEQFIN}
  f_i^{(eq)}(\bm{x}_b, t)=
  \begin{cases}
     \rho_0-(1-\omega_0)\frac{\varepsilon p(\bm{x}_f,t)}{c_s^2}+\rho_0s_0(\bm{u}(\bm{x}_b,t))+\rho_0r_0(\bm{u}(\bm{x}_f,t)),\quad i=0\\
     \omega_i\frac{\varepsilon p(\bm{x}_f,t)}{c_s^2}+\rho_0s_i(\bm{u}(\bm{x}_b,t))+\rho_0r_i(\bm{u}(\bm{x}_f,t)),\qquad\qquad\qquad\; i\neq0
  \end{cases}
\end{equation}
where $\bm{x}_f$ is the nearest neighbor fluid node away from $\bm{x}_b$, and
\begin{equation}
   s_i(\bm{u}(\bm{x}_b,t))=\omega_i\left[\frac{\bm{c}_i\cdot\bm{u}(\bm{x}_b,t)}{c_s^2}+\frac{\bm{u}(\bm{x}_b,t)\bm{u}(\bm{x}_b,t):(\bm{c}_i\bm{c}_i-c_s^2\bm{I})}{2\varepsilon c_s^4}\right],\;\; r_i(\bm{u}(\bm{x}_f,t))=\omega_i\frac{A\delta_t\bm{S}(\bm{x}_f,t):(\bm{c}_i\bm{c}_i-c_s^2\bm{I})}{2c_s^2}.
\end{equation}
Note that $\bm{S}(\bm{x}_f,t)$ can be calculated directly according to Eq. \eqref{ShearEqCom}. For the nonequilibrium part at $\bm{x}_b$, $f_i^{(neq)}(\bm{x}_b, t)$ is approximated by the nonequilibrium part of $f_i(\bm{x}_f, t)$
\begin{equation}
  f_i^{(neq)}(\bm{x}_b, t)=f_i(\bm{x}_f, t)-f_i^{(eq)}(\bm{x}_f, t).
\end{equation}
As a whole, the distribution function at the boundary node $\bm{x}_b$ is calculated by
\begin{equation}\label{NEESVel}
   f_i(\bm{x}_b, t)=f_i^{(eq)}(\bm{x}_b, t)+\left[f_i(\bm{x}_f, t)-f_i^{(eq)}(\bm{x}_f, t)\right].
\end{equation}

Thermal boundary conditions with the NEES can be treated in a similar way. Two cases of temperature boundary conditions are taken into account. For the case that the temperature $T(\bm{x}_b,t)$ at $\bm{x}_b$ is only known, determination of $\nabla T(\bm{x}_b,t)$ is needed for $g_i^{(eq)}(\bm{x}_b, t)$ (see Eq. \eqref{TeEDFEQ}). By approximating $\nabla T(\bm{x}_b,t)$ with $\nabla T(\bm{x}_f,t)$ and combining the velocity boundary condition, the equilibrium part $g_i^{(eq)}(\bm{x}_b, t)$ is specified as
\begin{equation}\label{TeEBounDFEQ}
\begin{split}
  g_i^{(eq)}=\omega_i T(\bm{x}_b,t) \left[\sigma+\frac{\bm{c}_i\cdot\bm{u}(\bm{x}_b,t)}{c_s^2}+\frac{\bm{u}(\bm{x}_b,t)\bm{u}(\bm{x}_b,t):(\bm{c}_i\bm{c}_i-c_s^2\bm{I})}{2\varepsilon c_s^4}\right]\\
  + \varpi_iT(\bm{x}_b,t)\frac{\varepsilon p(\bm{x}_f,t)}{c_s^2\rho_0}+\sigma\omega_i B \delta_t\bm{c}_i\cdot\nabla T(\bm{x}_f,t).
\end{split}
\end{equation}
As such, the temperature distribution function $g_i(\bm{x}_b,t)$ is calculated by
\begin{equation}\label{TNEESAp}
   g_i(\bm{x}_b, t)=g_i^{(eq)}(\bm{x}_b, t)+\left[g_i(\bm{x}_f, t)-g_i^{(eq)}(\bm{x}_f, t)\right].
\end{equation}
For the other case when the temperature gradient $\nabla T(\bm{x}_b,t)$ is known, the local temperature $T(\bm{x}_b,t)$ at the boundary node $\bm{x}_b$ can be taken from the following second-order finite difference approximation
\begin{equation}\label{TNEESApGra}
  T(\bm{x}_b,t)=\frac{4T(\bm{x}_f,t)-T(\bm{x}_{ff},t)-2\Delta\cdot \nabla T(\bm{x}_b,t)}{3},
\end{equation}
where $x_{ff}$ is the nearest fluid node of $\bm{x}_f$ such that $\Delta=\bm{x}_f-\bm{x}_b=\bm{x}_{ff}-\bm{x}_f$. Hence, the EDF $g_i^{(eq)}(\bm{x}_b, t)$ at $\bm{x}_b$ can be approximated by
\begin{align}
  g_i^{(eq)}(\bm{x}_b,t)&=\omega_i \frac{4T(\bm{x}_f,t)-T(\bm{x}_{ff},t)-2\Delta\cdot \nabla T(\bm{x}_b,t)}{3} \left[\sigma+\frac{\bm{c}_i\cdot\bm{u}(\bm{x}_b, t)}{c_s^2}+\frac{\bm{u}(\bm{x}_b, t)\bm{u}(\bm{x}_b, t):(\bm{c}_i\bm{c}_i-c_s^2\bm{I})}{2\varepsilon c_s^4}\right]\notag\\
   &\qquad+\varpi_iT(\bm{x}_b, t)\frac{\varepsilon p(\bm{x}_b, t)}{c_s^2\rho_0}+\sigma\omega_i B \delta_t\bm{c}_i\cdot\nabla T(\bm{x}_b, t),
\end{align}
and then the distribution function $g_i(\bm{x}_b, t)$ at $\bm{x}_b$ is calculated according Eq. \eqref{TNEESAp}. Finally, we would like to point out that other boundary schemes can also be developed in a similar way for the velocity and thermal boundary conditions in the present LBGK model.

\section{Results and discussions}
\label{Sec:Numer}
In this section, we shall validate the proposed LBGK model by some two-dimensional convective heat transfer problems in porous media, and compare the simulation results with analytical and previous numerical results. To ensure the numerical results to reach the steady state, the following convergent criterion is used in the simulations
\begin{equation}
  \frac{\sum_{\bm{x}}\parallel \phi(\bm{x},t)-\phi(\bm{x},t-100\delta_t)\parallel}{\sum_{\bm{x}}\parallel \phi(\bm{x},t)\parallel}<1.0\times10^{-6},
\end{equation}
where $\phi$ is $\bm{u}$ or $T$, and its norm is computed respectively with the $L^2$ and $L^1$ -norm. Unless otherwise specified, values of the following parameters are used: $\rho_0=1.0$, $J_e=1.0$, $\sigma=1.0$, and $\alpha_e/\alpha_f=1.0$ ($\alpha_f$ is the thermal diffusivity of the fluid), and the dimensionless relaxation times $\tau_f$ and $\tau_T$ are set to be unity. Due to the positivity of the viscosity and effective thermal diffusivity, it should be clear that values of $A$ and $B$ should satisfy the constraint of $A<\tau_f-1/2$ and $B<\tau_T-1/2$. As noted for the relaxation times elsewhere \cite{Liu14,Dixit06}, $A$ and $B$ can also be determined via $Pr$, $Ra$ and $Ma$, where $Ma$ is the Mach number and defined as $Ma=U/c_s$ ($U=\sqrt{g\beta\Delta T L}$ is the characteristic velocity), which are given by
\begin{equation}\label{EQA&B}
  A=\tau_f-\frac{1}{2}-Ma\frac{L}{\delta_x}\sqrt{\frac{3Pr}{Ra}}, \quad B=\tau_T-\frac{1}{2}-\frac{Ma}{\sigma}\frac{L}{\delta_x}\sqrt{\frac{3}{Pr Ra}}.
\end{equation}
To make the flows fully lie in the incompressible regime, the Mach number should be small and is set to be $0.1$ in this work.

In addition, the accuracy of the present model and the computing scheme for the shear rate and temperature gradient will also be investigated. Correspondingly, the following relative global error is used
\begin{equation}
   E(\Phi)=\frac{\sum_{\bm{x}}\parallel \Phi_a(\bm{x},t)-\Phi_n(\bm{x},t)\parallel}{\sum_{\bm{x}}\parallel \Phi_a(\bm{x},t)\parallel},
\end{equation}
where $\Phi$ represents a scalar or vectorial variable in terms of velocity $\bm{u}$ and temperature $T$, and the subscripts $a$ and $n$ denote the analytical and numerical results.

\subsection{Mixed convection in porous channel}
\begin{figure}[htb!]
\centering
\includegraphics[scale=1.15]{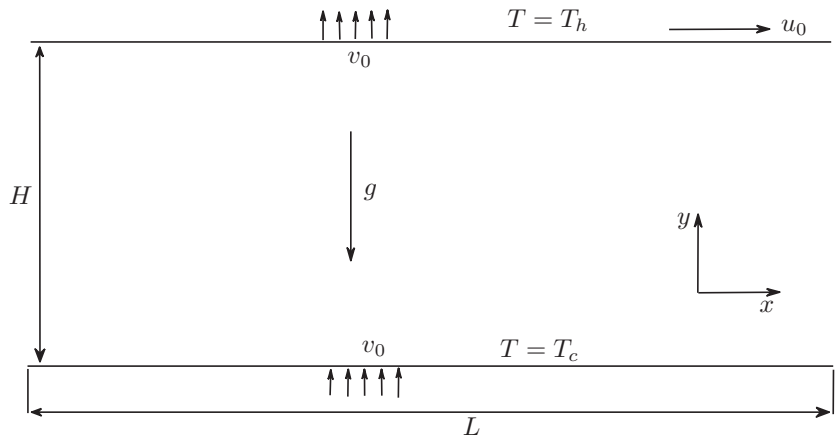}%
\caption{Schematic of mixed convective flow in a porous channel.}
\label{fig:SchTest1}
\end{figure}

The first test problem is the mixed convection flow in a channel filled with a porous medium. The schematic of this problem is shown in Fig. \ref{fig:SchTest1}. The channel of height $H$ and length $L$ is filled with a porous media of porosity $\varepsilon$. The top wall of the channel is held at a high temperature $T_h$ and moves with a uniform velocity $u_0$ along the $x$-direction, and the static bottom wall is held at a low temperature $T_c$. A constant normal flow of fluid is injected through the bottom wall and is withdrawn at the same rate from the top wall. For the injected fluid in the porous channel, it will be sheared by the identical fluid in the normal direction of injection.
If the nonlinear inertia force ($F_\varepsilon=0$) is neglected, the flow at the steady state obeys the following governing equations \cite{Guo05,Liu14}:
\begin{subequations}\label{TestGEq1}
  \begin{equation}
    \frac{u_y}{\varepsilon}\frac{\partial u_x}{\partial y}=\nu_e\frac{\partial^2u_x}{\partial y^2}-\frac{\varepsilon\nu_e}{K}u_x,
  \end{equation}
  \begin{equation}
     \frac{1}{\rho_0}\frac{\partial p}{\partial y}=g\beta(T-T_0)-\frac{\nu}{K}u_y+a_y,
  \end{equation}
  \begin{equation}
    u_y\frac{\partial T}{\partial y}=\nabla\cdot(\alpha_e\nabla T),
  \end{equation}
\end{subequations}
where $u_x$ and $u_y$ are the two components of fluid velocity $\bm{u}$, $T_0=(T_h+T_c)/2$ is the average temperature, and $a_y$ represents the external force in the $y$-direction which is given by
\begin{equation}
  a_y=\frac{\nu}{K}v_0-g\beta\Delta T\left[\frac{\text{exp}(yv_0/\alpha_e)-1}{\text{exp}(Hv_0/\alpha_e)-1}\right].
\end{equation}
The analytical solutions of $\bm{u}$ and $T$ in Eq. \eqref{TestGEq1} are identified as
\begin{subequations}
  \begin{equation}\label{EqVelocity}
     u_x=u_0\text{exp}\left[\zeta_1\left(\frac{y}{H}-1\right)\right]\frac{\text{sinh}(\zeta_2\cdot y/H)}{\text{sinh}(\zeta_2)},\qquad u_y=v_0,
  \end{equation}
  \begin{equation}\label{EqTemper}
     T=T_0+\Delta T\frac{\text{exp}(PrRe\cdot y/H)-1}{\text{exp}(PrRe)-1},
  \end{equation}
\end{subequations}
where $Re$ is the Reynolds number defined by $Re=Hv_0/\nu$, $\Delta T=T_h-T_c$ is the temperature difference, and the two parameters $\zeta_1$ and $\zeta_2$ are respectively given by
\begin{equation}
  \zeta_1=\frac{Re}{2\varepsilon J_e}, \qquad \zeta_2=\frac{1}{2\varepsilon J_e}\sqrt{Re^2+\frac{4\varepsilon^3J_e}{Da}}.
\end{equation}
From Eqs. \eqref{EqVelocity} and \eqref{EqTemper}, the following exact solutions to the gradients of velocity and temperature can also be obtained
\begin{subequations}
  \begin{equation}\label{EqSoluVelocity}
     \frac{\partial u_x}{\partial y}=\frac{u_0}{\text{sinh}(\zeta_2)}\text{exp}\left[\zeta_1\left(\frac{y}{H}-1\right)\right]
     \left[\frac{\zeta_1}{H}\text{sinh}(\zeta_2\cdot y/H)+\frac{\zeta_2}{H}\text{cosh}(\zeta_2\cdot y/H)\right],
  \end{equation}
  \begin{equation}\label{EqSoluTemper}
     \frac{\partial T}{\partial y}=\Delta T\frac{PrRe}{H}\frac{\text{exp}(PrRe\cdot y/H)}{\text{exp}(PrRe)-1}.
  \end{equation}
\end{subequations}

\begin{figure}
\begin{tabular}{cc}
\includegraphics[width=0.49\textwidth,height=0.28\textheight]{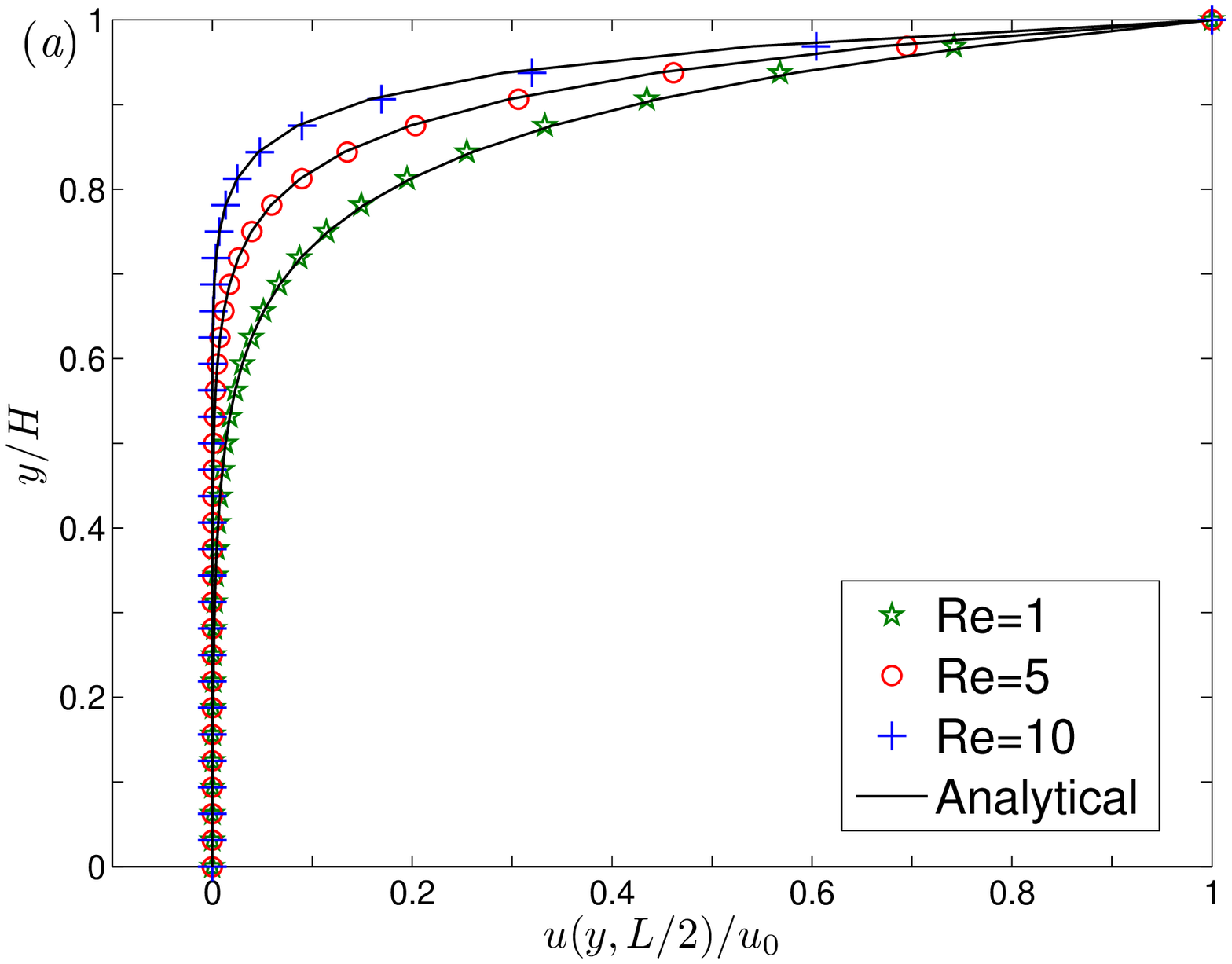}&
\includegraphics[width=0.49\textwidth,height=0.28\textheight]{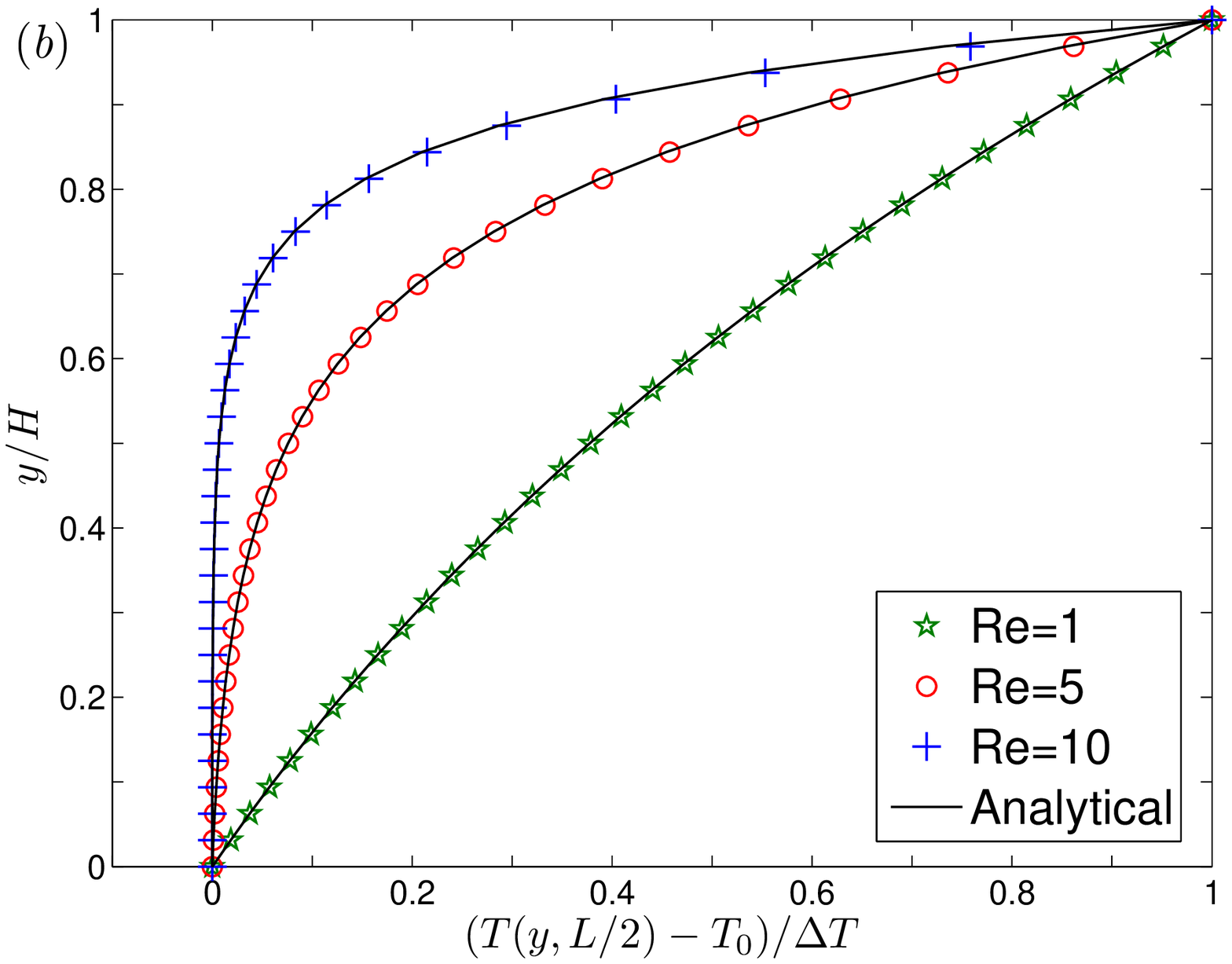}\\
\includegraphics[width=0.49\textwidth,height=0.28\textheight]{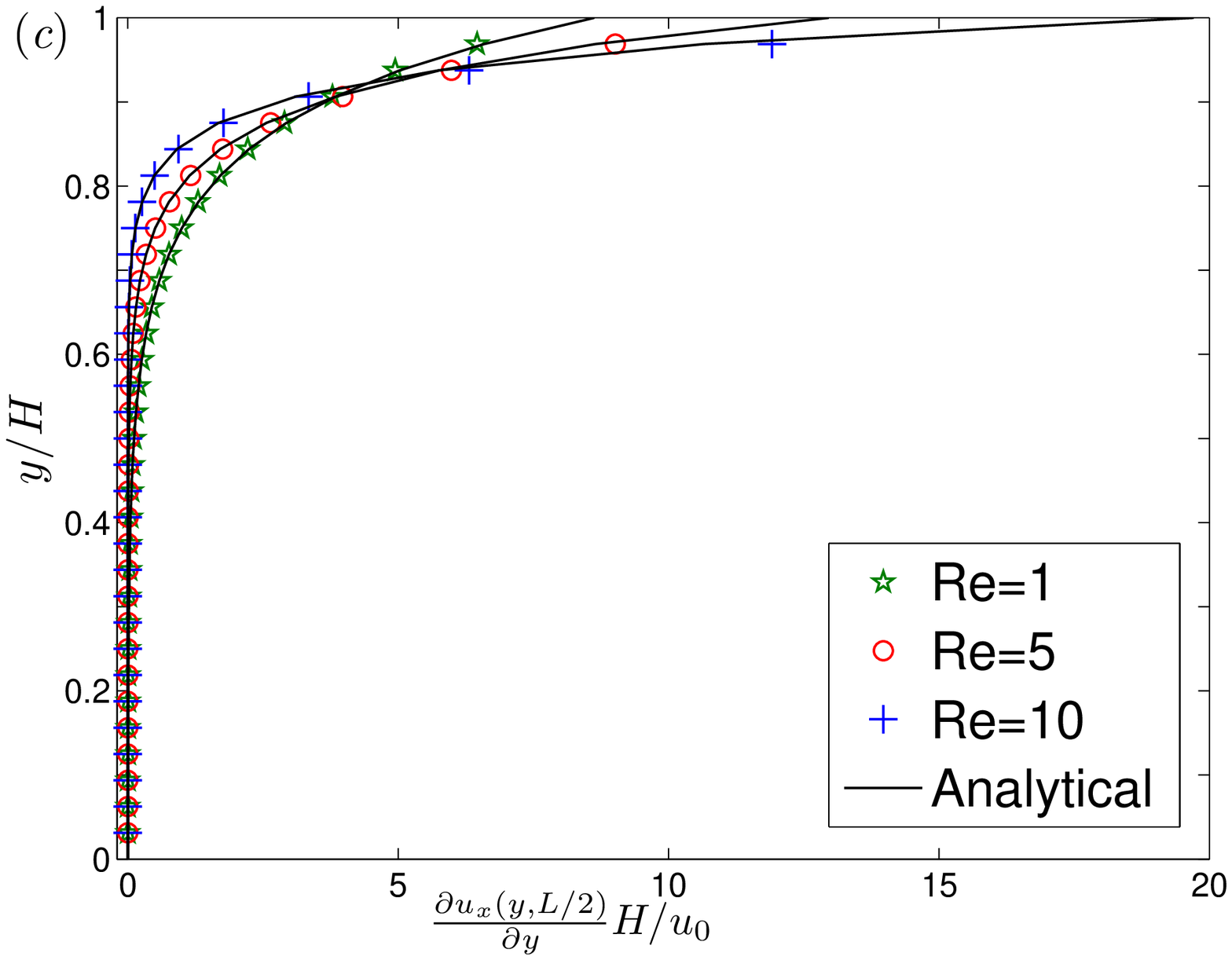}&
\includegraphics[width=0.49\textwidth,height=0.28\textheight]{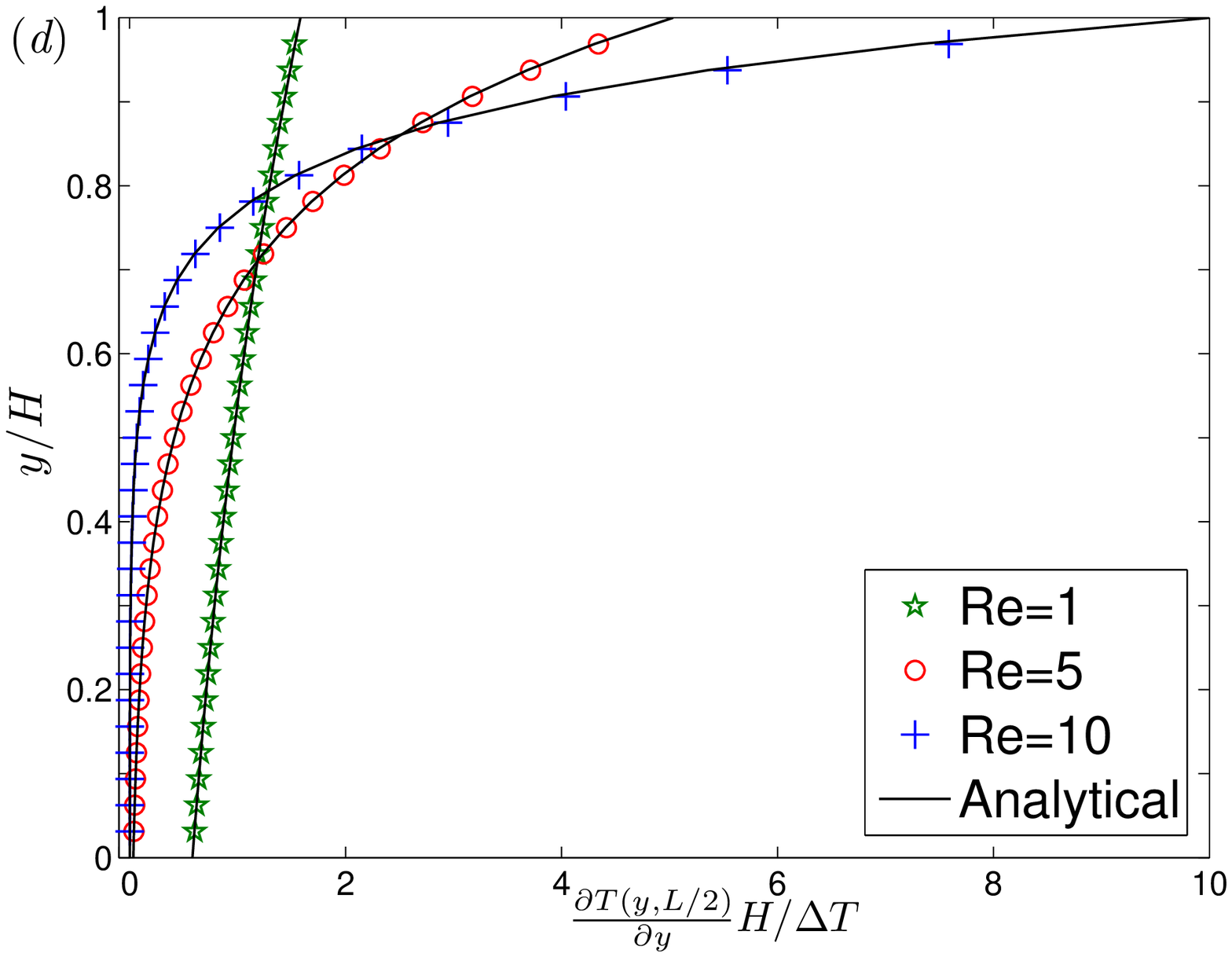}\\
\end{tabular}
\caption{Profiles of (a) velocity; (b) temperature; (c) velocity gradient and (d) temperature gradient for different $Re$ at $\varepsilon=0.6,~Ra=100,~Pr=1,~Da=0.01$ with a mesh size of $32\times 32$. Solid lines: analytical solutions; Symbols: numerical results.}
\label{MixedConFig}
\end{figure}

In the simulations, the computational domain is chosen to be $H\times L=1.0\times 1.0$, and some physical parameters are listed as follows: $\varepsilon=0.6$, $Ra=100$, $Pr_e=1.0$, $Da=0.01$, $u_0=v_0=0.01$. The lattice size is $N_x\times N_y=32\times32$, and the values of $A$ and $B$ are determined by $A=\tau_f-0.5-Hv_0/(c_s^2\delta_t Re)$, and $B=\tau_T-0.5-(\tau_f-0.5-A)/(\sigma Pr Je)$ (or $B=\tau_T-0.5-Hv_0/(\sigma RePrc_s^2\delta_t)$) here. Periodic boundary conditions are employed at the entrance and outlet, and the NEES presented in Sec. \ref{Sec:NEES} is applied to the top and bottom walls for the velocity and temperature boundary conditions. In Fig \ref{MixedConFig}, the comparisons of profiles for the velocity and temperature together with their gradients are summarized under different Reynolds numbers at $Ra=100$, $Pr=1$, $\varepsilon=0.6$ and $Da=0.01$. As can be observed, the present numerical results of velocity and temperature and their gradients agree well with the analytical solutions.

\begin{figure}
\begin{tabular}{cc}
\includegraphics[width=0.5\textwidth,height=0.28\textheight]{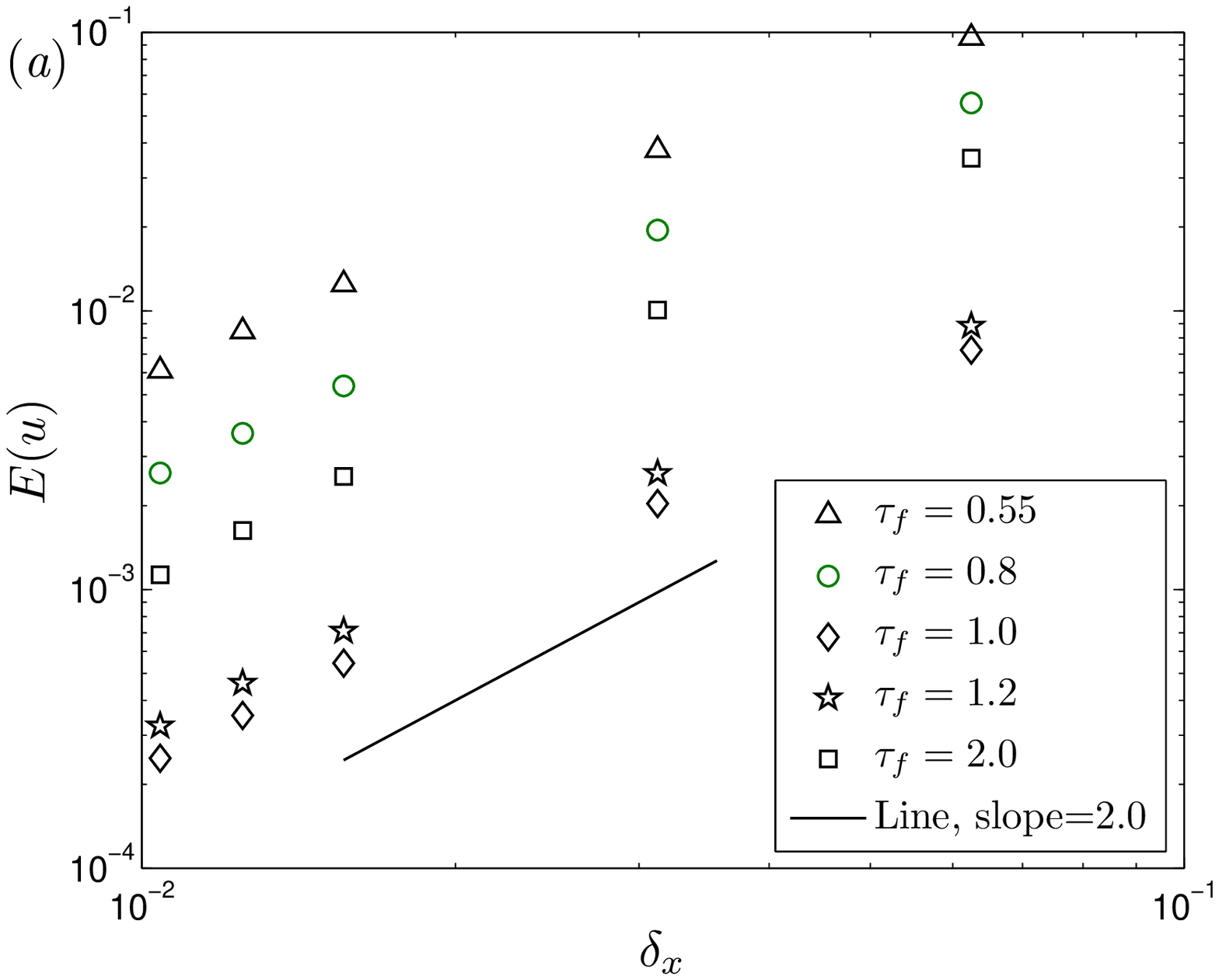}&
\includegraphics[width=0.5\textwidth,height=0.28\textheight]{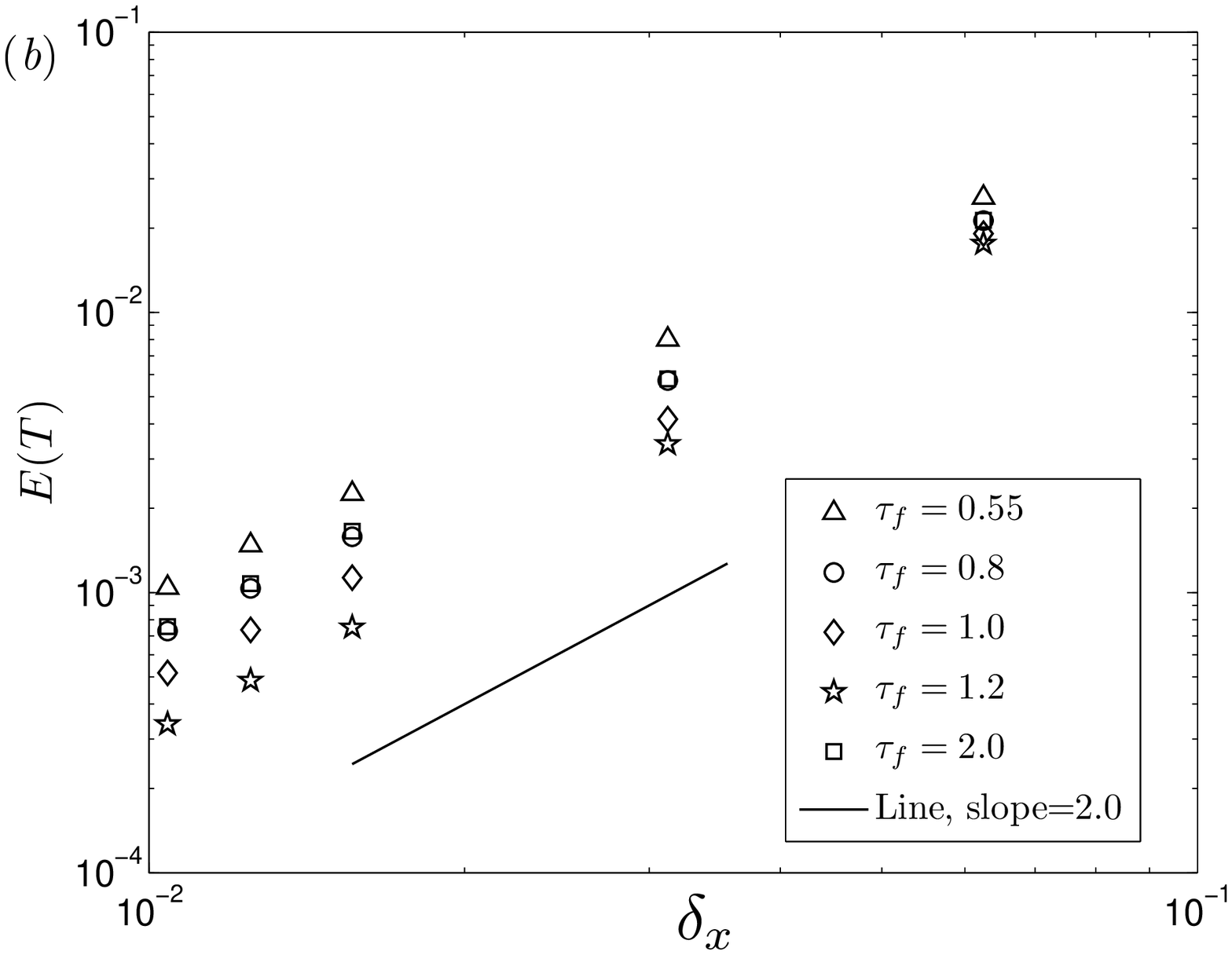}\\
\end{tabular}
\caption{Relative errors of (a) velocity and (b) temperature fields against lattice spacing $\delta_x$ at $Re=5,~\varepsilon=0.6,~Ra=100,~Pr=1,~Da=0.01$. The slope of the inserted line is $2.0$, which indicates that the present models for the flow and temperature fields are of second-order accuracy in space.}
\label{AccurFig}
\end{figure}

\begin{figure}
\begin{tabular}{cc}
\includegraphics[width=0.5\textwidth,height=0.28\textheight]{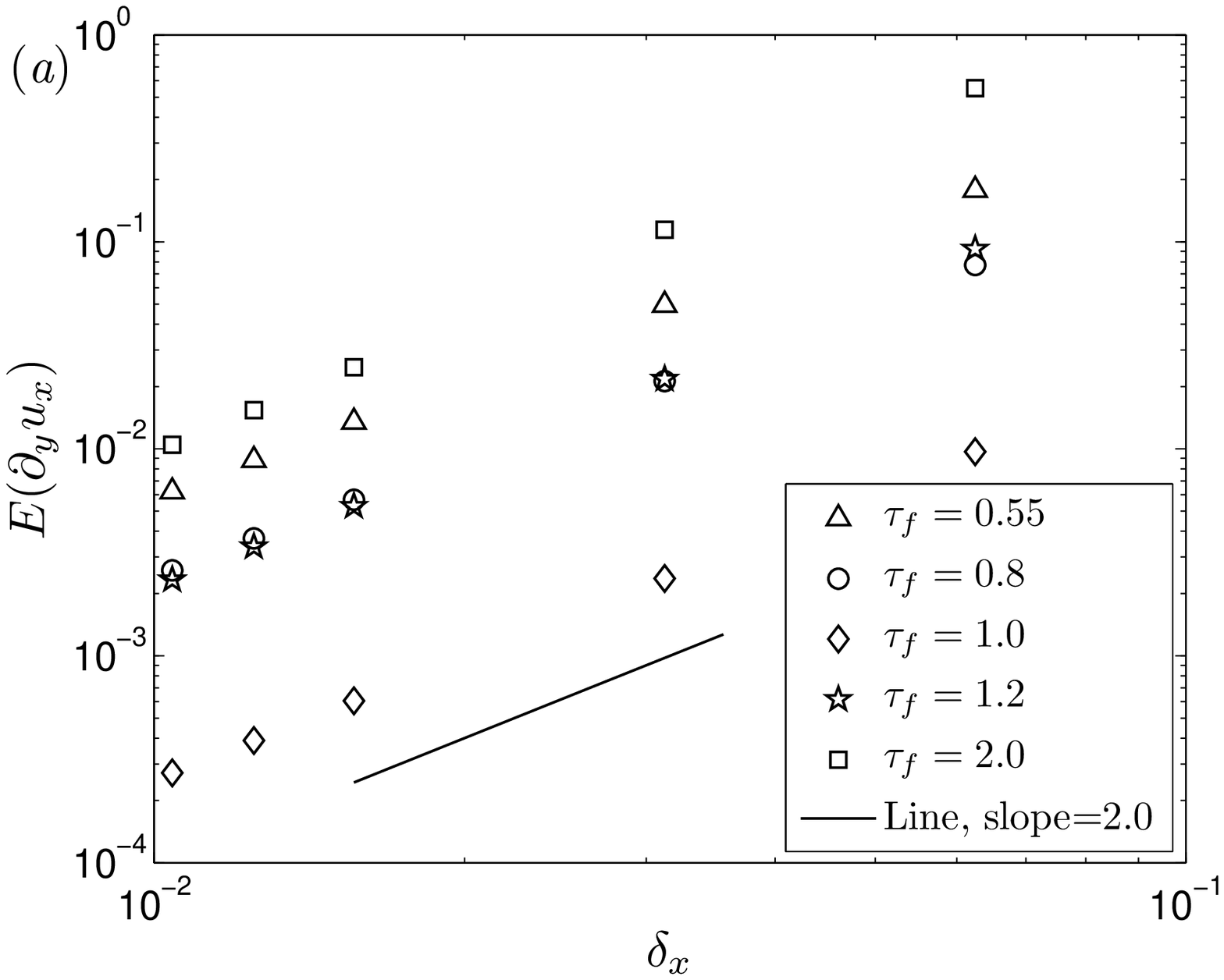}&
\includegraphics[width=0.5\textwidth,height=0.28\textheight]{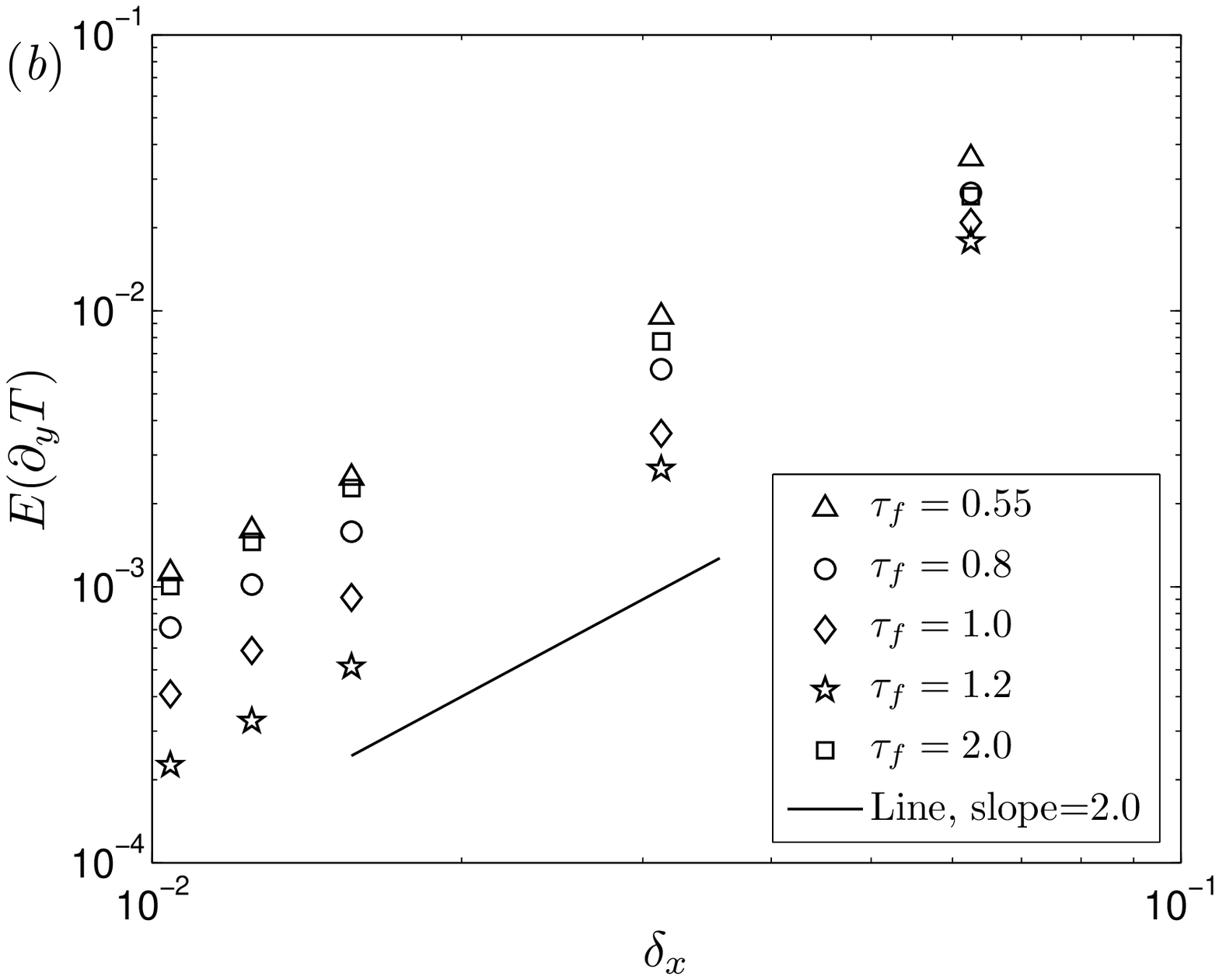}\\
\end{tabular}
\caption{Relative errors of (a) $\frac{\partial u_x}{\partial y}$ and (b) $\frac{\partial T}{\partial y}$ against lattice spacing $\delta_x$ at $Re=5,~\varepsilon=0.6,~Ra=100,~Pr=1,~Da=0.01$. The slope of the inserted line is $2.0$, which indicates that the present models for the flow and temperature fields are of second-order accuracy in space.}
\label{AccurGradFig}
\end{figure}

In what follows, the spatial accuracies of the present model for the flow and temperature equations are tested by computing the relative errors with different lattice sizes. In Fig. \ref{AccurFig}, the relative errors in velocity and temperature fields are presented at $Re=5,~\varepsilon=0.6,~Ra=100,~Pr=1,~Da=0.01$. The results are obtained based on variational lattice spacings from $1/16$ to $1/96$, and five values of relaxation times ($\tau_f=0.55,~0.8,~1.0,~1.2,~2.0$) are used. As seen from the figure, the present model for the flow and temperature fields are second-order accurate in space.
In addition, the local computing schemes for the shear rate tensor and the temperature gradient are evaluated in terms of spatial accuracy. Note that the analytical solutions for the fluid velocity and the temperature have the forms as Eqs. \eqref{EqVelocity} and \eqref{EqTemper}. Thus, the component of $\frac{\partial u_x}{\partial y}$ (and $\frac{\partial T}{\partial y}$) in $\bm{S}$ (and $\nabla T$) is only computed for its relative errors with different lattice sizes, which is plotted in Fig. \ref{AccurGradFig}. The results shown in Fig. \ref{AccurGradFig} clearly indicate that the present local schemes for computing the shear rate tensor and the temperature gradient are of second-order accuracy in space.

\begin{figure}
\begin{tabular}{cc}
\includegraphics[width=0.52\textwidth,height=0.325\textheight]{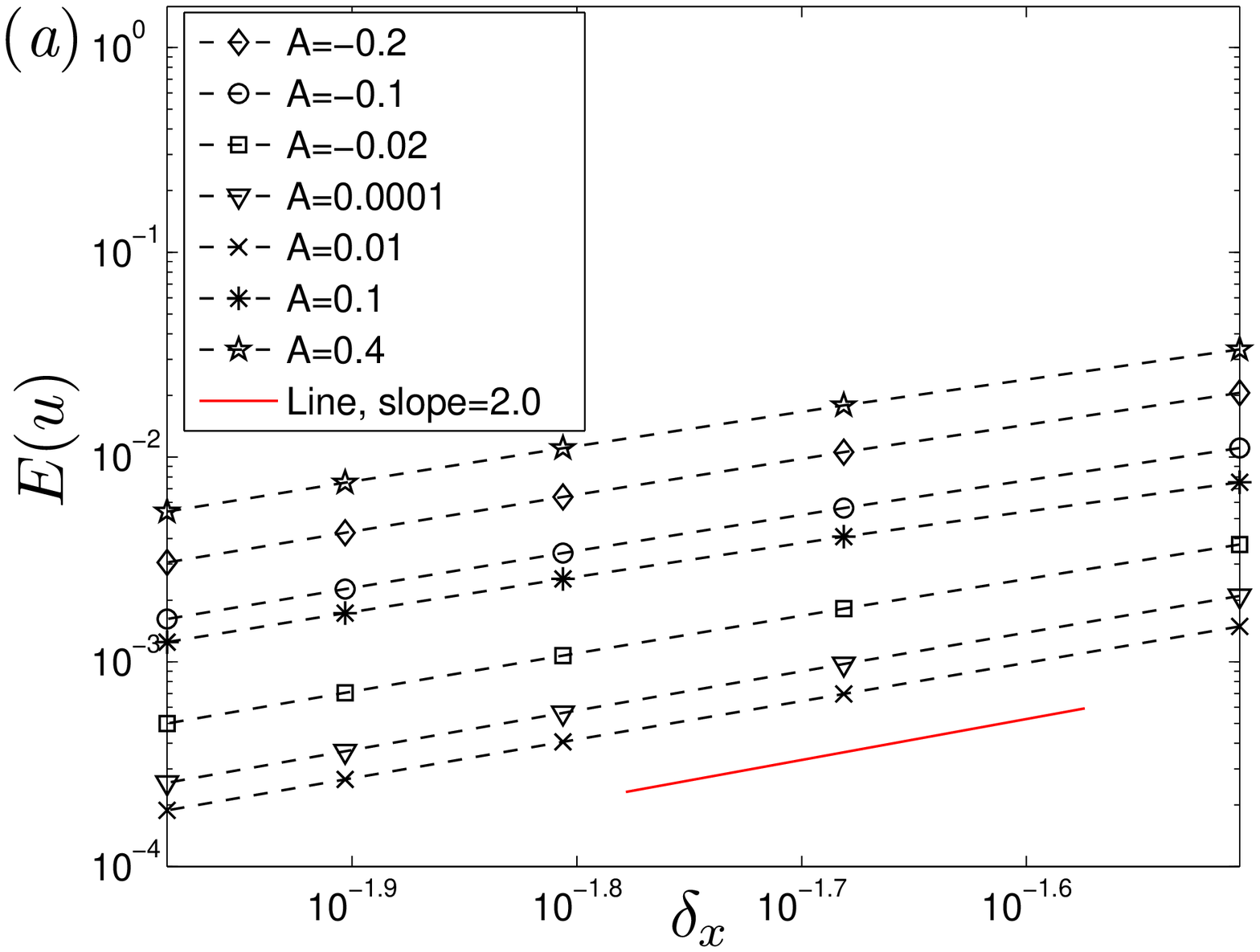}&
\includegraphics[width=0.52\textwidth,height=0.325\textheight]{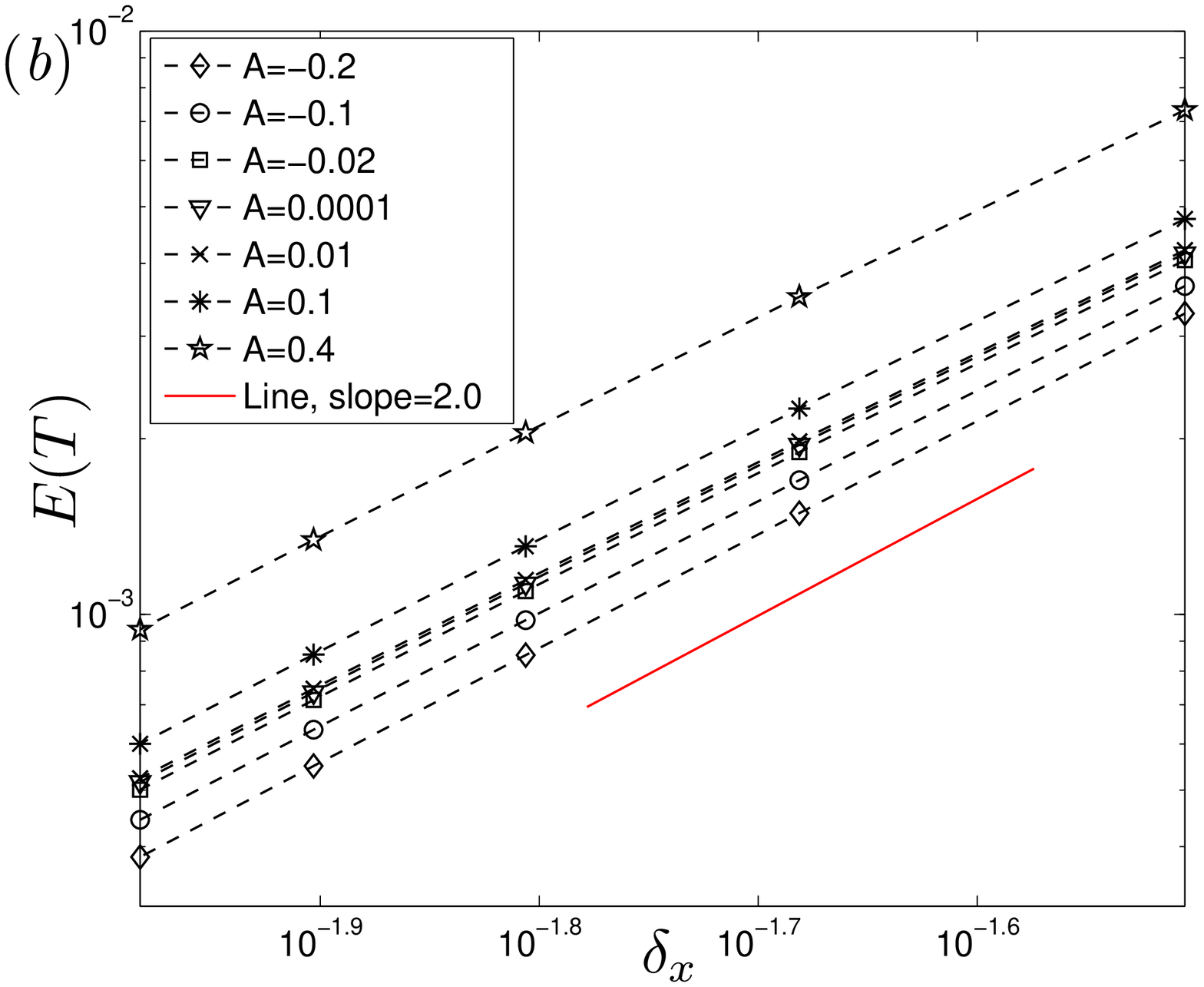}\\
\includegraphics[width=0.52\textwidth,height=0.325\textheight]{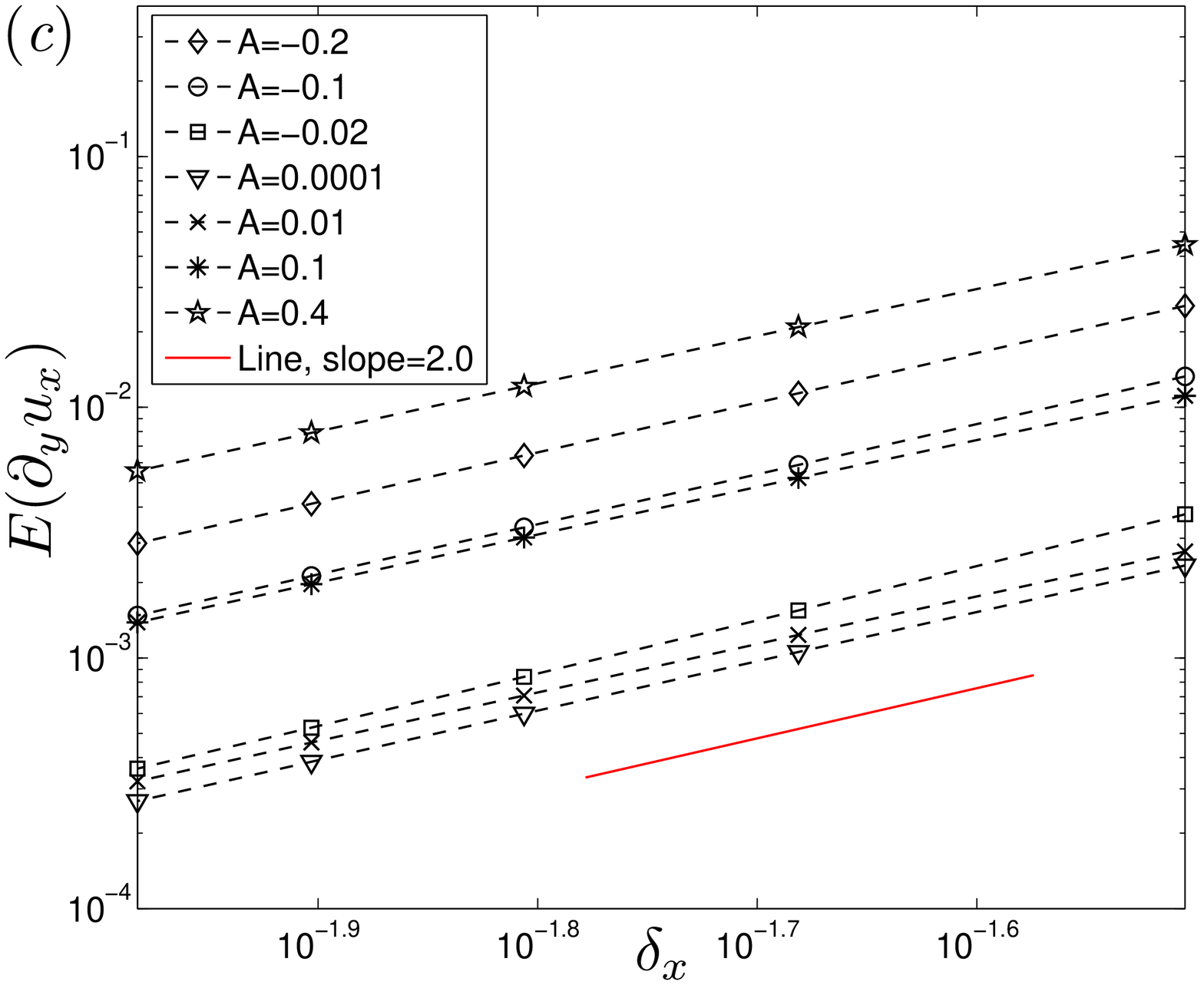}&
\includegraphics[width=0.52\textwidth,height=0.325\textheight]{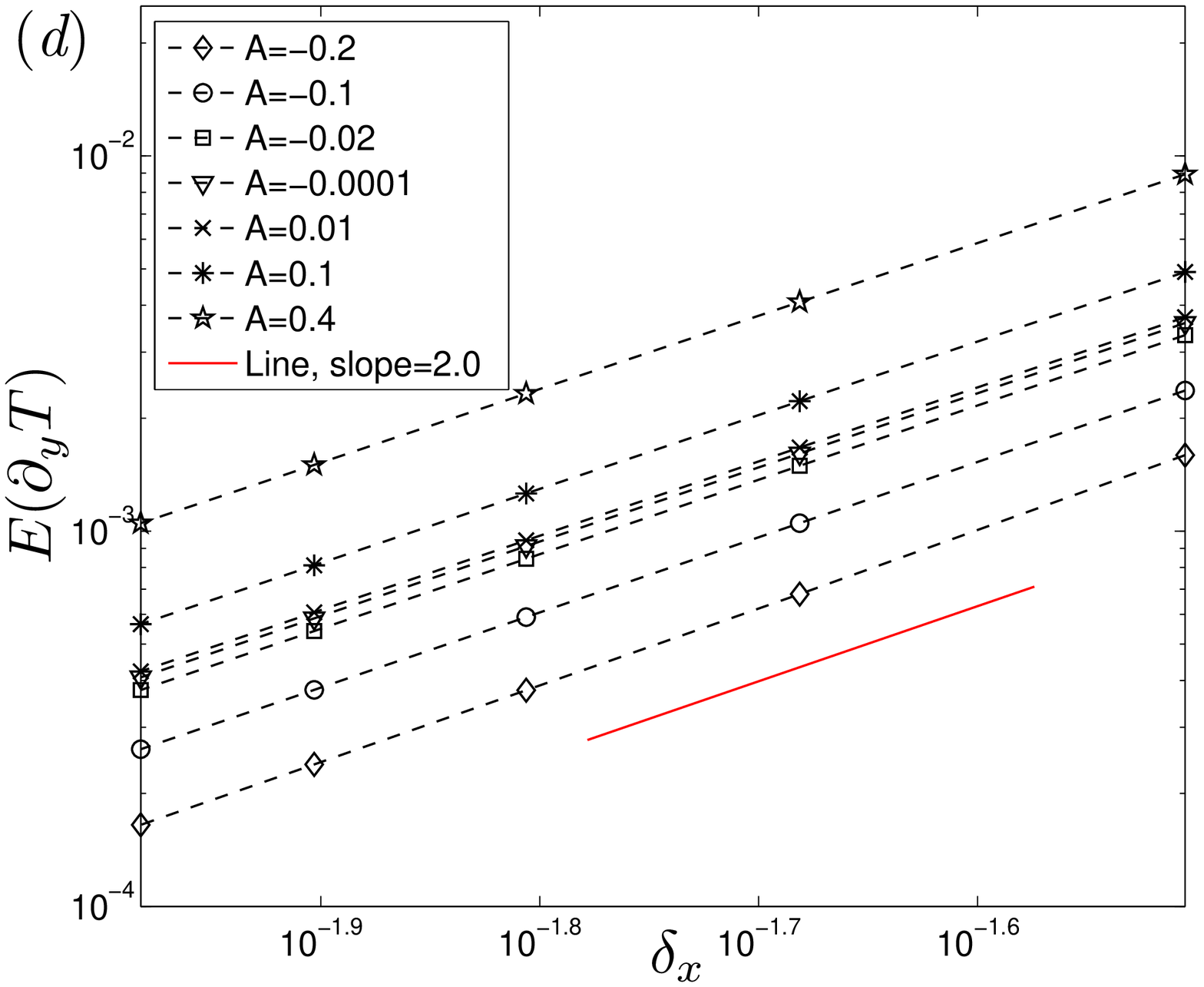}\\
\end{tabular}
\caption{Relative errors of (a) $\frac{\partial u_x}{\partial y}$ and (b) $\frac{\partial T}{\partial y}$ against lattice spacing $\delta_x$ at $Re=5,~\varepsilon=0.6,~Ra=100,~Pr=1,~Da=0.01$. The slope of the inserted line is $2.0$, which indicates that the present models for the flow and temperature fields are of second-order accuracy in space.}
\label{AccurGradABFig}
\end{figure}

\begin{figure}[htb!]
\centering
\includegraphics[scale=0.5]{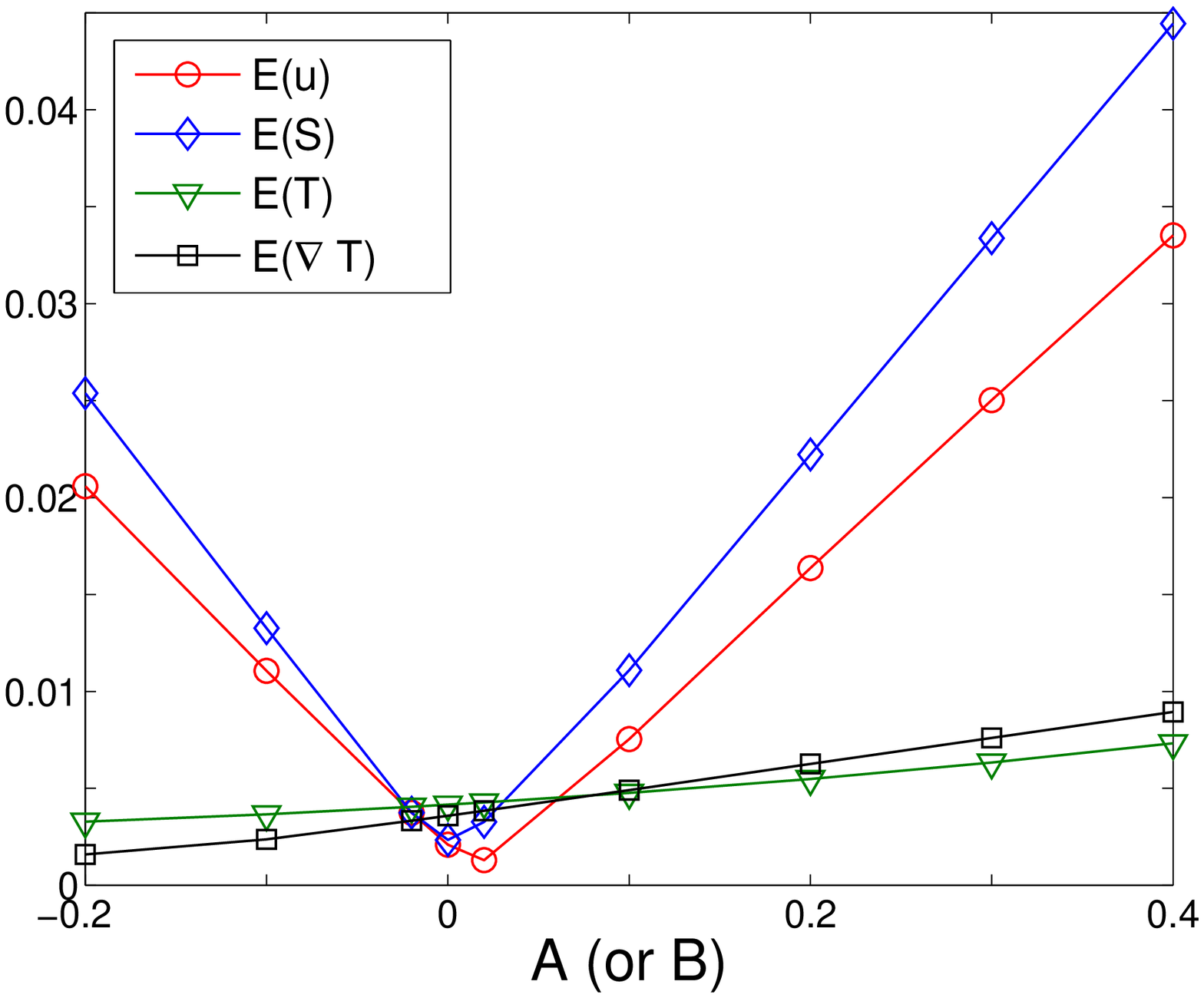}%
\caption{Relative errors of the flow velocity and the shear rate together with the temperature and its gradient at different values of $A$ and $B$.}
\label{fig:AvarInf}
\end{figure}

As can be seen from Figs. \ref{AccurFig} and \ref{AccurGradFig}, the relaxation time has an influence on the computed relative errors of velocity and temperature
and their spatial gradients, and smaller errors are obtained with the relaxation time near unity. This indicates that more accurate results can be obtained when the relaxation time is around unity. Next, we turn to investigate the influences of $A$ and $B$ on the spatial accuracy of the present model. In the simulations, the relaxation times are fixed at $\tau_f=\tau_T=1.0$, the values of $A$ and $B$ are varied in the range of $-0.2\leq A, B \leq 0.4$, and other parameters are identical to those used before. Fig. \ref{AccurGradABFig} shows the computed relative errors under different lattice sizes.
As indicated in the figure, the variation of $A$ and $B$ do not influence the spatial second-order convergence rate of the present model and the computing schemes for the shear rate and temperature gradient, but it affects the accuracy of the present model and the local scheme. Further, it is found that the influence of $A$ on the accuracy of flow field is different from that of $B$ on the temperature field. For instance, as $A=-0.2$ and $B=-0.2$, the relative errors of velocity and shear rate are relatively large among the whole computed results (see Figs. \ref{AccurGradABFig} $(\textit{a})$ and $(\textit{c})$), while the relative errors of temperature and its spatial gradient are the smallest (see Figs. \ref{AccurGradABFig} $(\textit{b})$ and $(\textit{d})$). To reveal the influence of these two parameters on the accuracy of computed results more clearly, the four relative errors against different values of $A$ and $B$ are plotted in Fig. \ref{fig:AvarInf} where $\delta_x=1/32$.
As can be seen, the relative errors of temperature and its gradient increase with increasing $B$. In contrast, the relative errors of velocity and shear rate exhibit two different trends as $A$ increases, that is, they decrease when $A$ is approximately smaller than zero, and turn to increase when $A$ is larger than zero. Based on these results, it can be recognized that, to derive more accurate results, $A$ and $B$ as well as the relaxation times should be also assigned with appropriate values, and especially $A$ should be around zero.

\begin{figure}
  \centering
  \subfigure[LBGK \cite{Guo05}, $t=46000\delta_t$, $\tau_f=0.5009$]{
    \label{figVis:subfig:a} 
    \includegraphics[width=0.48\textwidth,height=0.325\textheight]{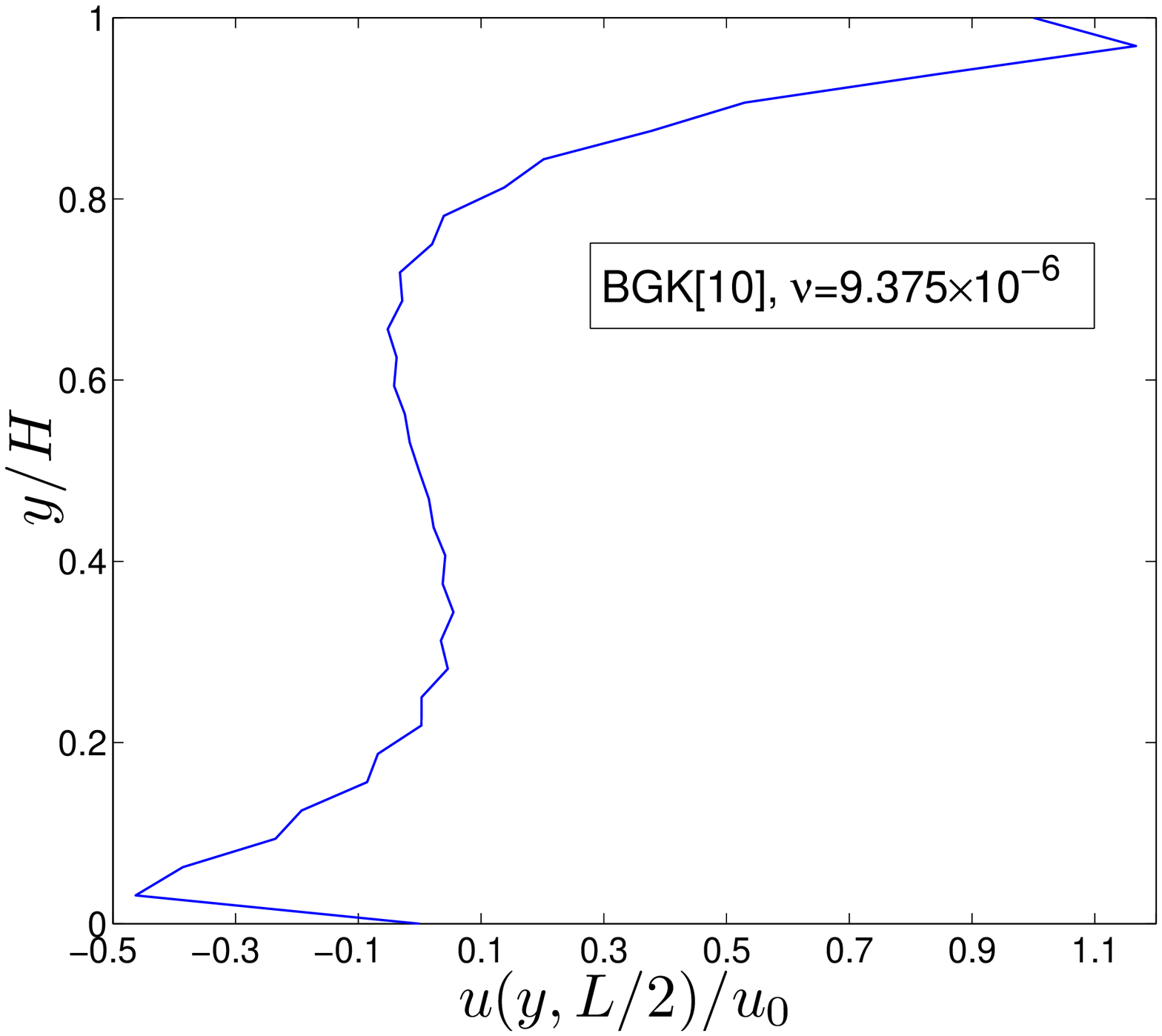}}
  \hspace{5pt}
  \subfigure[Present LBGK, $t=46000\delta_t$, $\tau_f=1.0$]{
    \label{figVis:subfig:b} 
    \includegraphics[width=0.48\textwidth,height=0.325\textheight]{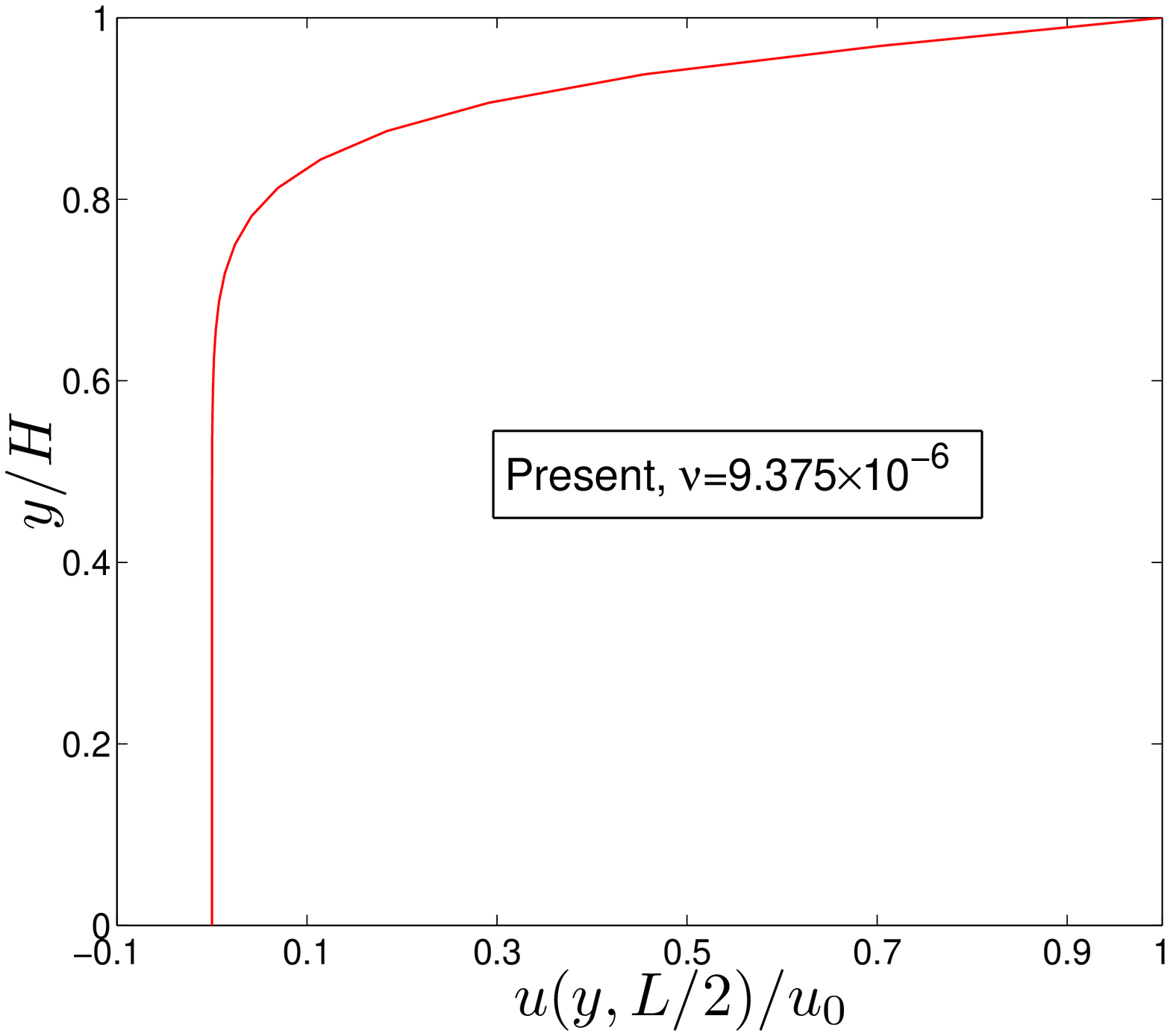}}
    \subfigure[Present LBGK, steady-state, $\tau_f=1.0$]{
    \label{figVis:subfig:c} 
    \includegraphics[width=0.48\textwidth,height=0.325\textheight]{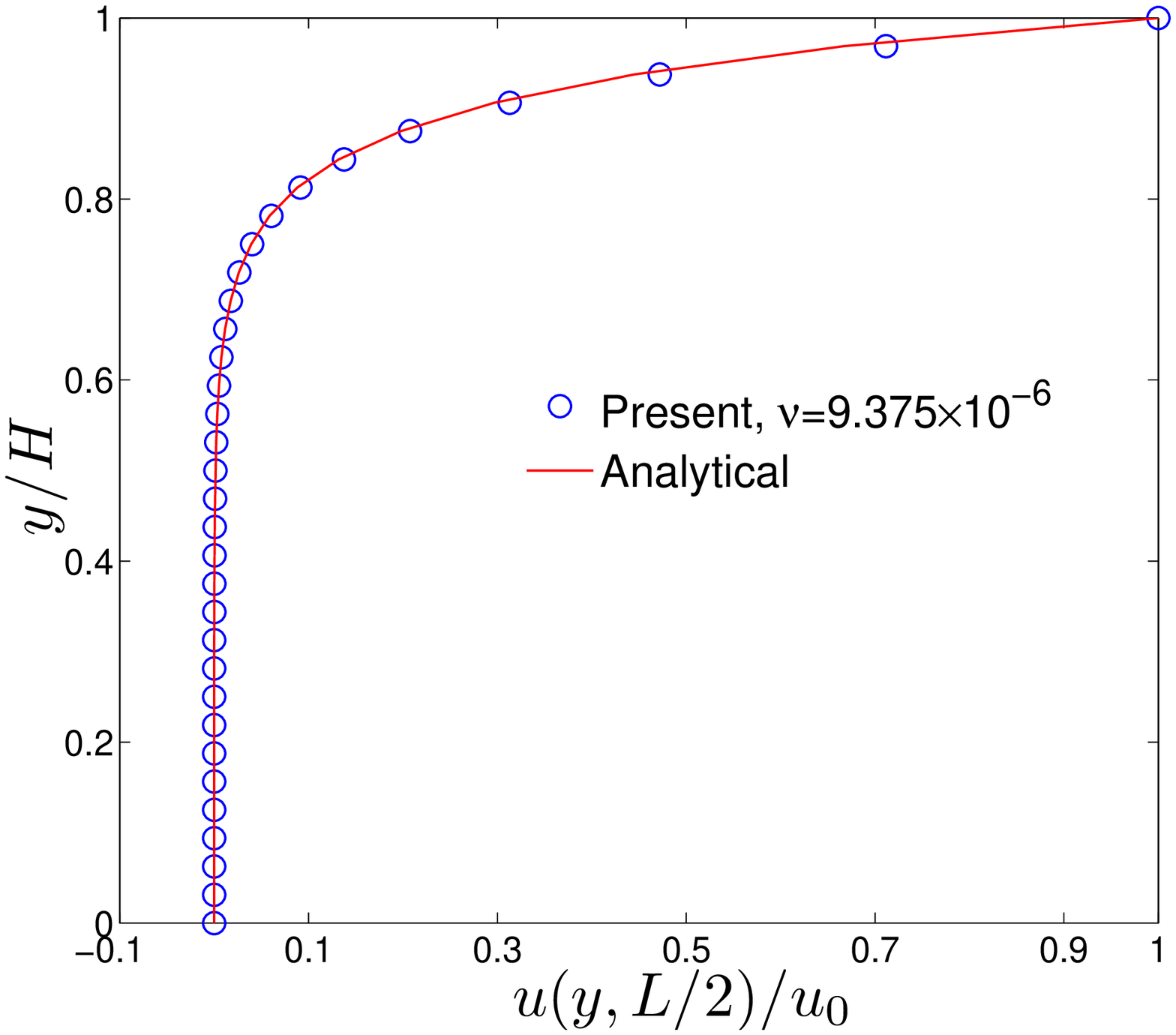}}
  \hspace{5pt}
  \subfigure[Present LBGK, steady-state, $\tau_f=1.0$]{
    \label{figVis:subfig:d} 
    \includegraphics[width=0.48\textwidth,height=0.325\textheight]{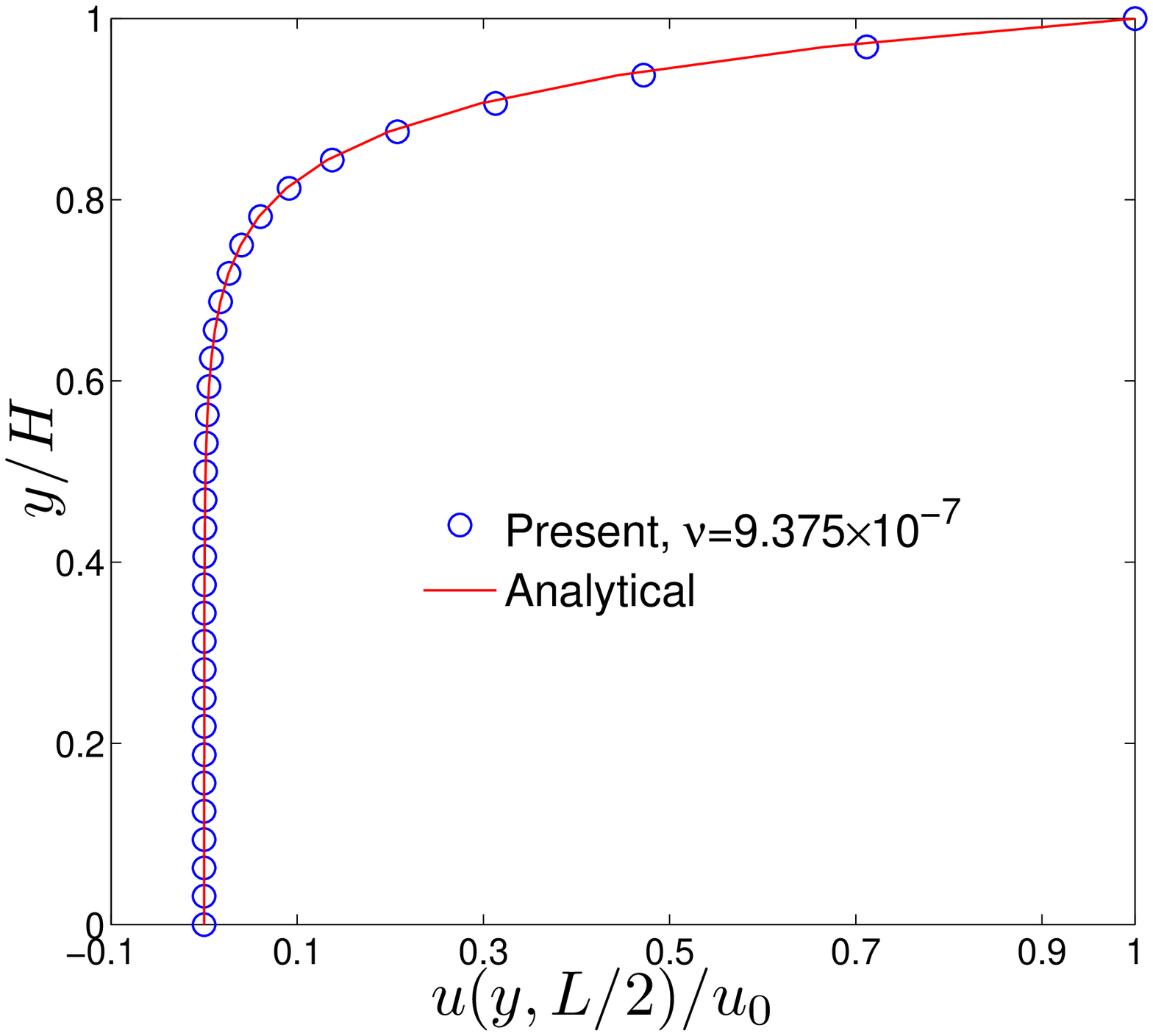}}
  \caption{Velocity profiles of the mixed convection flow at $Re=5, Ra=100, Pr=1, \varepsilon=0.6, Da=0.01$ with low viscosities $\nu=9.375\times10^{-6}$ and $\nu=9.375\times10^{-7}$. The numerical results are predicted by the present LBGK model and the standard LBGK model \cite{Guo05} on a $N_x\times N_y =32\times32$ mesh size.}
  \label{fig:VisCompare} 
\end{figure}

As noted previously, due to the presence of two additional parameters $A$ and $B$, the present modified LBGK model would be more stable than the standard LBGK model. Now, we will confirm this speculation by comparing the results predicted by the present model with those by the LBGK model \cite{Guo05} at low viscosities. In Fig. \ref{fig:VisCompare}, the computation results with two viscosities $\nu=9.375\times10^{-6}$ and $\nu=9.375\times10^{-7}$ are shown at $Re=5, Ra=100, Pr=1, \varepsilon=0.6, Da=0.01, \delta_x=1/32$. From Fig. \ref{figVis:subfig:a}, one can see that the LBGK model \cite{Guo05} becomes unstable at $\nu=9.375\times10^{-6}$ ($\tau_f=0.5009$) when $t=46000\delta_t$, while the present LBGK model ($\tau_f=1.0$) is stable (see Fig. \ref{figVis:subfig:b}) and finally yields excellently agreeable results to the analytical solution (see Fig. \ref{figVis:subfig:c}).
This demonstrates that the present modified LBGK model is more stable than the standard LBGK model. To strengthen this statement, we further performed simulations with smaller values of viscosity, for example $\nu=9.375\times10^{-7}$ ($\tau_f=0.50009$ in the LBGK model \cite{Guo05}), and find that the present LBGK model can still give excellent agreement results (see Fig. \ref{figVis:subfig:d}). Additionally, since the Prandtl number is unity in the simulations, we would like to point out that better numerical stability of the present LBGK model at low effective thermal diffusivities is actually also verified. These results further demonstrate the enhanced numerical stability of the present LBGK model over the standard LBGK model, which can be attributed to the independent determination of relaxation times with the viscosity and effective thermal diffusivity.

\subsection{Natural convection in a porous cavity}
\begin{figure}[htb!]
\centering
\includegraphics[scale=0.65]{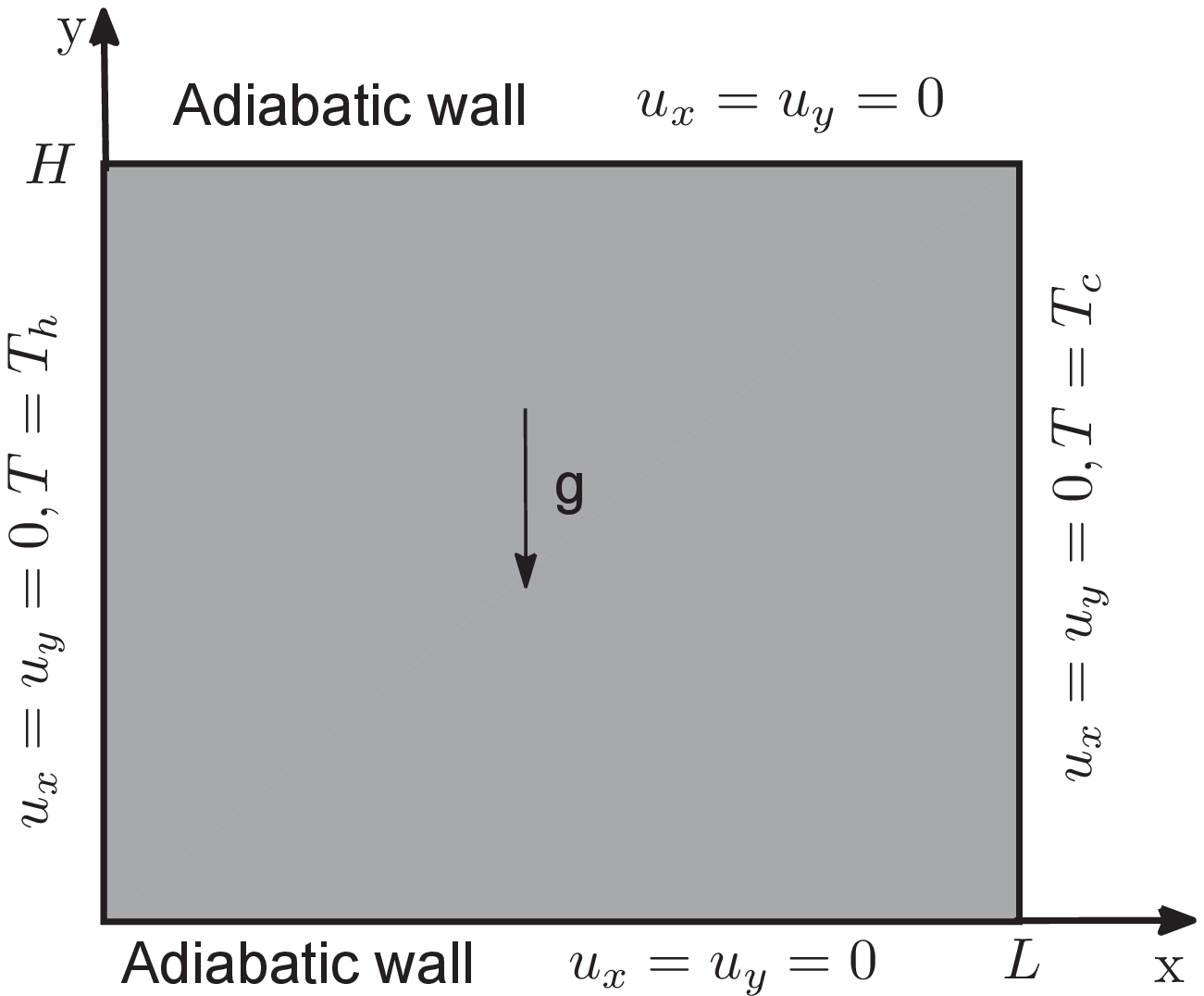}%
\caption{Schematic of natural convection in a fluid-saturated porous cavity.}
\label{fig:natuCovSch}
\end{figure}

Natural convection in a square cavity saturated with porous media has been extensively studied as a benchmark flow by many researchers \cite{Guo05,Seta06,Beckermann88,Lauriat89,Nithiarasu97}. The schematic of the problem is shown in Fig. \ref{fig:natuCovSch}, where the upper and lower walls of the cavity are adiabatic, while the left and right walls are isothermal
but held at different temperatures $T_h$ and $T_c$ ($T_h>T_c$), respectively. At all walls of the cavity, the zero velocity boundary condition is imposed. The height and width of the cavity are $H$ and $L$, the temperature difference is $\Delta T=T_h-T_c$, and the reference temperature is $T_0=(T_h+T_c)/2$. The average Nusselt number $\overline{Nu}$ on the left (or right) vertical wall is defined as
\begin{equation}\label{Eq:AvNuss}
   \overline{Nu}=\frac{1}{H}\int_0^H Nu(y)dy,
\end{equation}
where $Nu(y)=-L(\partial T/\partial x)_{wall}/\Delta T$ is the local Nusselt number.

We first apply the present modified LBGK model to the case in which $\varepsilon \rightarrow 1$ and $Da$ tends to infinity with $10^3\leq Ra\leq 10^6$. As noted before, the convection heat transfer problem is actually reduced to the case without porous media. In the simulations, we use the grid size of $100\times 100$ for $Ra=10^3$, $150\times 150$ for $Ra=10^4$, and $200\times 200$ for $Ra=10^5, 10^6$, respectively. For comparison with previous results, some quantities are concerned here: the maximum horizontal velocity component $u_{max}$ at the mid-width ($x=L/2$) and its location $y_{max}$, the maximum vertical velocity component $v_{max}$ at the mid-height ($y=H/2$) and its location $x_{max}$, the maximum Nusselt number $Nu_{max}$ and the corresponding location $y_{Nu}$, and the average Nusselt number $\overline{Nu}$ along the cold wall. In Table \ref{Tab:Square}, the present results are listed for $\varepsilon=0.9999$, $Da=10^8$, $Pr=0.71$, and $Ra_I=0$. Here, the computed velocities are normalized by the reference velocity $\alpha/L$, and the $x$- and $y$-coordinates are normalized by $L$ and $H$, respectively. The benchmark data from Refs. \cite{Davis83,Hortmann90} and the LB results given in Refs. \cite{Liu14,GuoB02} are included in the Table for quantitative comparison. It can be found that our numerical results are in excellent agreement with those reported data.
\begin{table}
  \caption{Comparisons of the numerical results by the present modified LBGK model with the benchmark solutions \cite{Davis83,Hortmann90} and the reported LB data \cite{Liu14,GuoB02} ($\varepsilon=0.9999, Da=10^8, Ra_I=0, Pr=0.71$, and grid sizes: $100\times100$ for $Ra=10^3$, $150\times150$ for $Ra=10^4$, and $200\times200$ for $Ra=10^5, Ra=10^6$).}
  \vspace{0.4em}
  \label{Tab:Square}
  \centering
  \begin{tabular*}
   {16cm}{@{\extracolsep{\fill}}lllllllll}
 \toprule[0.06em]
 $Ra (N_x\times N_y)$ &{}  &$u_{max}$   &$y_{max}$   &$v_{max}$   &$x_{max}$    &$Nu_{max}$   &$y_{Nu}$   &$\overline{Nu}$\\ \midrule[0.04em]
 $10^3$               &Ref. \cite{Davis83}     &3.649    &0.813     &3.697     &0.178       &-          &-             &- \\
 $(100\times 100)$    &Ref. \cite{GuoB02}      &3.6554   &0.8125    &3.6985    &0.1797      &1.5004     &0.90625       &1.1168\\
 {}                   &Present                 &3.6348    &0.8100    &3.7007    &0.1800     &1.5089     &0.9100        &1.1181\\
 $10^4$               &Ref. \cite{Hortmann90}     &16.1802    &0.8265    &19.6295    &0.1193     &3.5309     &0.8531     &2.2448\\
 $(150\times 150)$    &Ref. \cite{Liu14}          &16.1623    &0.8200    &19.6074    &0.1200     &3.5311     &0.8533     &2.2436\\
 {}                   &Present                    &16.1629    &0.8267    &19.6164    &0.1200     &3.5405     &0.8600     &2.2465\\
 $10^5$               &Ref. \cite{Hortmann90}     &34.7399    &0.8558     &68.6396    &0.0657     &7.7201    &0.9180     &4.5216\\
 $(200\times 200)$    &Ref. \cite{Liu14}          &34.7118    &0.8550     &68.4537    &0.0650     &7.7427    &0.9200     &4.5197\\
 {}                   &Present                    &34.9841    &0.8550     &68.4220    &0.0650     &7.7525    &0.9250     &4.5182\\
 $10^6$               &Ref. \cite{Hortmann90}     &64.8367    &0.8505     &220.461     &0.0390     &17.5360    &0.9608     &8.8251\\
 $(200\times 200)$    &Ref. \cite{Liu14}          &65.1052    &0.8500     &219.379     &0.0400      &17.5857    &0.9600     &8.7731\\
 {}                   &Present                    &65.0494    &0.8500     &218.2964    &0.0400     &17.6373    &0.9650     &8.7792\\
 \bottomrule[0.06em]
\end{tabular*}
\end{table}

The cases with the influence of porous media are further examined under different Darcy numbers and porosities.
\begin{figure}[htbp]
\centering
\includegraphics[width=0.35\textwidth,height=0.25\textheight]{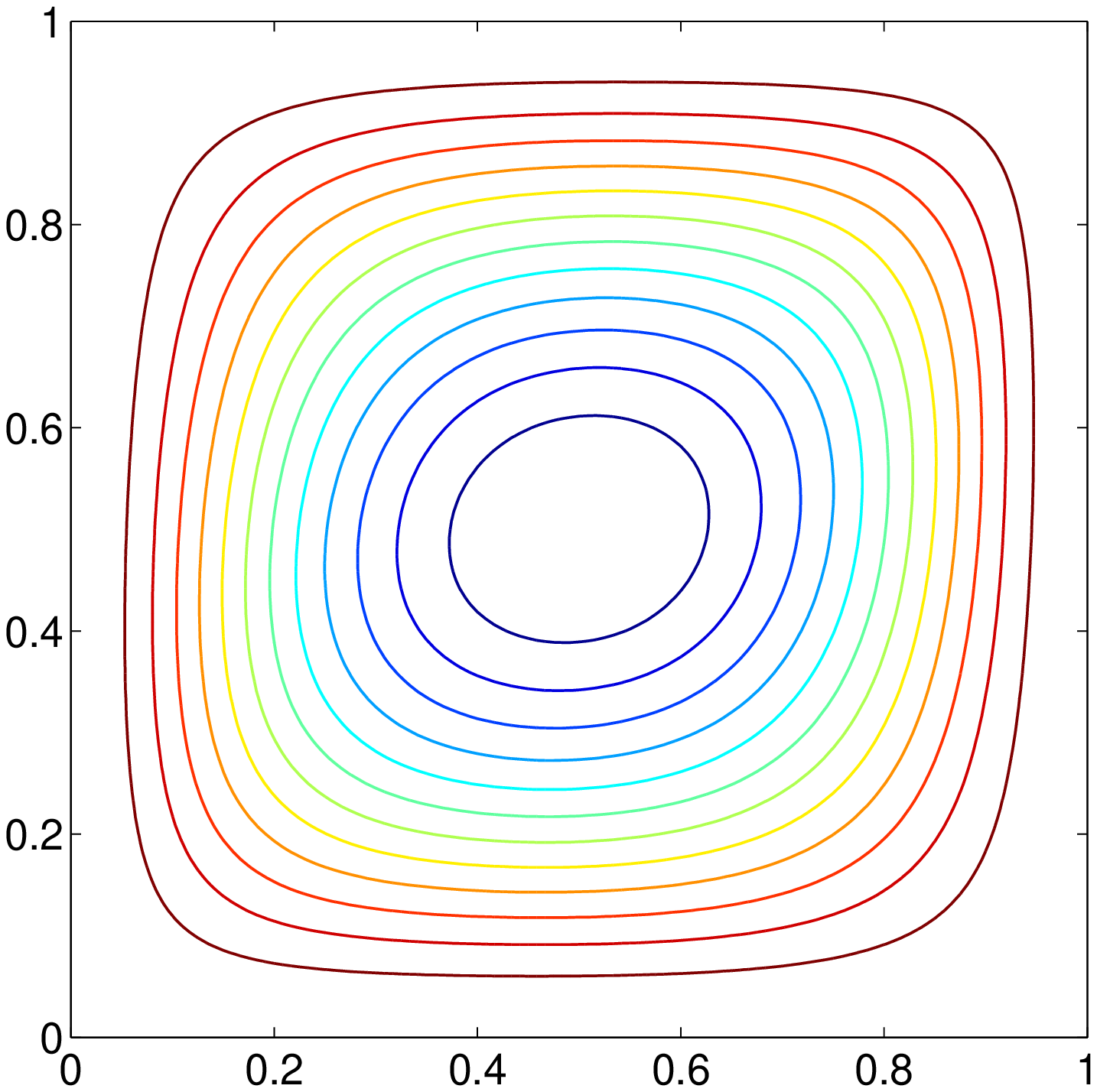}
\includegraphics[width=0.35\textwidth,height=0.25\textheight]{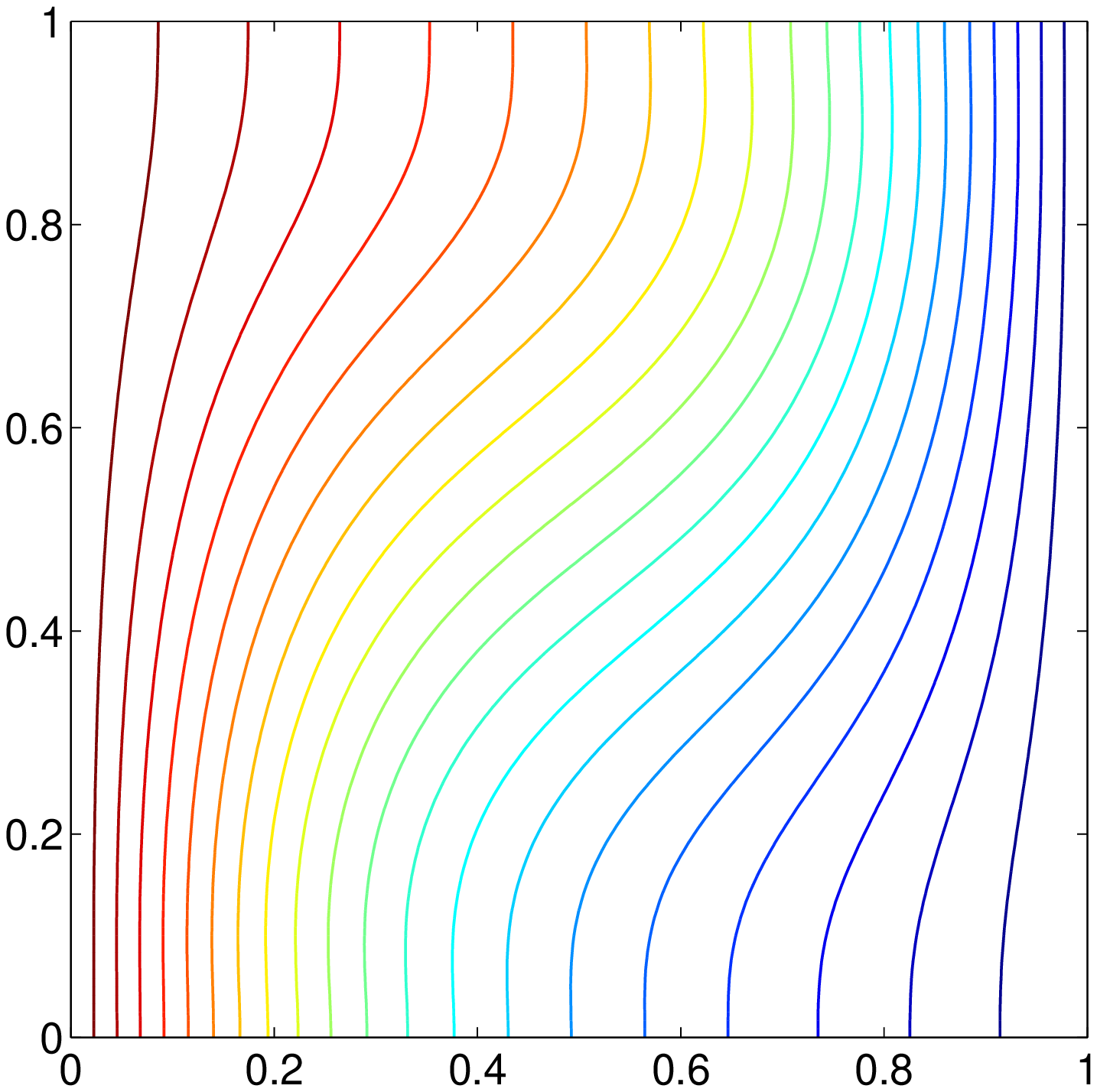}\\
\vspace{-10pt}(a) $Da=10^{-2}, Ra=10^4$\\ \vspace{10pt}
\includegraphics[width=0.35\textwidth,height=0.25\textheight]{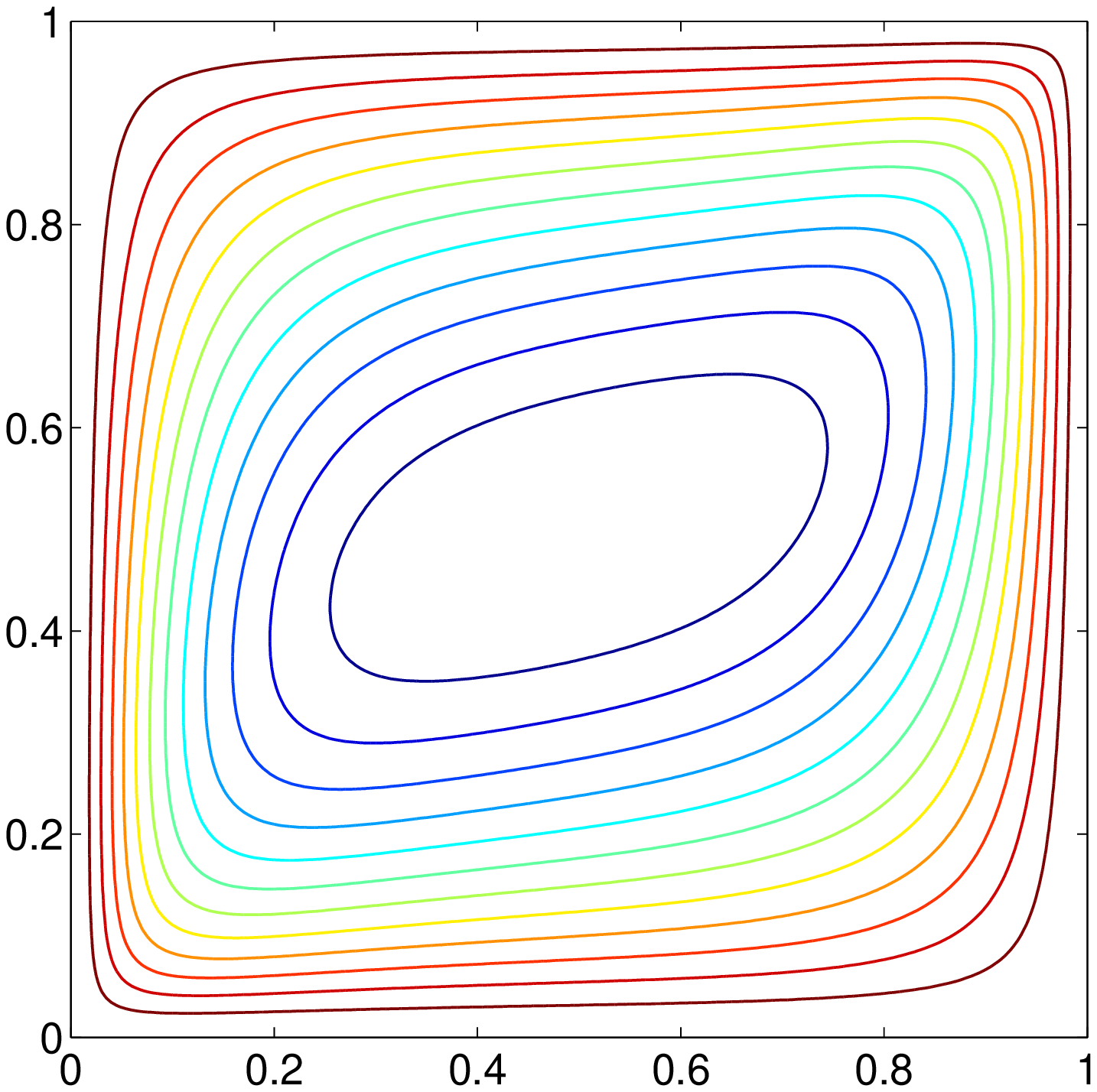}
\includegraphics[width=0.35\textwidth,height=0.25\textheight]{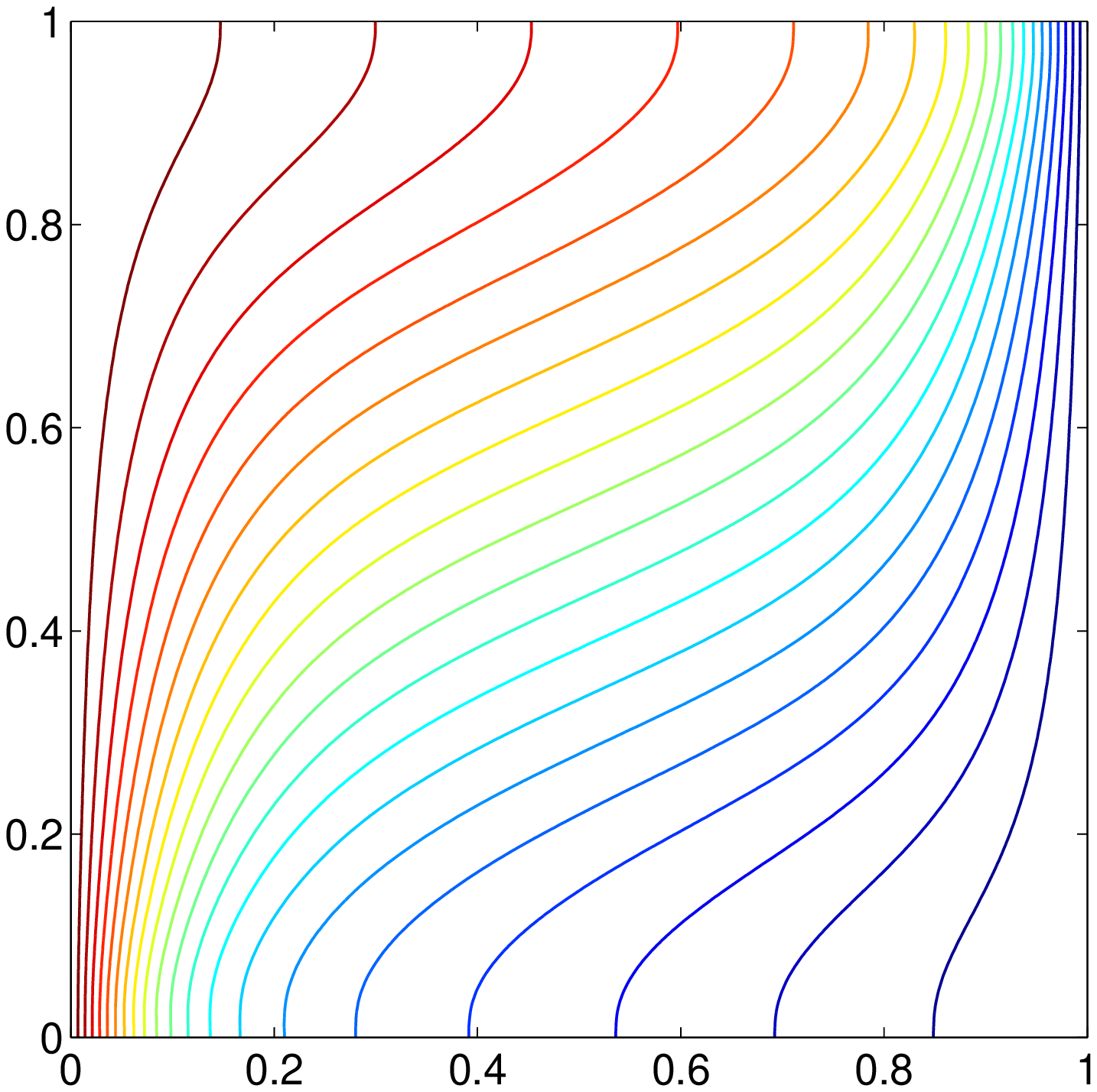}\\
\vspace{-10pt}(b) $Da=10^{-4}, Ra=10^6$\\  \vspace{10pt}
\includegraphics[width=0.35\textwidth,height=0.25\textheight]{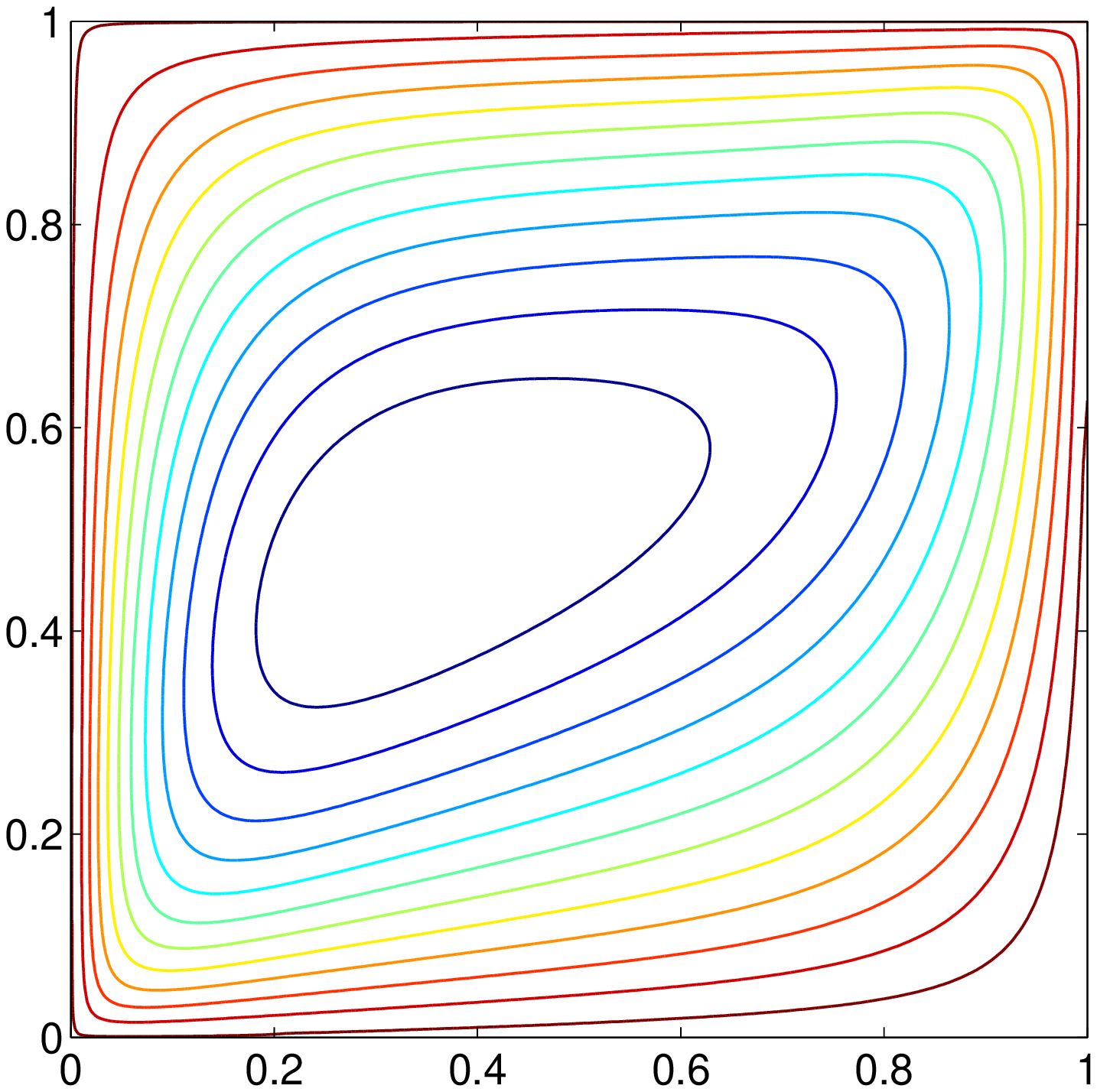}
\includegraphics[width=0.35\textwidth,height=0.25\textheight]{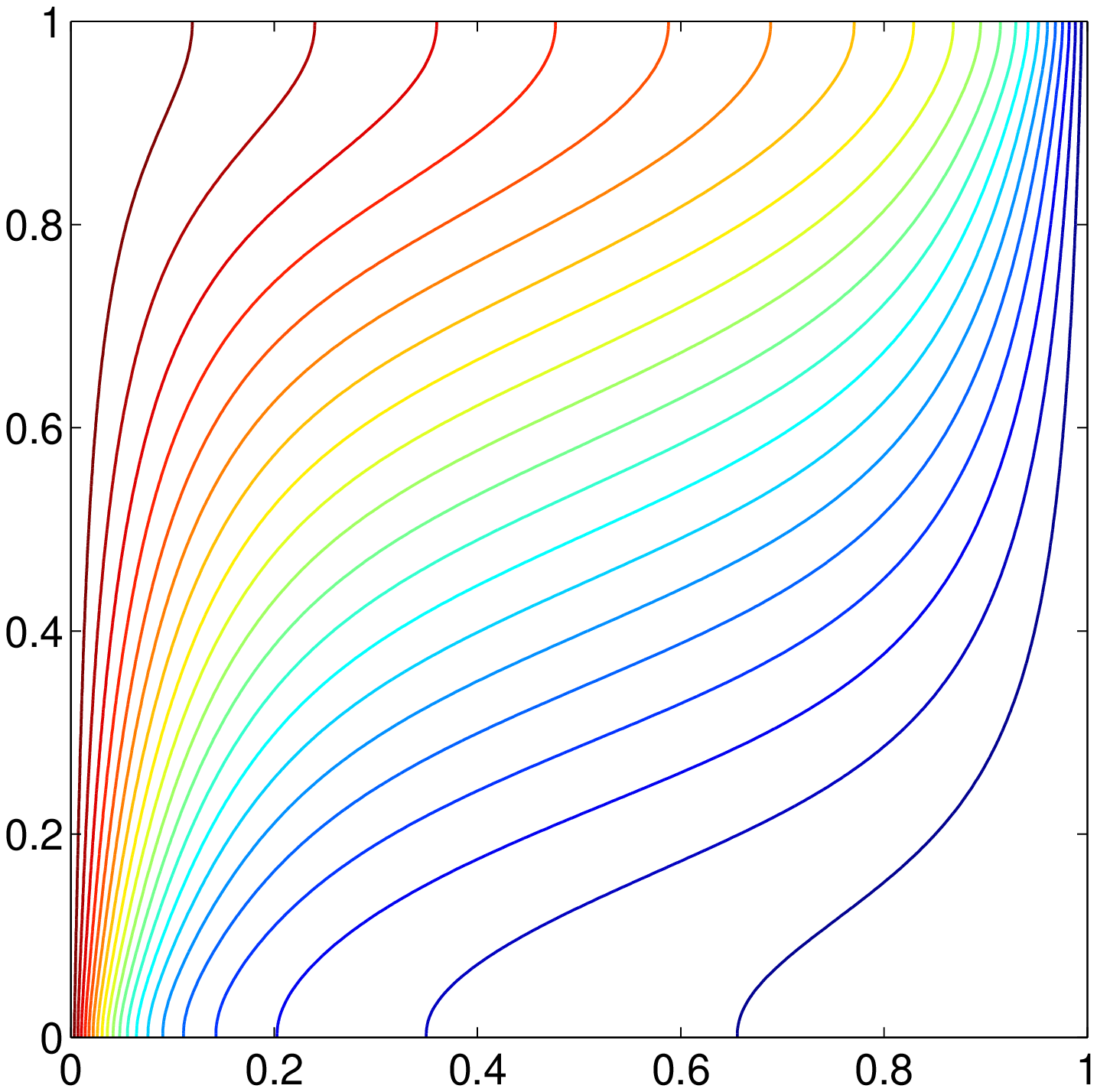}\\
\vspace{-10pt}(c) $Da=10^{-6}, Ra=10^8$\\  \vspace{10pt}
\caption{Streamlines (left) and isotherms (right) of the natural convection in a cavity for $\varepsilon=0.4$, $Ra_I=0$, and $Pr=1.0$: (a) $Da=10^{-2}, Ra=10^4$; (b) $Da=10^{-4}, Ra=10^6$; (c) $Da=10^{-6}, Ra=10^8$ respectively with a mesh size of $120\times120$, $200\times 200$ and $250\times 250$.}
\label{fig:NatualPo}
\end{figure}
In Fig. \ref{fig:NatualPo}, the streamlines and isotherms predicted by the present model are plotted for the Darcy-Rayleigh number $Ra^*=Da Ra=100$ and $\varepsilon=0.4$. In the simulations, the results at $Da=10^{-2}$, $Da=10^{-4}$ and $Da=10^{-6}$ are obtained with a mesh size of $120\times120$, $200\times 200$ and $250\times 250$, respectively. As can be observed from the figure, for the same $Ra^*$ the velocity and thermal boundary layers near the hot and cold walls become thinner as $Da$ decreases. At higher values of $Da$ ($Da=10^{-2}$), the convective mixing is more vigorous inside the cavity, and the isotherms are more sparse near the corners. On the other hand, we also find that as $Ra$ increases with a fixed Darcy number, the patters of isotherms change from almost vertical to be horizontal in the center of the cavity while being vertical only in the thin boundary layers near the hot and cold walls. This indicates that the dominant heat transfer mechanism is varied from conduction to convection. All of these observations and findings are in good agreement with those reported in Refs \cite{Liu14,Nithiarasu97,GuoB02}. To quantify the results, the average Nusselt numbers of the left vertical wall from the present calculations are compared with those previous results, which are listed in Table \ref{Tab:Square1}. It is clearly seen that the present results agree well with the well-documented numerical results in previous studies.
\begin{table}
 \caption{Comparisons of the average Nusselt numbers for various $Ra$, $\varepsilon$ and $Da$ with $Pr=1.0$ and $Ra_I=0$ (grid sizes: $120\times120$ for $Da=10^{-2}$, $200\times200$ for $Da=10^{-4}$, and $250\times250$ for $Da=10^{-6}$).}
  \vspace{0.4em} \label{Tab:Square1}
  \centering
  \begin{tabular*}
   {18cm}[gtb!]{@{\extracolsep{\fill}}cllllllllll}
   \toprule[0.06em]
   \multicolumn{2}{c}{} &\multicolumn{3}{c}{$\varepsilon=0.4$} &\multicolumn{3}{c}{$\varepsilon=0.6$}    &\multicolumn{3}{c}{$\varepsilon=0.9$}\\
   \cmidrule(lr){3-5} \cmidrule(lr){6-8} \cmidrule(lr){9-11}
   $Da (N_x\times N_y)$    &$Ra$    &Ref. \cite{Nithiarasu97}    &Ref. \cite{Liu14}    &Present         &Ref. \cite{Nithiarasu97} &Ref. \cite{Liu14}  &Present &Ref. \cite{Nithiarasu97}    &Ref. \cite{Liu14}    &Present\\ \midrule
    $10^{-2}$              &$10^3$      &1.010     &1.007     &1.008        &1.015     &1.012     &1.012       &1.023     &1.017     &1.018\\
   $(120\times 120)$       &$10^4$      &1.408     &1.362     &1.365        &1.530     &1.494     &1.498       &1.640     &1.628     &1.641\\
                           &$10^5$      &2.983     &3.009     &3.012        &3.555     &3.460     &3.463       &3.910     &3.939     &3.946\\
   $10^{-4}$               &$10^5$      &1.067     &1.067     &1.067        &1.071     &1.069     &1.069       &1.072     &1.073     &1.073\\
   $(200\times 200)$       &$10^6$      &2.550     &2.630     &2.618        &2.725     &2.733     &2.734       &2.740     &2.796     &2.818\\
                           &$10^7$      &7.810     &7.808     &7.811        &8.183     &8.457     &8.506       &9.202     &9.352     &9.351\\
   $10^{-6}$               &$10^7$      &1.079     &1.085     &1.089        &1.079     &1.089     &1.094       &1.08      &1.090     &1.102\\
   $(250\times 250)$       &$10^8$      &2.970     &2.949     &3.014        &2.997     &2.957     &3.035       &3.00      &3.050     &3.068\\
                           &$10^9$      &11.460    &11.610    &11.733       &11.790    &12.092    &12.149      &12.02     &12.341    &12.399\\
   \bottomrule
   \end{tabular*}
 \end{table}

\subsection{Thermal convection of a heat-generating fluid in a porous cavity with isothermally cooled walls}
In the above tests, the thermal convection problems in a fluid-saturated porous medium is studied but without internal heat generation. To further show the potential of the present LBGK model, the natural convection flows in a porous cavity with internal heat generation are tested. This kind of thermal convection flows have been studied numerically by many researchers \cite{Liu14,Khanafer98}, and the convective flows induced by internal heat generation in a porous cavity with isothermally cooled walls are considered in this subsection.

\begin{figure}[htb!]
\centering
\includegraphics[scale=0.6]{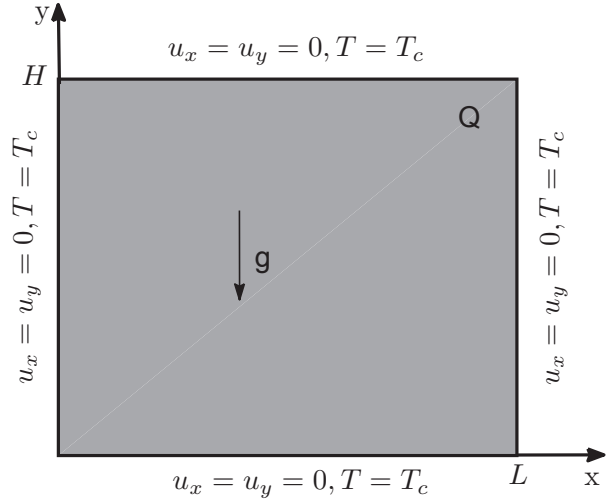}%
\caption{Schematic diagram of a porous cavity with internal heat-generating fluid.}
\label{fig:natIHGCovSch}
\end{figure}

The schematic illustration of simulated problems are shown in Fig. \ref{fig:natIHGCovSch}.
The walls of the porous cavity are maintained at a constant temperature $T_c$ and subjected to no-slip boundary conditions. The internal heat source term is $Q$, and the temperature difference $\Delta T$ is defined as $\Delta T= QL^2/\alpha_e$, by which $g\beta$ is determined as $g\beta=Ra_I\nu\alpha_e/(\Delta T L^3)$. In simulations, the temperature difference $\Delta T$ is fixed at $10$, the reference temperature $T_0$ is assigned to be $T_c$, $Pr$ is set to be $7$, $Ra=0$, and $Ra_I=6.4\times 10^5$. In this test, the dimensionless parameters $A$ and $B$ are determined respectively by $A=\tau_f-0.5-Ma \sqrt{3Pr/Ra_I}L/\delta_x$ and $B=\tau_T-0.5-Ma\sqrt{3/(Pr Ra_I)}L/(\sigma\delta_x)$, and the square cavity are covered by a uniform lattice of $N_x\times N_y=120\times 120$.

\begin{figure}[htbp]
\centering
\includegraphics[width=0.35\textwidth,height=0.25\textheight]{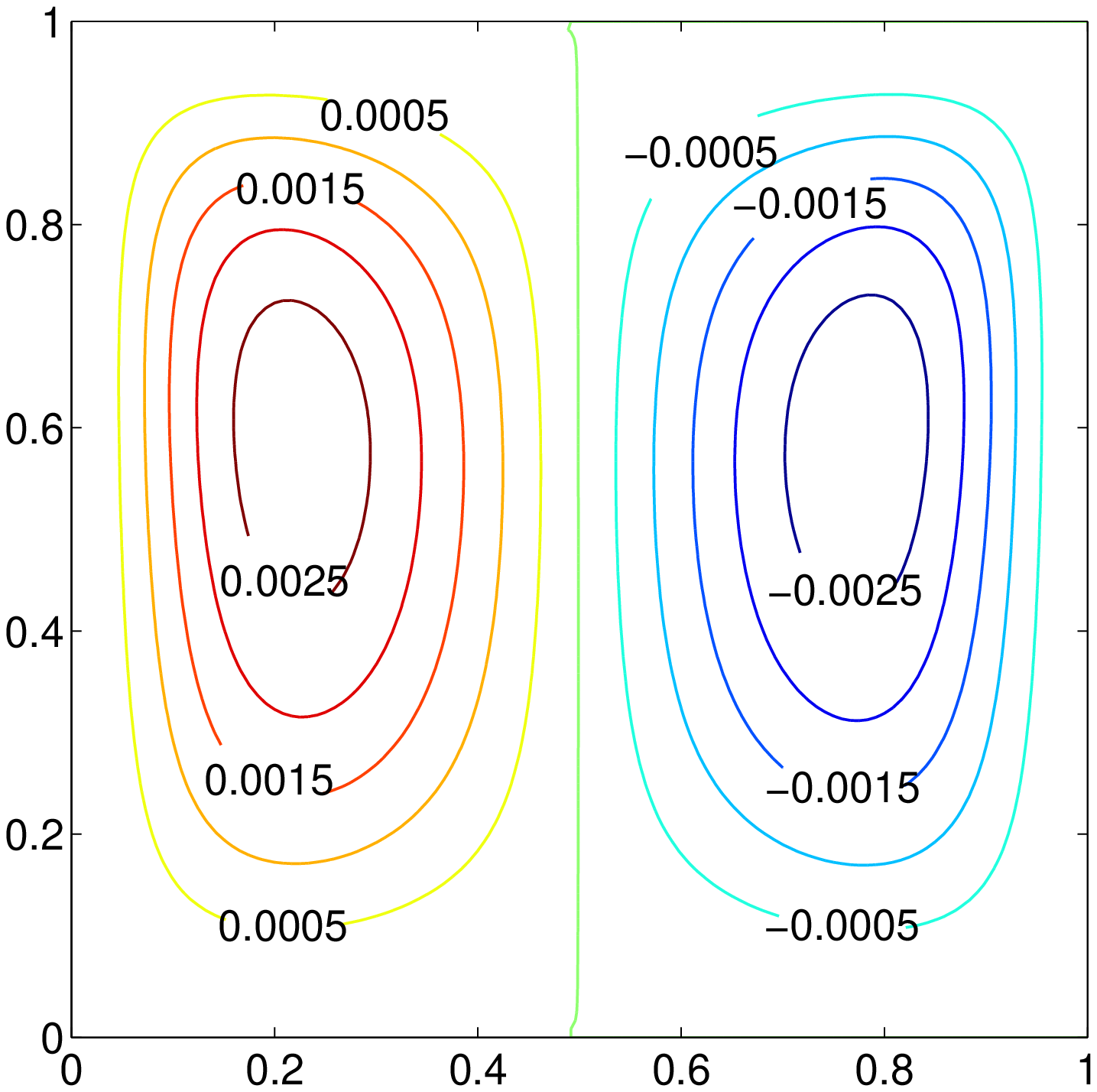}
\includegraphics[width=0.35\textwidth,height=0.25\textheight]{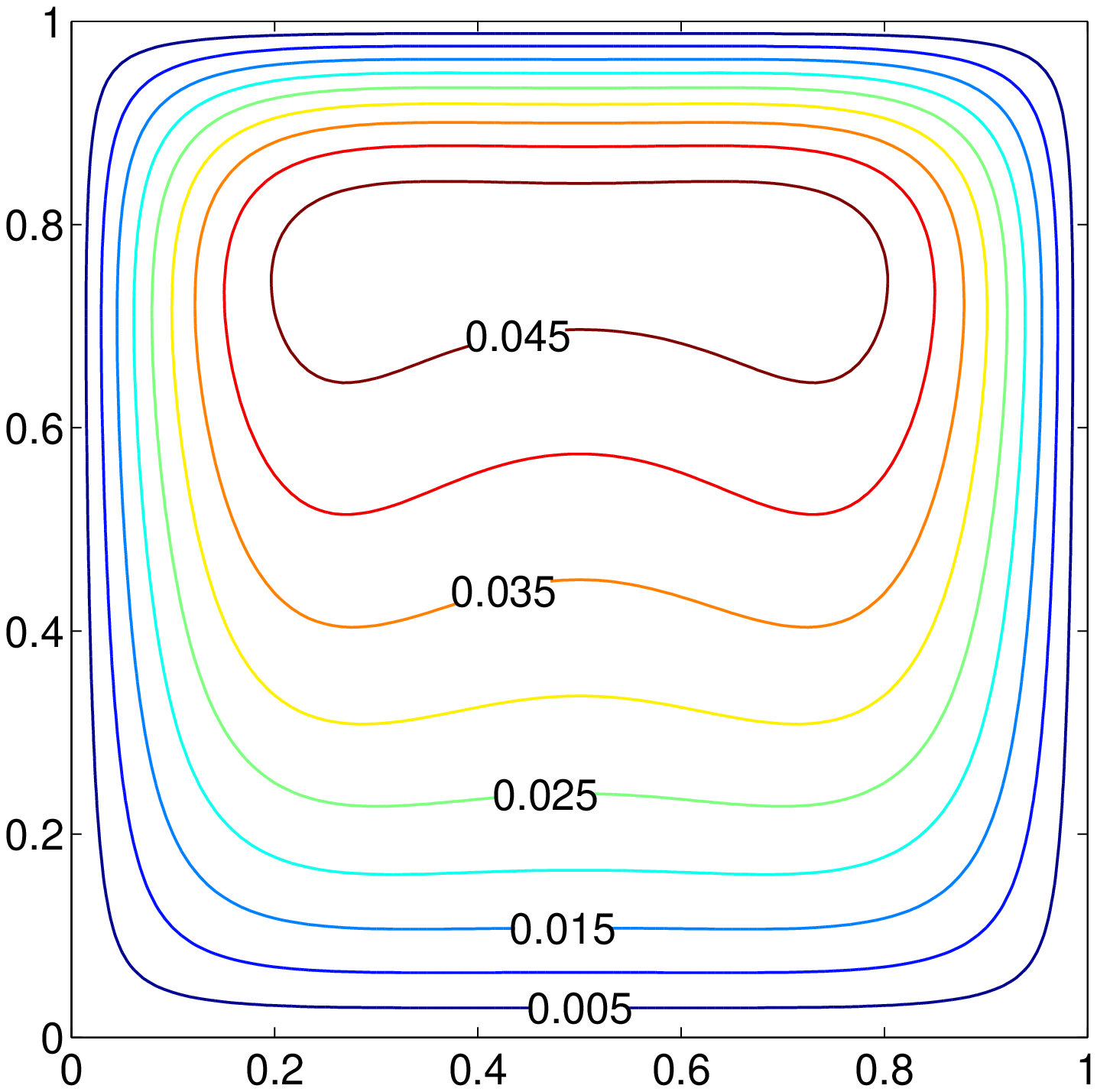}\\
\vspace{-10pt}(a) $Da=\infty$  \\ \vspace{10pt}
\includegraphics[width=0.35\textwidth,height=0.25\textheight]{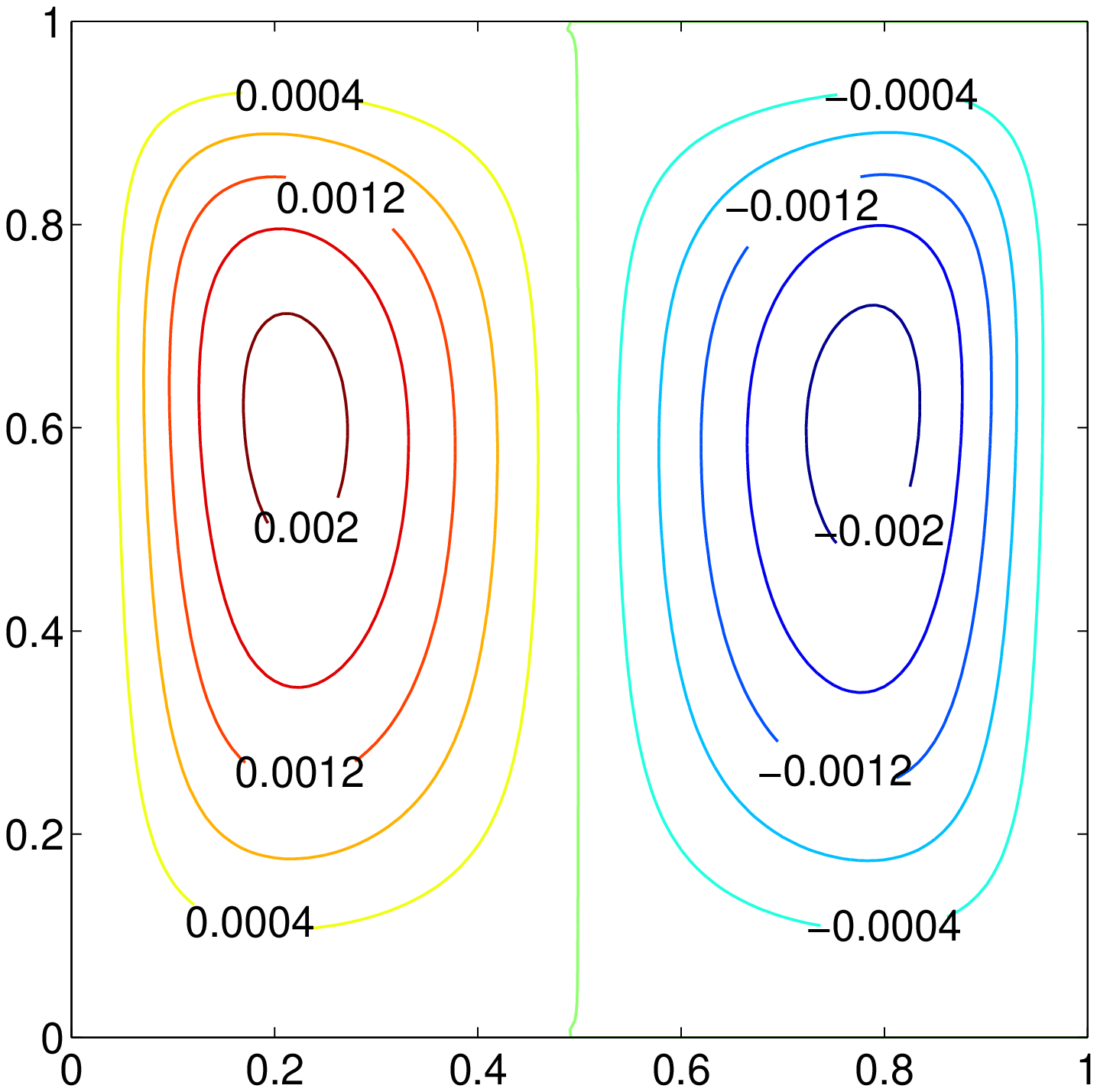}
\includegraphics[width=0.35\textwidth,height=0.25\textheight]{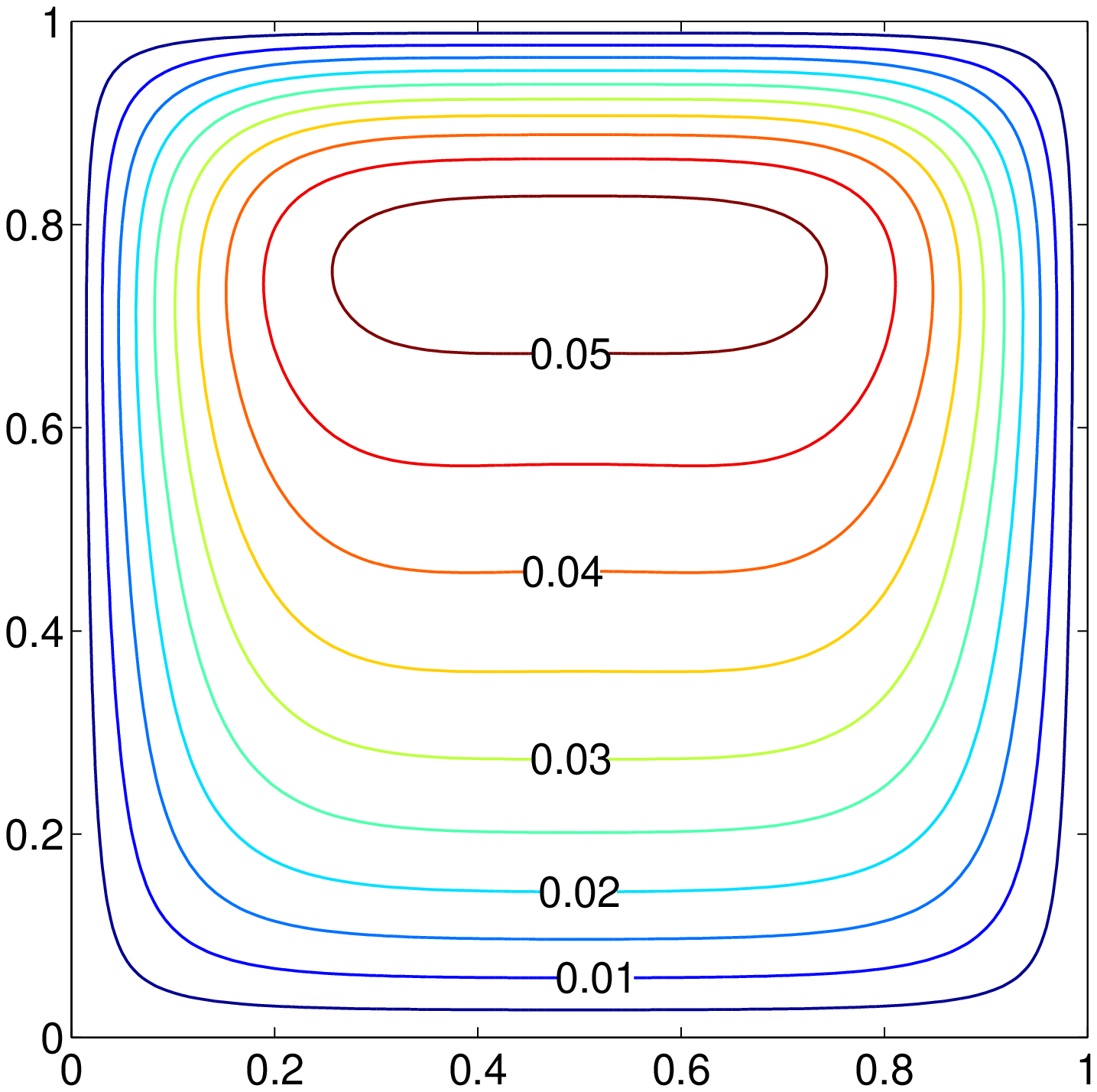}\\
\vspace{-10pt}(b)  $Da=10^{-2}$ \\  \vspace{10pt}
\includegraphics[width=0.35\textwidth,height=0.25\textheight]{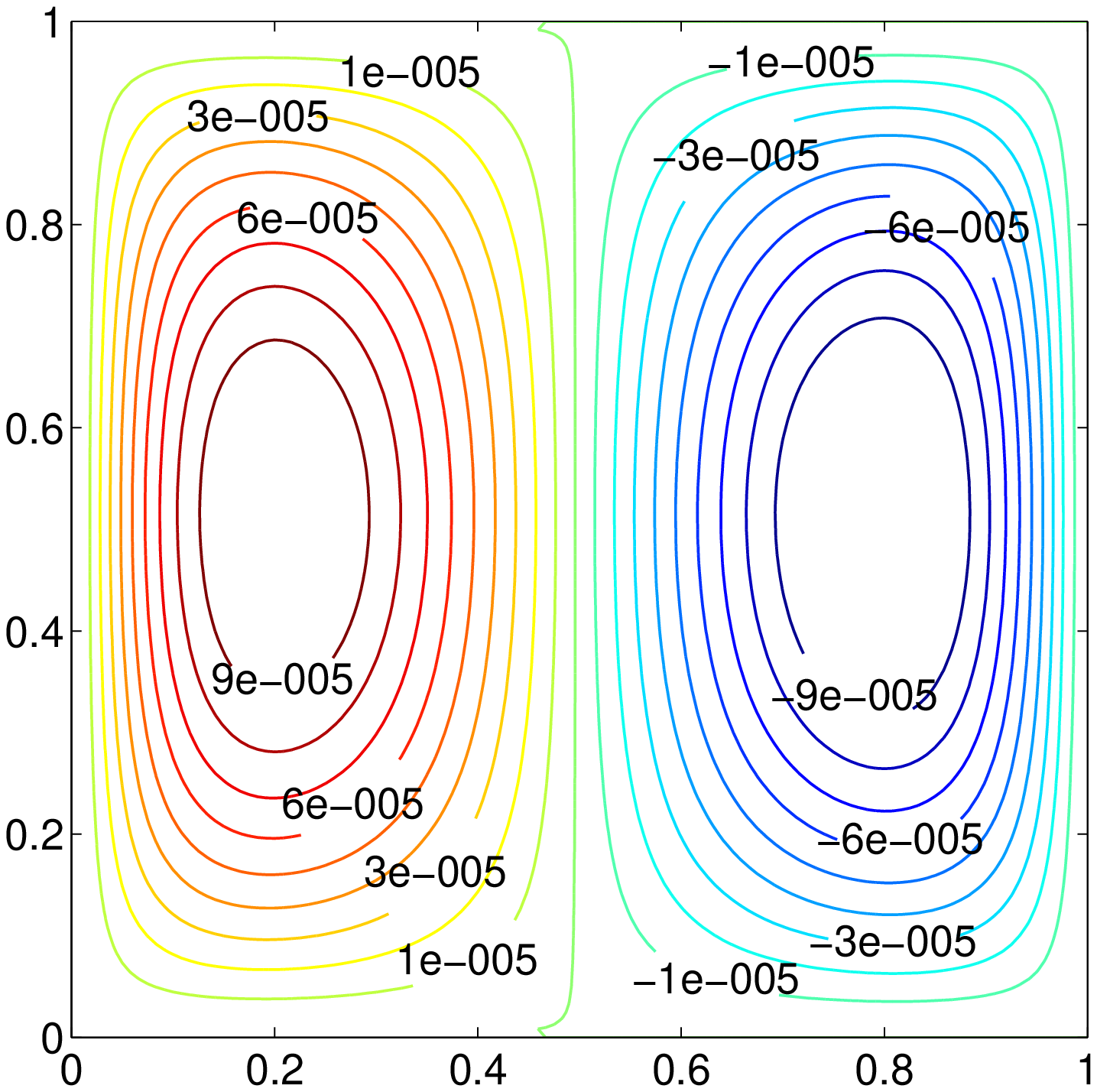}
\includegraphics[width=0.35\textwidth,height=0.25\textheight]{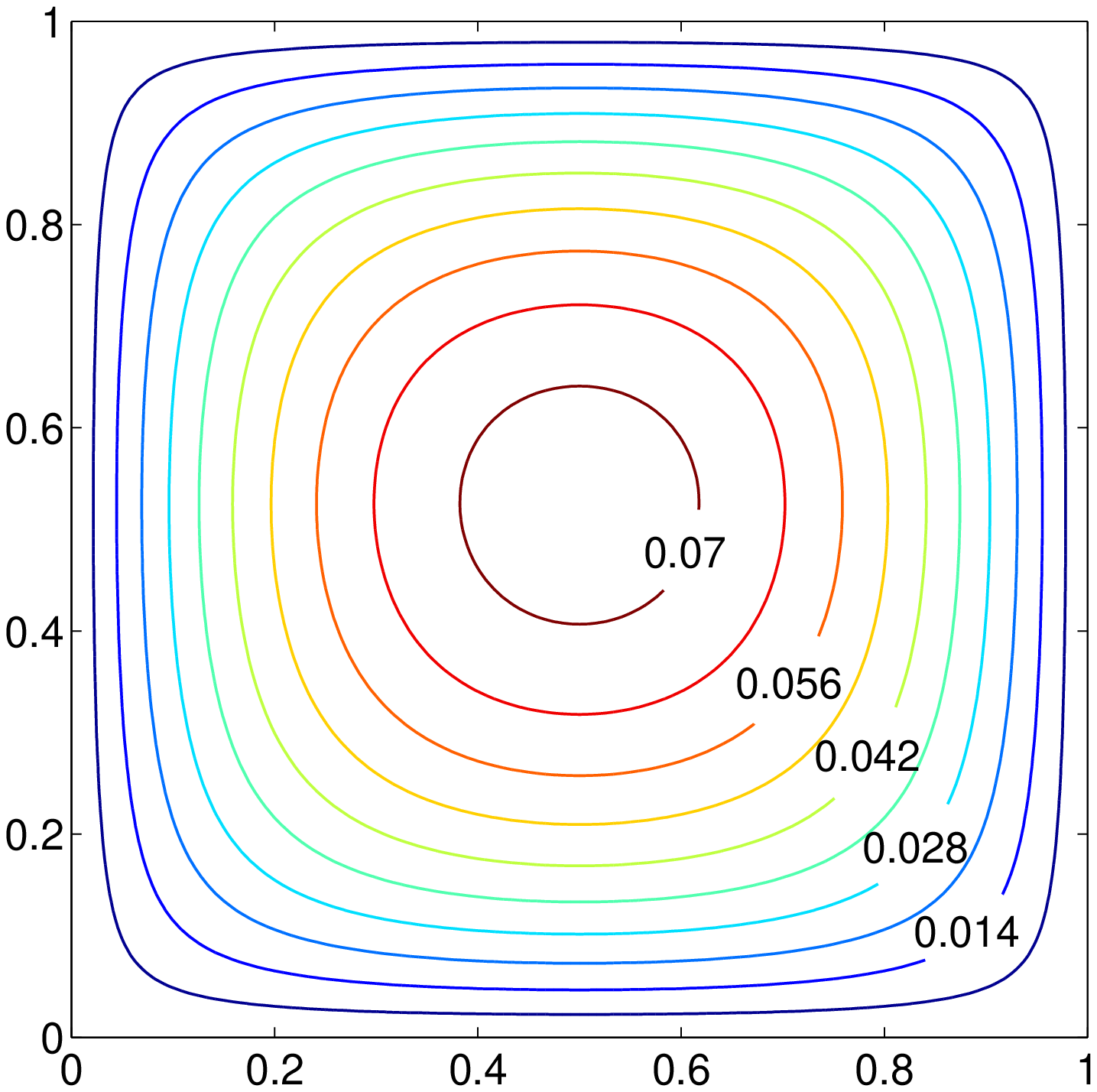}\\
\vspace{-10pt}(c) $Da=10^{-4}$\\  \vspace{10pt}
\caption{Streamlines (left) and isotherms (right) for $Ra=0, Ra_I=6.4\times10^5$, $\varepsilon=1.0, F_\varepsilon=0$, and $Pr=1.0$: (a) $Da=\infty$; (b) $Da=10^{-2}$; (c) $Da=10^{-4}$ (gird size: $120\times120$).}
\label{fig:NatualIHGPo}
\end{figure}

In Fig. \ref{fig:NatualIHGPo}, the streamlines and isotherms predicted by the present model with the Brinkman-extended Darcy model ($\varepsilon=1, F_\varepsilon=0$) are shown for different $Da$.
We note that these presented results are qualitatively consistent with those in previous numerical studies \cite{Liu14,Khanafer98}. To make a quantitative comparison, the maximum dimensionless stream function $\psi_{max}$ (normalized by $L\sqrt{g\beta\Delta TL}$) and the maximum dimensionless temperature $\theta_{max}$ of $\theta=(T-T_c)/\Delta T$ are computed and listed in Table \ref{Tab:SquareIHG}.
\begin{table}
  \caption{Comparisons of $\psi_{max}$ and $\theta_{max}$ between the present results (grid size: $120\times120$) and those published solutions \cite{Liu14,Khanafer98} at $Ra=0, Ra_I=6.4\times10^5$, $\varepsilon=1.0, F_\varepsilon=0$, $Pr=1.0$.}
  \vspace{0.4em}
  \label{Tab:SquareIHG}
  \centering
  \begin{tabular*}
   {16cm}{@{\hspace{8pt}\extracolsep{\fill}}lllll}
 \toprule[0.06em]
 $Da (N_x\times N_y=120\times120)$      &{}      &Ref. \cite{Khanafer98}     &Ref. \cite{Liu14}        &Present \\ \midrule
 $\infty$                  &$\psi_{max}$         &$2.91\times10^{-3}$        &$2.86\times10^{-3}$      &$2.83\times10^{-3}$      \\
 {}                        &$\theta_{max}$       &$4.75\times10^{-2}$         &$4.79\times10^{-2}$     &$4.79\times10^{-2}$      \\
 $10^{-2}$                 &$\psi_{max}$         &$2.21\times10^{-3}$        &$2.17\times10^{-3}$      &$2.14\times10^{-3}$    \\
 {}                        &$\theta_{max}$       &$5.22\times10^{-2}$        &$5.26\times10^{-2}$      &$5.26\times10^{-2}$     \\
 $10^{-4}$                 &$\psi_{max}$         &$1.10\times10^{-4}$        &$1.06\times10^{-4}$      &$1.04\times10^{-4}$     \\
 {}                        &$\theta_{max}$       &$7.35\times10^{-2}$        &$7.34\times10^{-2}$      &$7.35\times10^{-2}$      \\
 \bottomrule[0.06em]
\end{tabular*}
\end{table}
From the comparisons shown in the Table, it is clear that our numerical results are in excellent agreement with those reported data in Refs. \cite{Liu14,Khanafer98}. In addition, we can observe that as $Da$ decreases the maximum dimensionless stream function $\psi_{max}$ decreases, while the maximum dimensionless temperature $\theta_{max}$ increases. This different trend can be attributed to the fact that as $Da$ decreases the flow is resisted increasingly by the presence of porous medium and hence the fluid velocity in the cavity decreases, while the buoyancy force is retarded to decrease the convective heat transfer \cite{Khanafer98}, which is also indicated in Fig. \ref{fig:NatualIHGPo}.

\subsection{Thermal convection in a porous cavity with internal heat generation}
In this subsection, the thermal convection flow in a porous cavity with internal heat generation and external sidewall heating is studied. The flow geometry and boundary conditions are identical to those shown in Fig. \ref{fig:natuCovSch}, but a volumetric internal heat source $Q$ is imposed over the domain. The temperature difference is $\Delta T=T_h-T_c$, the reference temperature is $T_0=(T_h+T_c)/2$, and the internal heat source $Q$ is determined by $Q=Ra_I\alpha_e\Delta T/(RaL^2)$. In the following simulations, the Prandtl number $Pr$ is set to be $0.7$, and the lattice size of $150\times 150$ and $200\times 200$ are employed for $Ra=10^5$ and $Ra=10^6$, respectively. The parameters $A$ and $B$ are determined according to Eq. \eqref{EQA&B}, and the average Nusselt number $\overline{Nu}$ are computed by Eq. \eqref{Eq:AvNuss}.

\begin{figure}[htbp]
\centering
\includegraphics[width=0.35\textwidth,height=0.25\textheight]{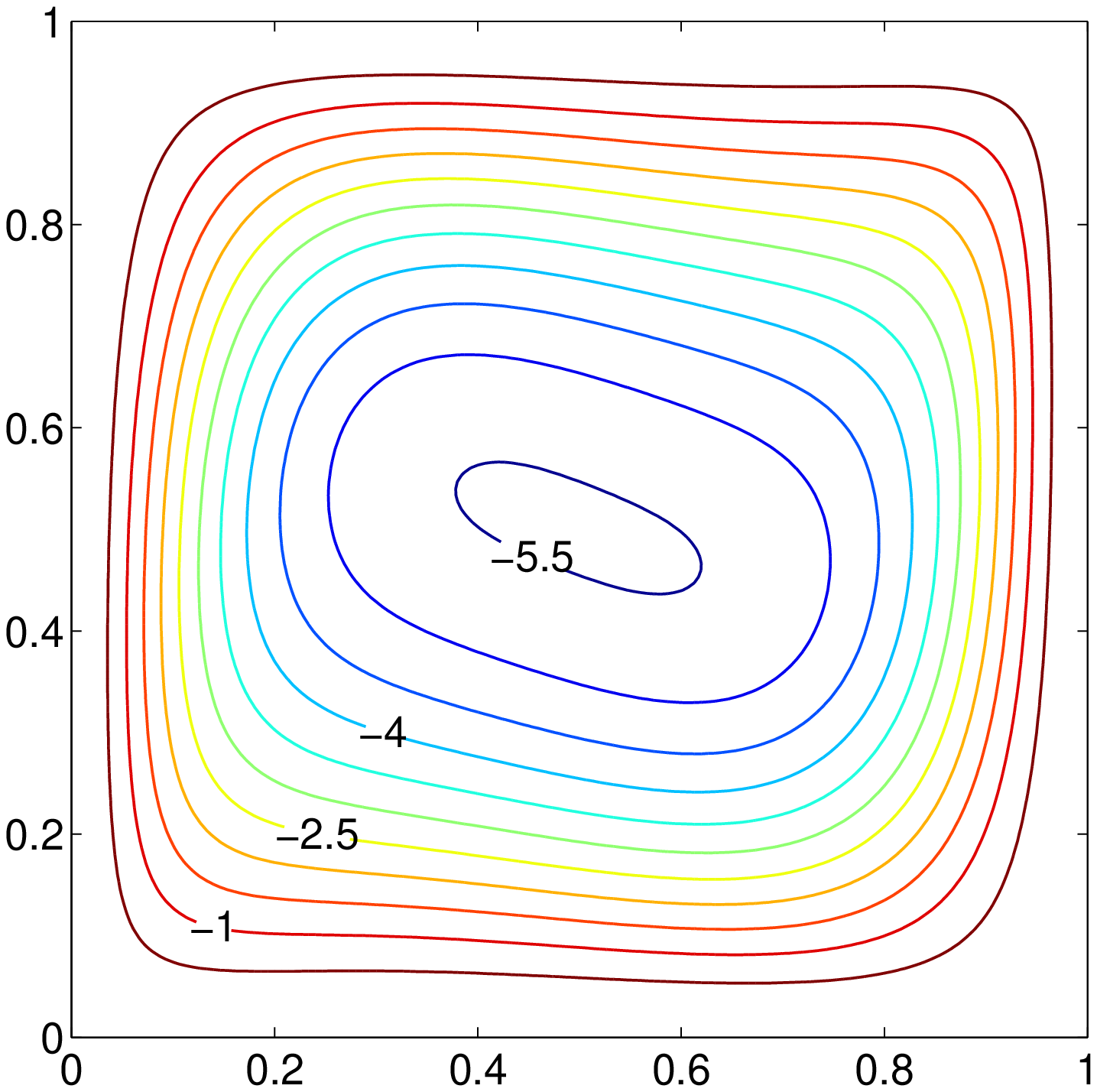}
\includegraphics[width=0.35\textwidth,height=0.25\textheight]{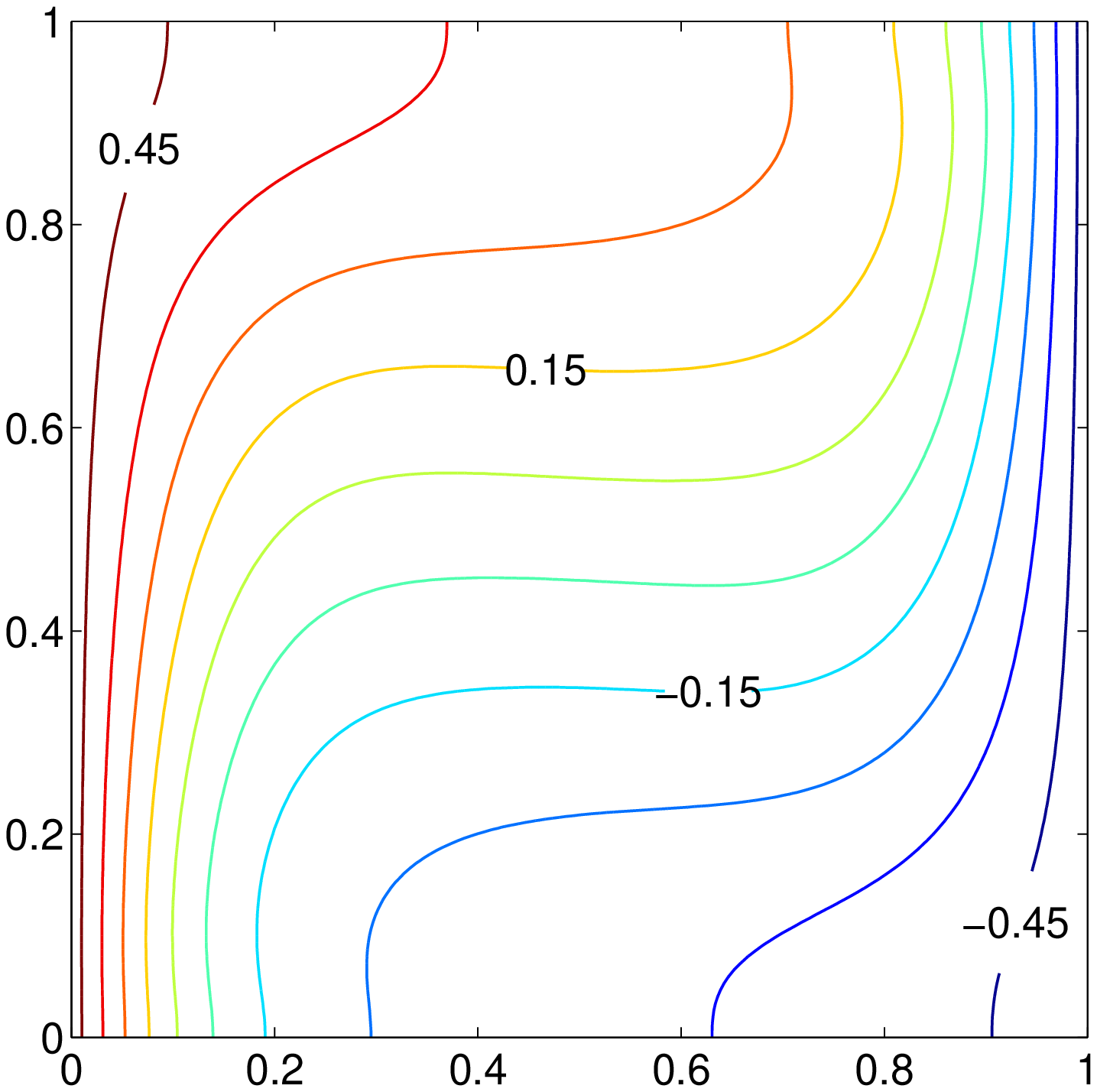}\\
\vspace{-10pt}(a) $Ra_I=10^3$  \\ \vspace{10pt}
\includegraphics[width=0.35\textwidth,height=0.25\textheight]{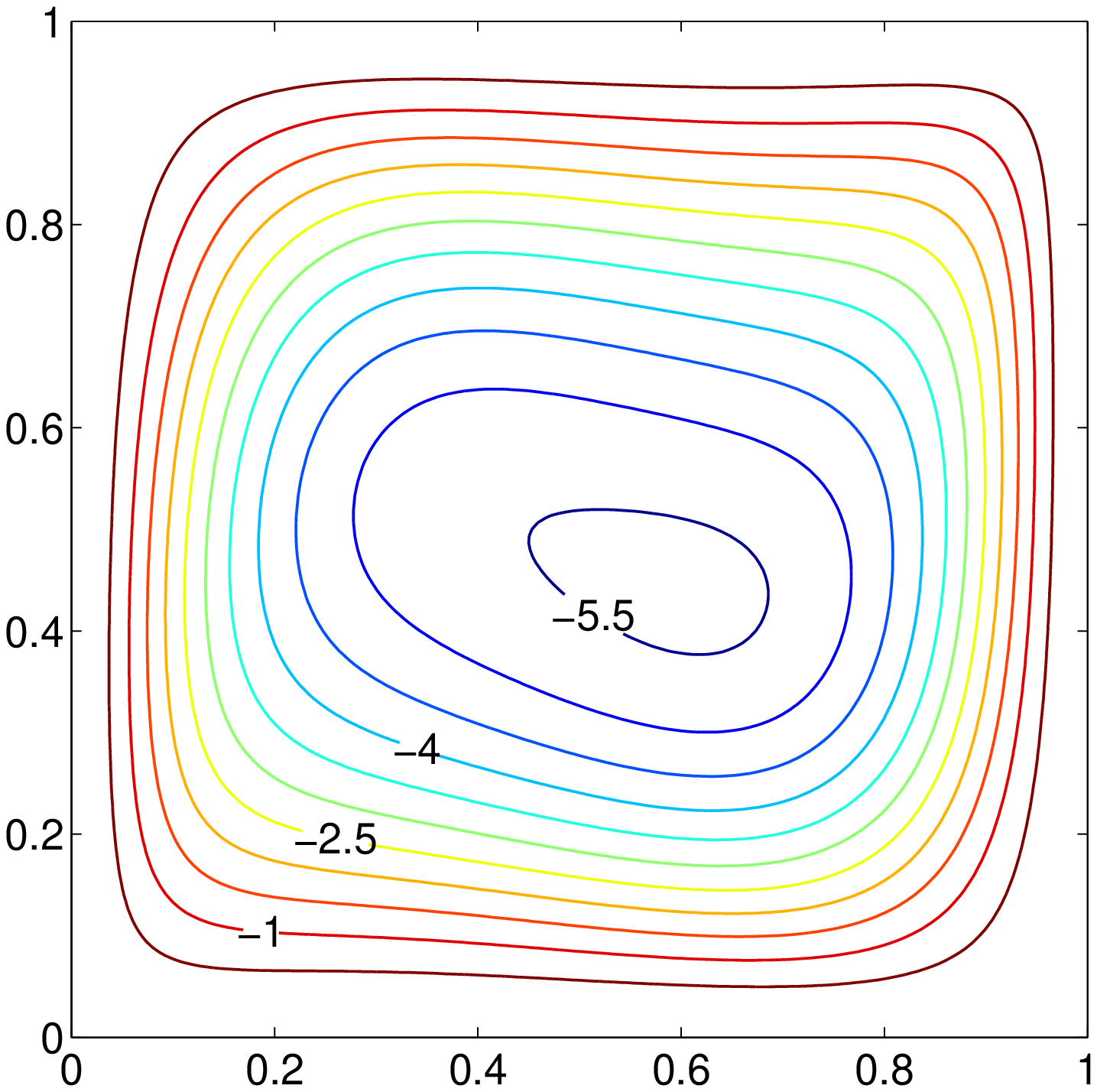}
\includegraphics[width=0.35\textwidth,height=0.25\textheight]{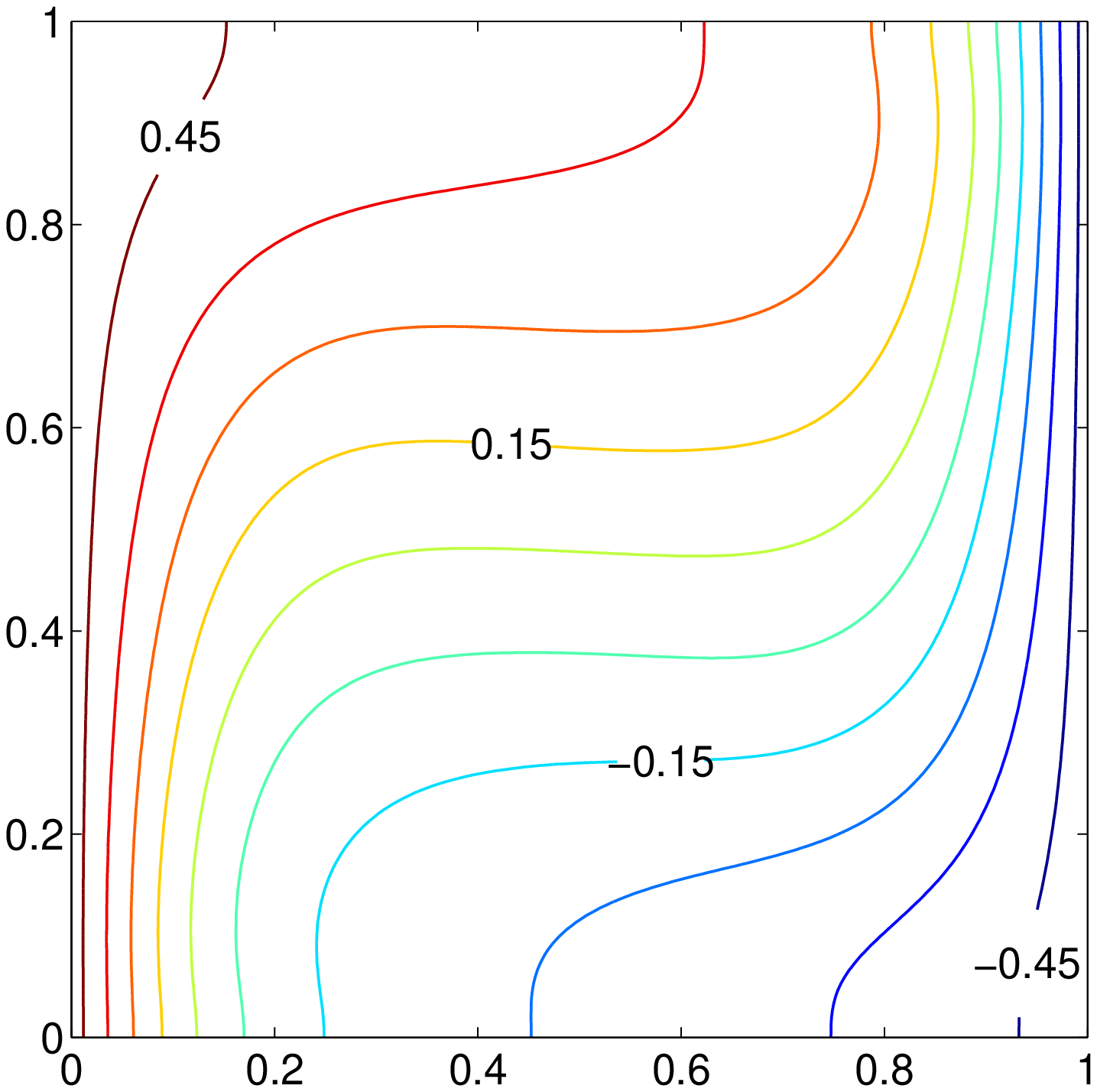}\\
\vspace{-10pt}(b)  $Ra_I=10^5$ \\  \vspace{10pt}
\includegraphics[width=0.35\textwidth,height=0.25\textheight]{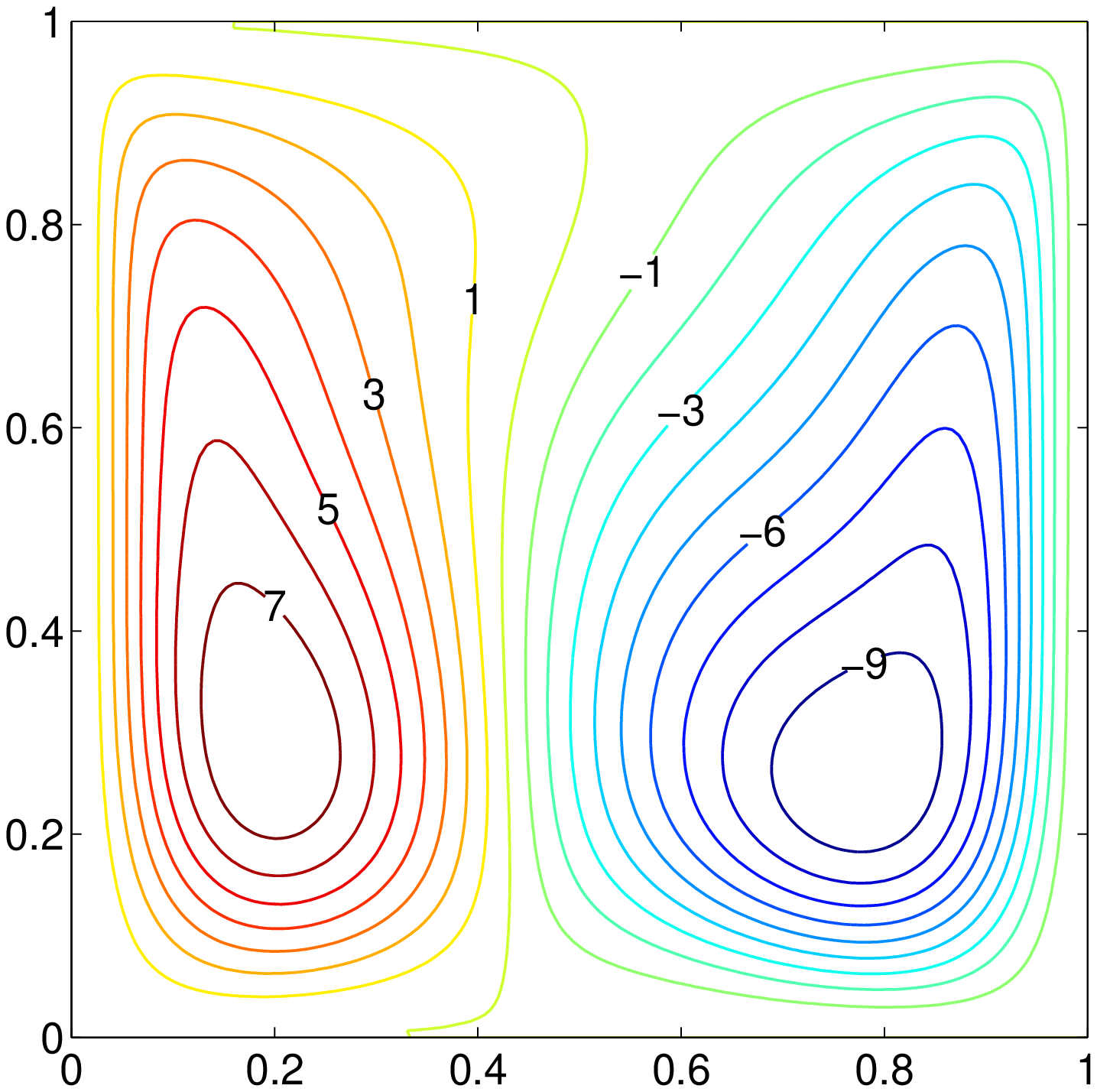}
\includegraphics[width=0.35\textwidth,height=0.25\textheight]{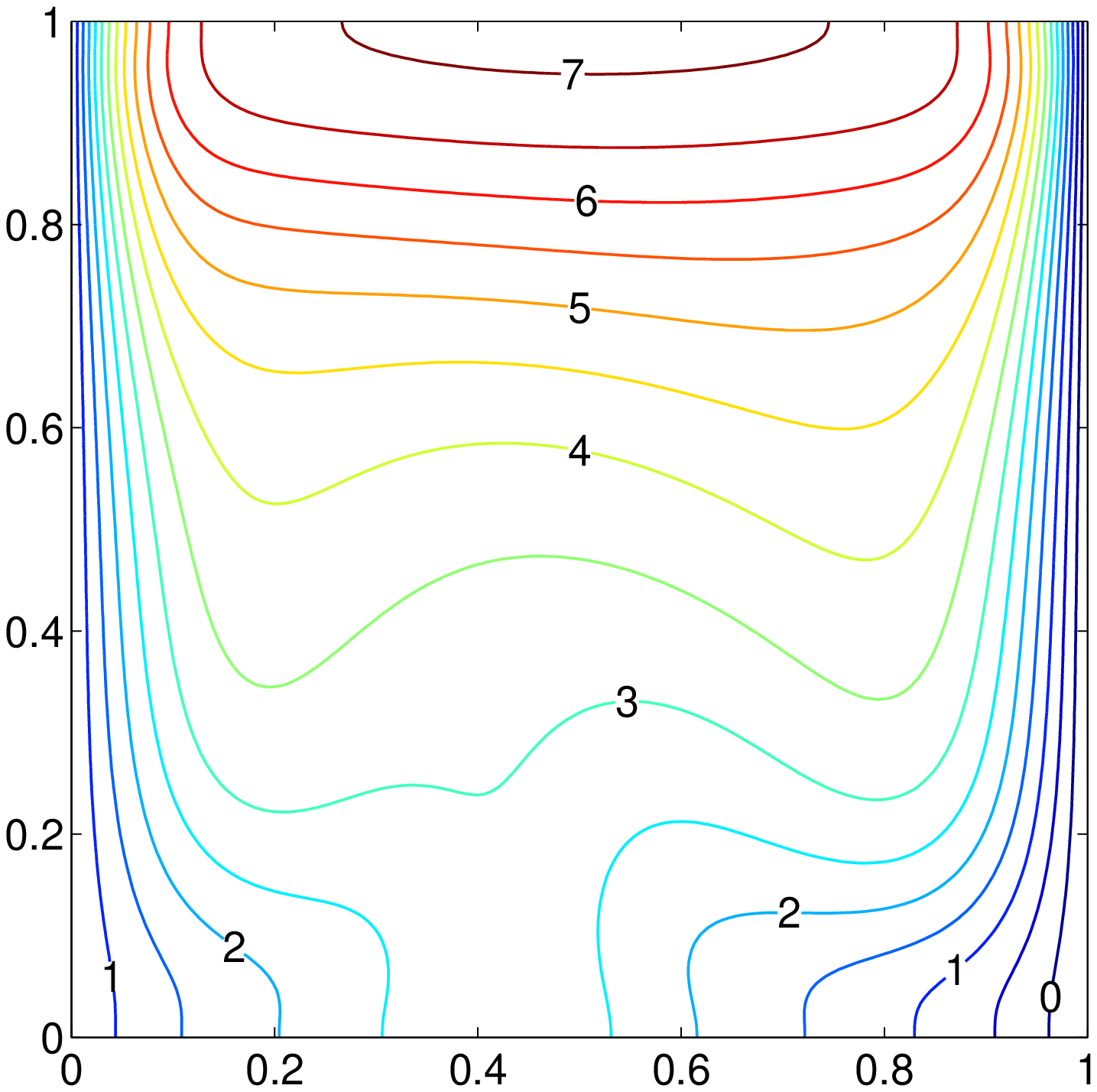}\\
\vspace{-10pt}(c)  $Ra_I=10^7$   \\  \vspace{10pt}
\caption{Streamlines (left) and isotherms (right) for $Ra=10^5$, $\varepsilon=0.4$, $Da=10^{-2}$, and $Pr=0.7$ (gird size: $150\times150$).}
\label{fig:NatualHGPor}
\end{figure}

Fig. \ref{fig:NatualHGPor} shows the streamlines and isotherms for different $Ra_I$ at $Ra=10^5, \varepsilon=0.4$, and $Da=10^{-2}$.
At $Ra_I=10^3$, the flow fields exhibit an elliptic vortex feature in the core of the cavity, and the isotherms are horizontal in the cavity center due to the dominated convection effect from the external sidewall-heating. As $Ra_I$ increases to $Ra_I=10^5$, the flow fields are in a similar status as those at $Ra_I=10^3$, but the flow circulation becomes more stronger. When $Ra_I$ is further increased to $10^7$, due to the increase of internal-heating effect, two vortices with counter rotations exist in the cavity, and the isotherms in the left region of the cavity exhibit an opposite curve to those in Figs. \ref{fig:NatualHGPor} $(\textrm{a})$ and $(\textrm{b})$.

\begin{figure}[htbp]
\centering
\includegraphics[width=0.35\textwidth,height=0.25\textheight]{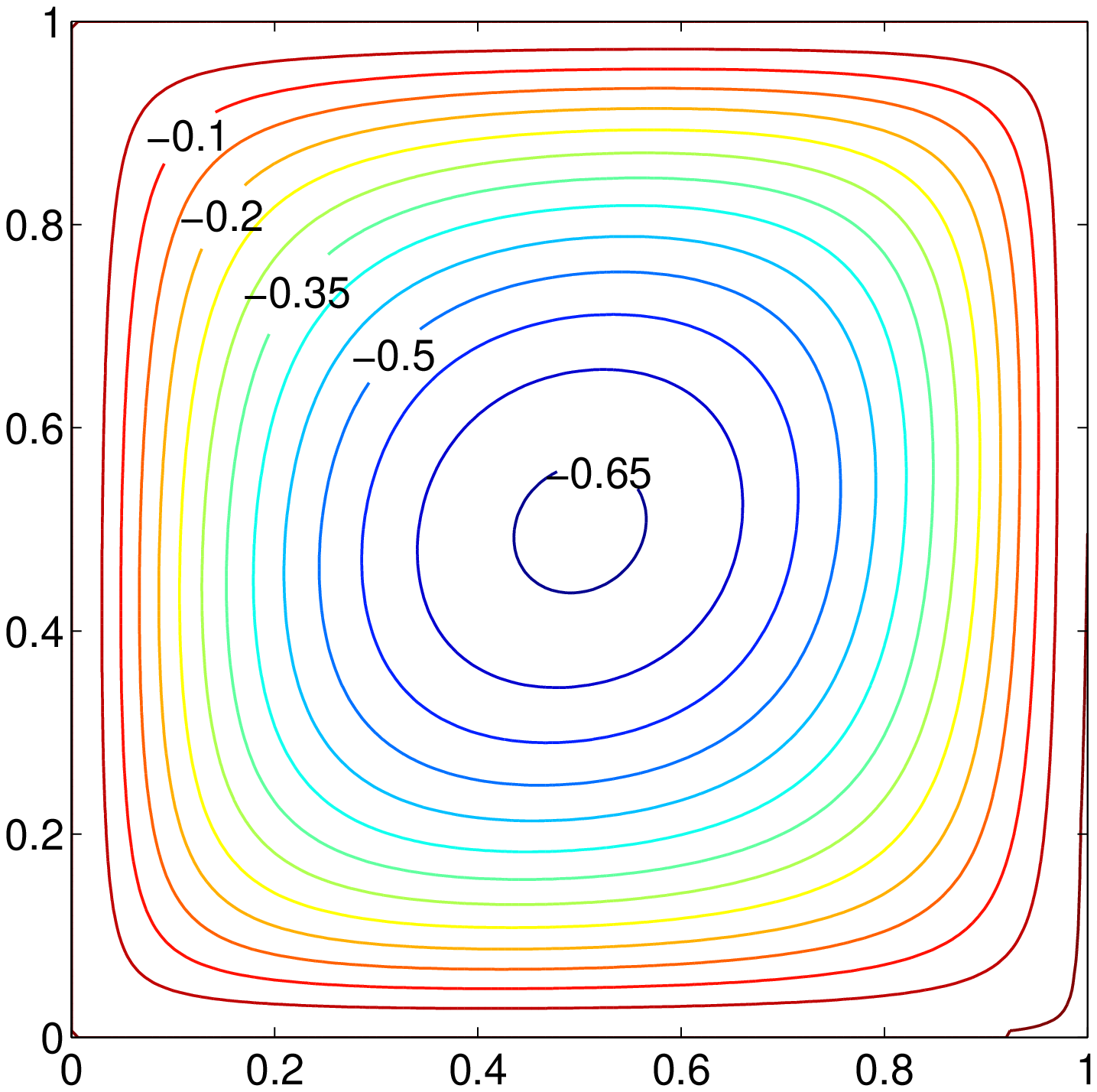}
\includegraphics[width=0.35\textwidth,height=0.25\textheight]{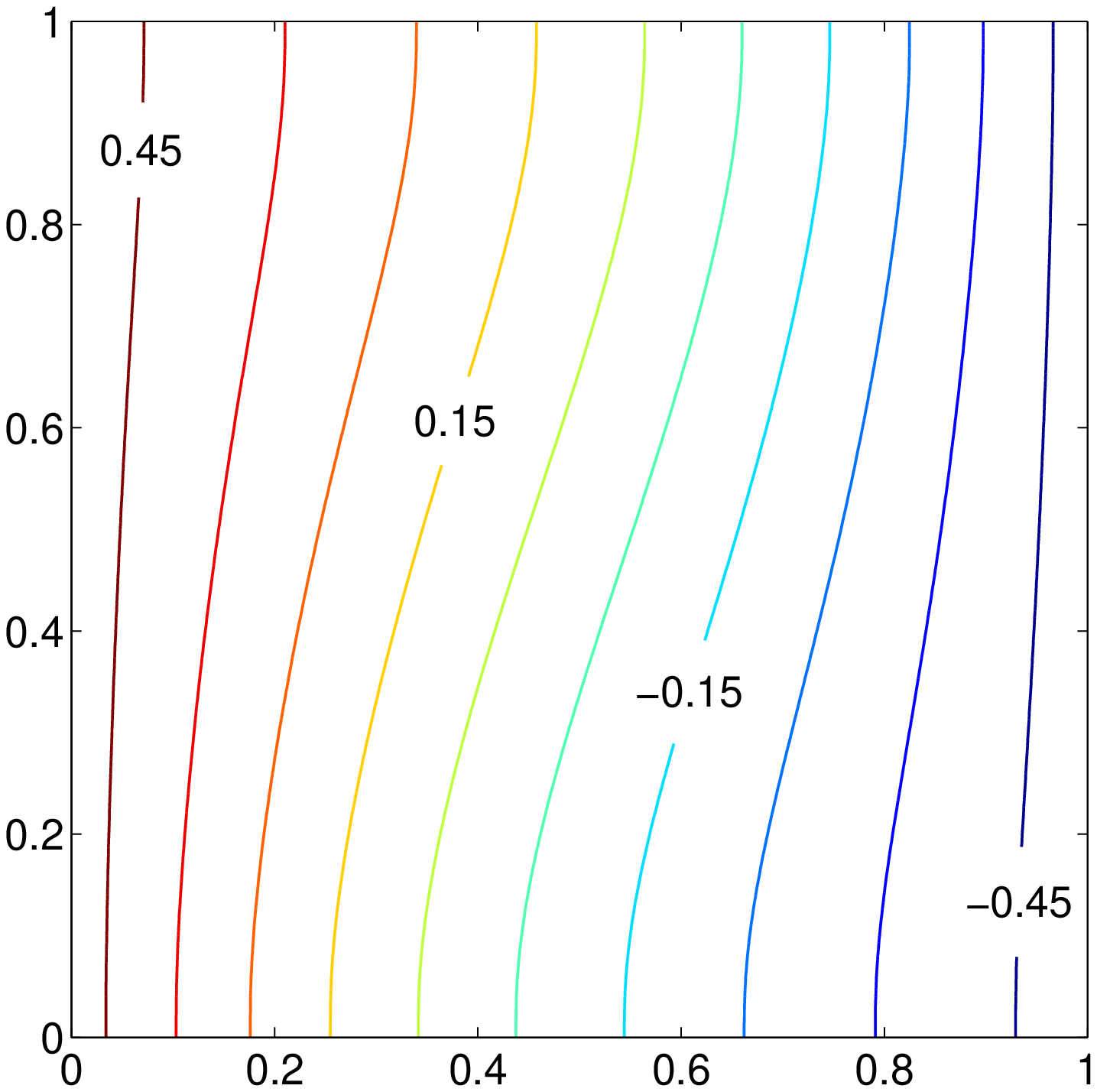}\\
\vspace{-10pt}(a) $Ra_I=10^3$  \\ \vspace{10pt}
\includegraphics[width=0.35\textwidth,height=0.25\textheight]{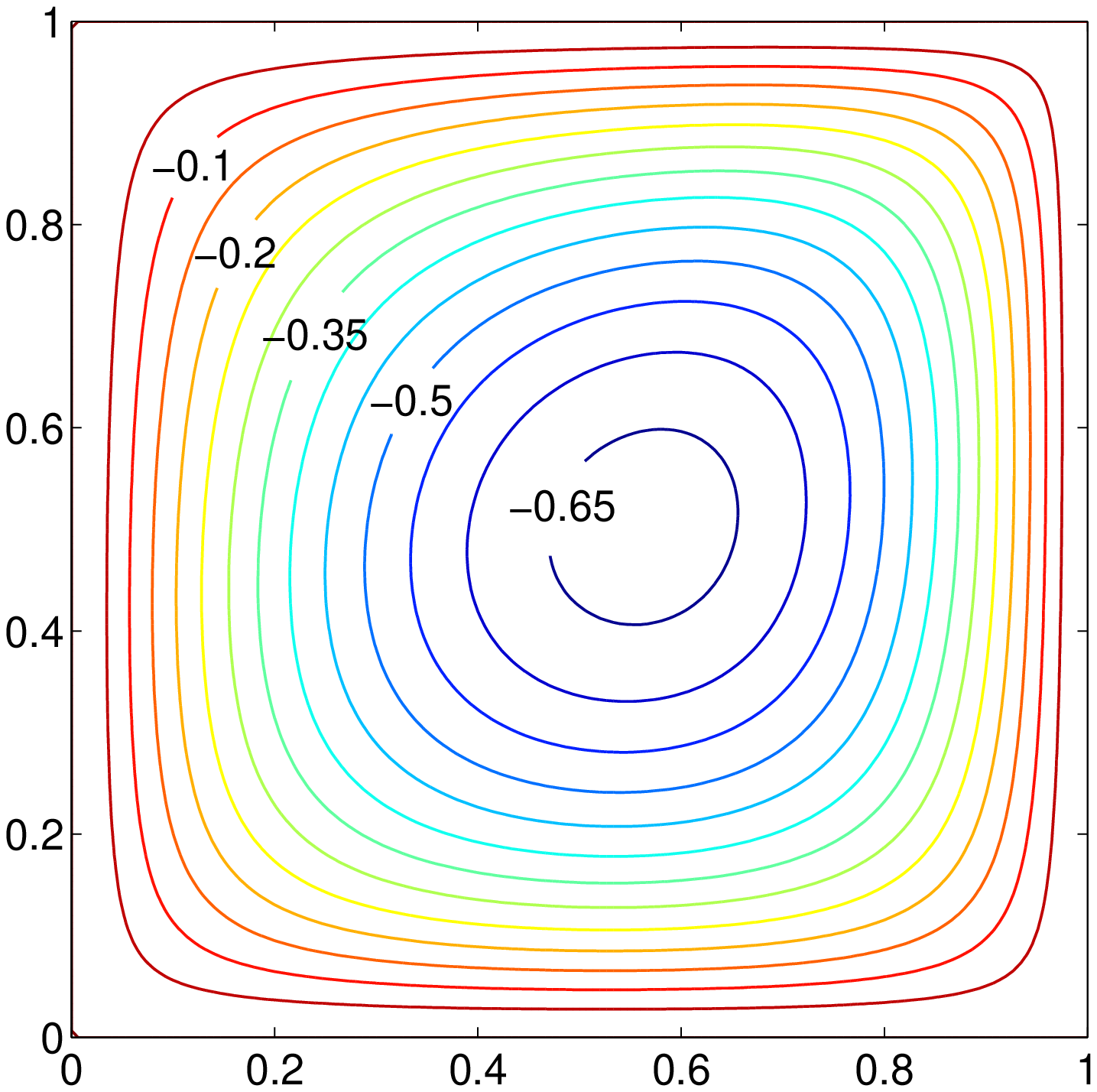}
\includegraphics[width=0.35\textwidth,height=0.25\textheight]{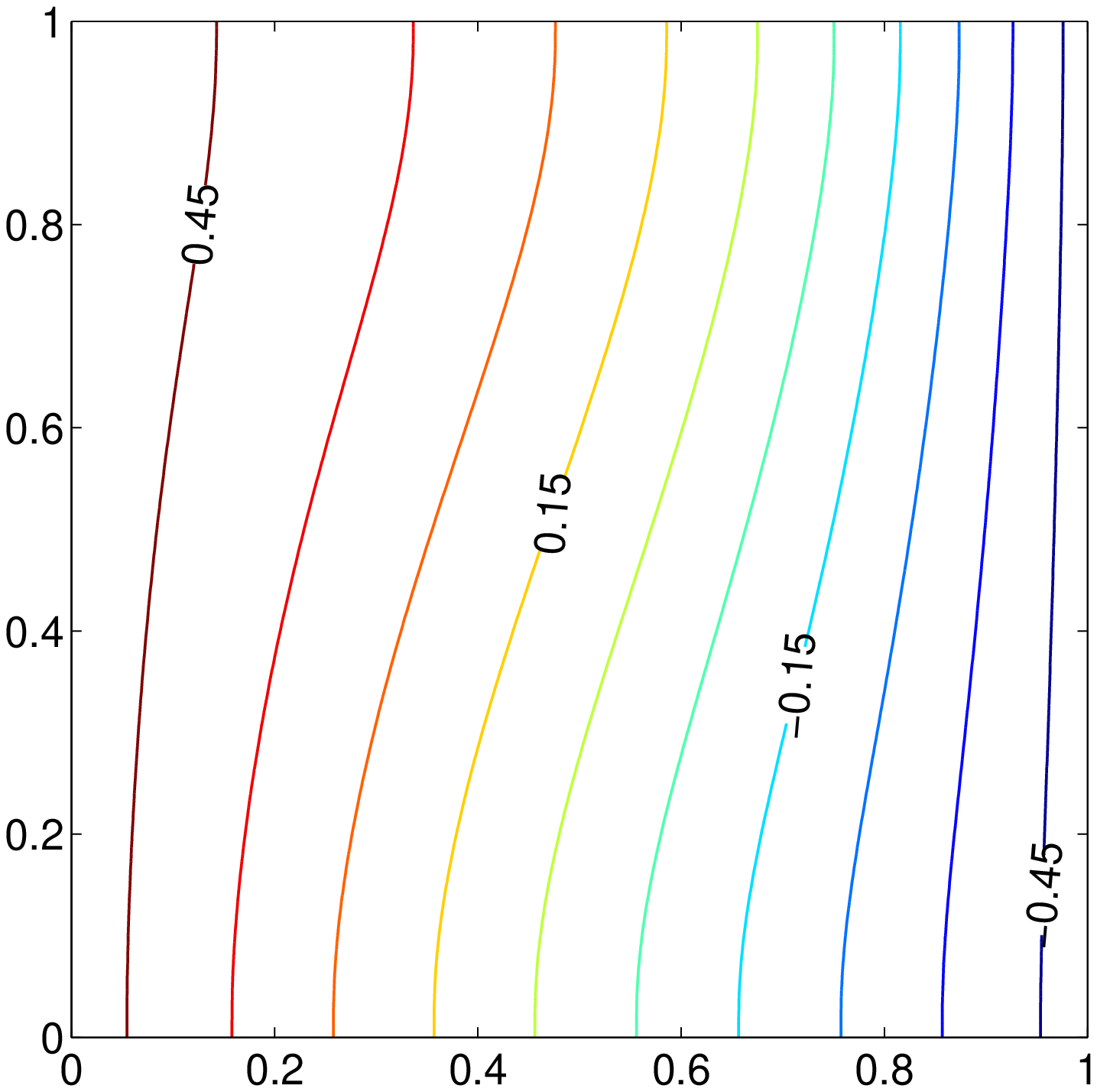}\\
\vspace{-10pt}(b)  $Ra_I=10^5$ \\  \vspace{10pt}
\includegraphics[width=0.35\textwidth,height=0.25\textheight]{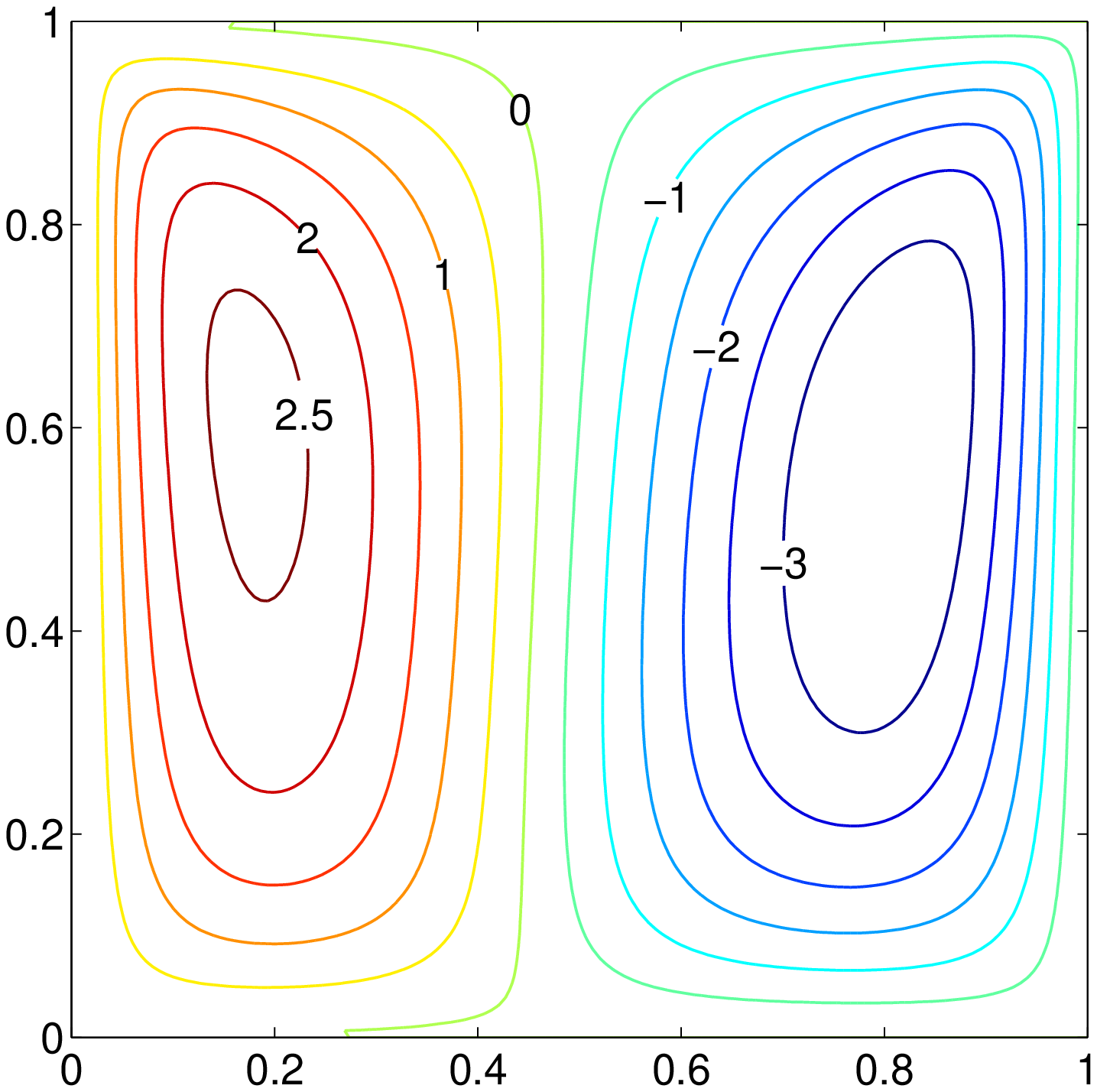}
\includegraphics[width=0.35\textwidth,height=0.25\textheight]{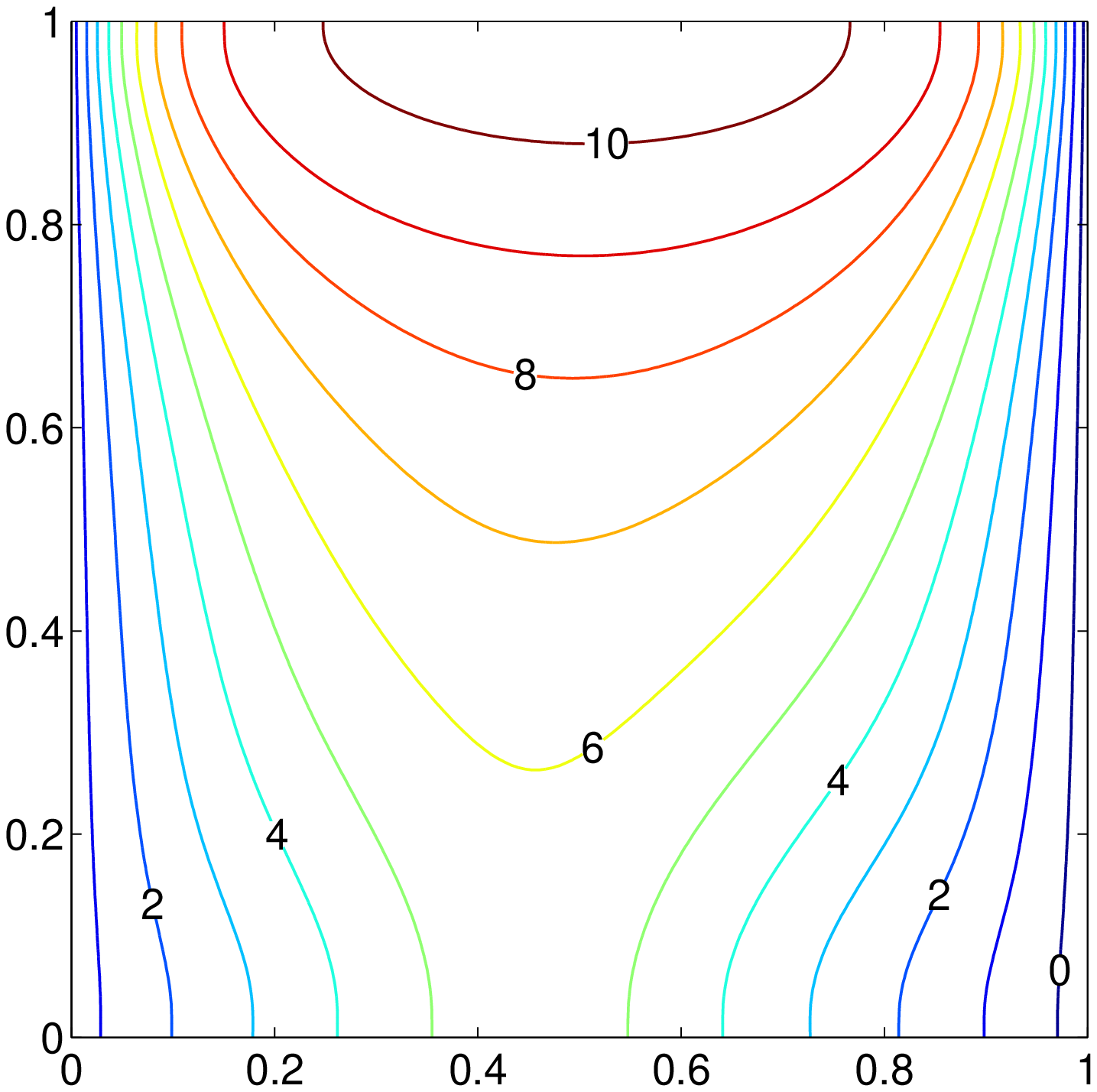}\\
\vspace{-10pt}(c)  $Ra_I=10^7$   \\  \vspace{10pt}
\caption{Streamlines (left) and isotherms (right) for $Ra=10^5$, $\varepsilon=0.4$, $Da=10^{-4}$, and $Pr=0.7$ (gird size: $150\times150$).}
\label{fig:NatualHGPor2}
\end{figure}

Next, the streamlines and isotherms for different $Ra_I$ at $Ra=10^5, \varepsilon=0.4$ but $Da=10^{-4}$ are depicted in Fig. \ref{fig:NatualHGPor2}.
Compared with the results of $Da=10^{-2}$ in Fig. \ref{fig:NatualHGPor}, the lower permeability of the porous medium reduces the strength of the flow field here. At $Ra_I=10^3$, the isotherms show a weak-convection structure, and the center vortex with weaker circulation can be observed from the streamlines. As $Ra_I$ increases to $10^7$, the flow field is dominated by changing from the external sidewall-heating to the internal-heating, and a convection temperature structure results from the larger internal-heating effect.

\begin{figure}[htbp]
\centering
\includegraphics[width=0.35\textwidth,height=0.25\textheight]{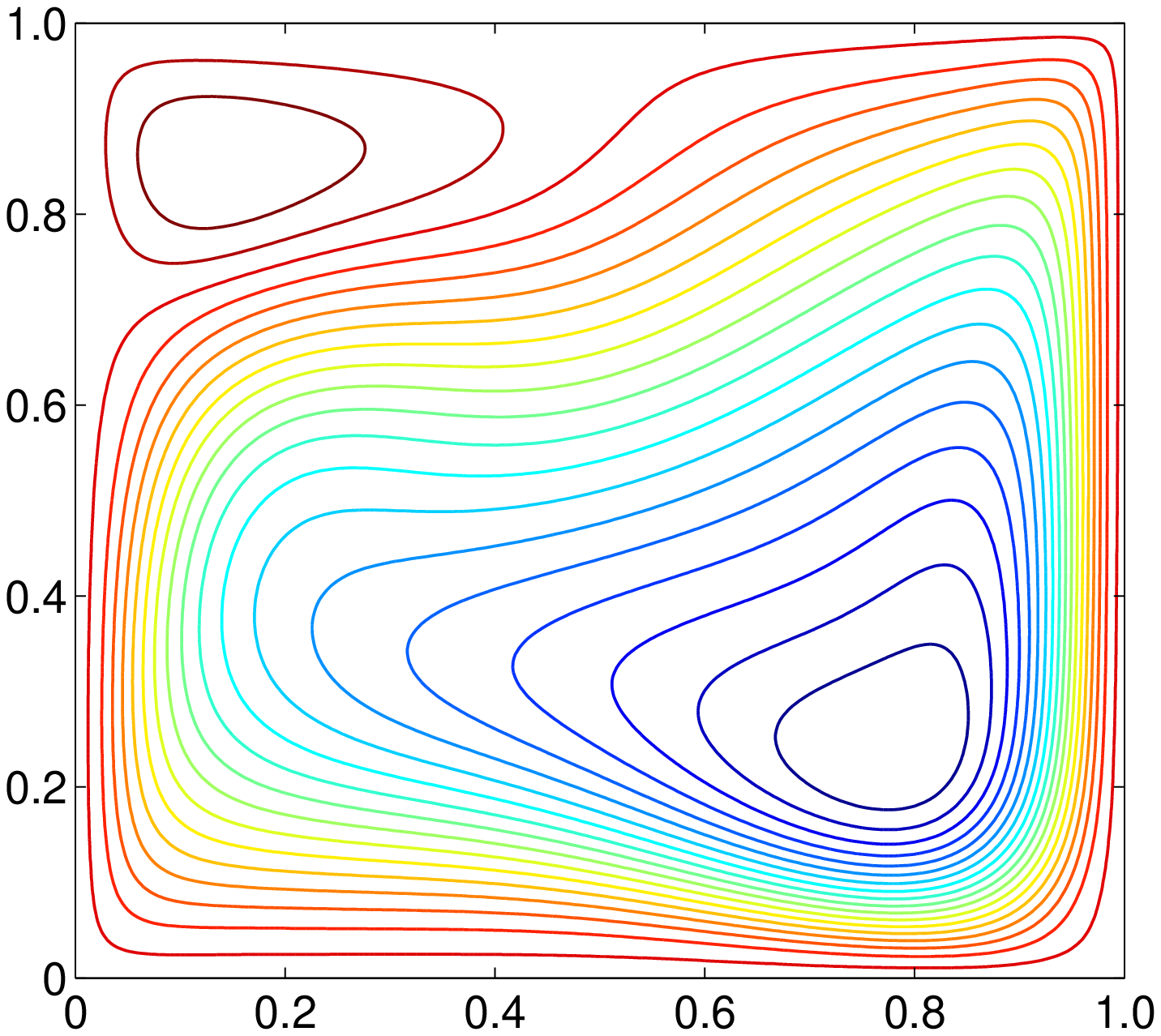}
\includegraphics[width=0.35\textwidth,height=0.25\textheight]{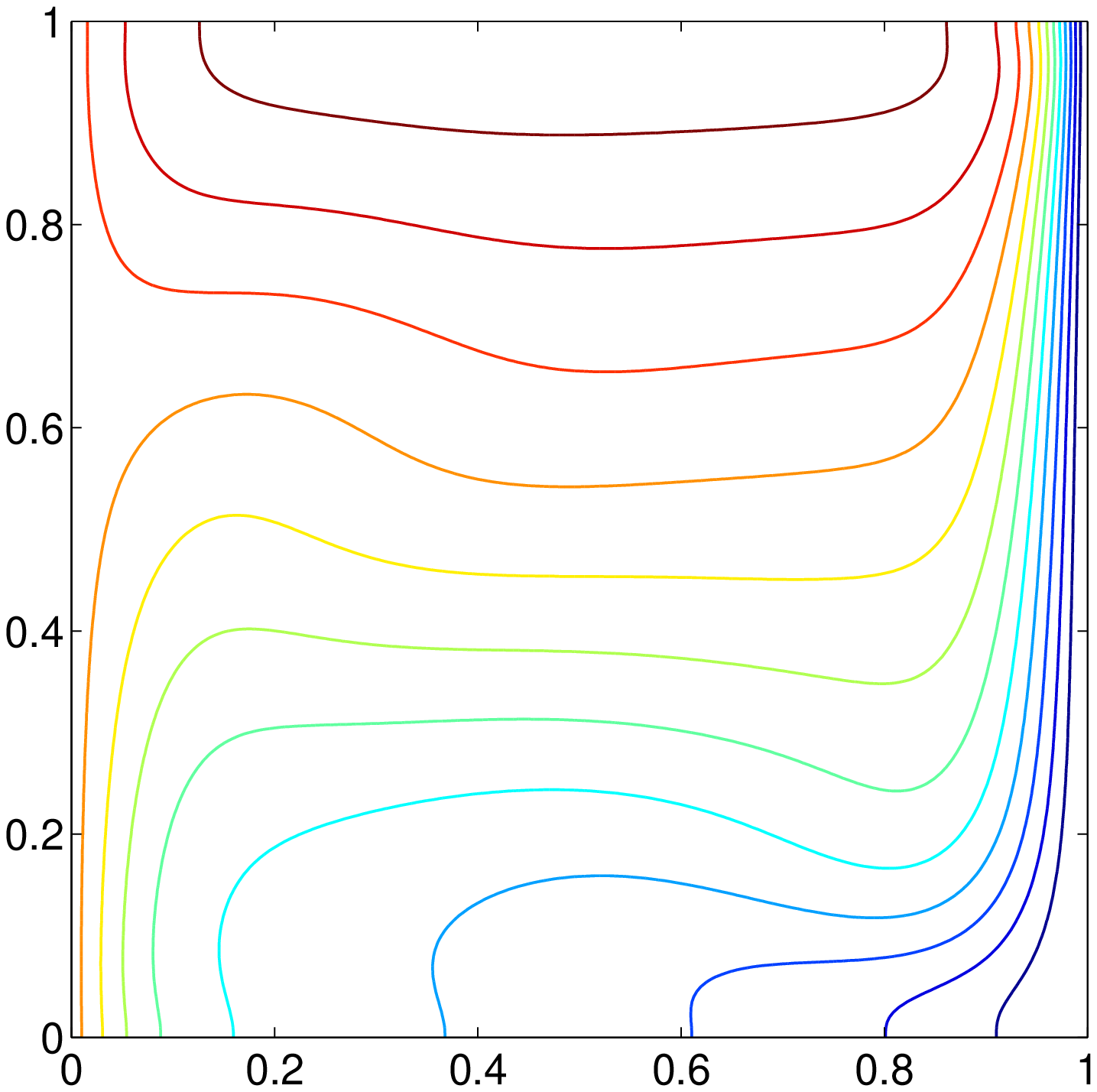}\\
\vspace{-10pt}(a) $Da=10^{-2}$  \\ \vspace{10pt}
\includegraphics[width=0.35\textwidth,height=0.25\textheight]{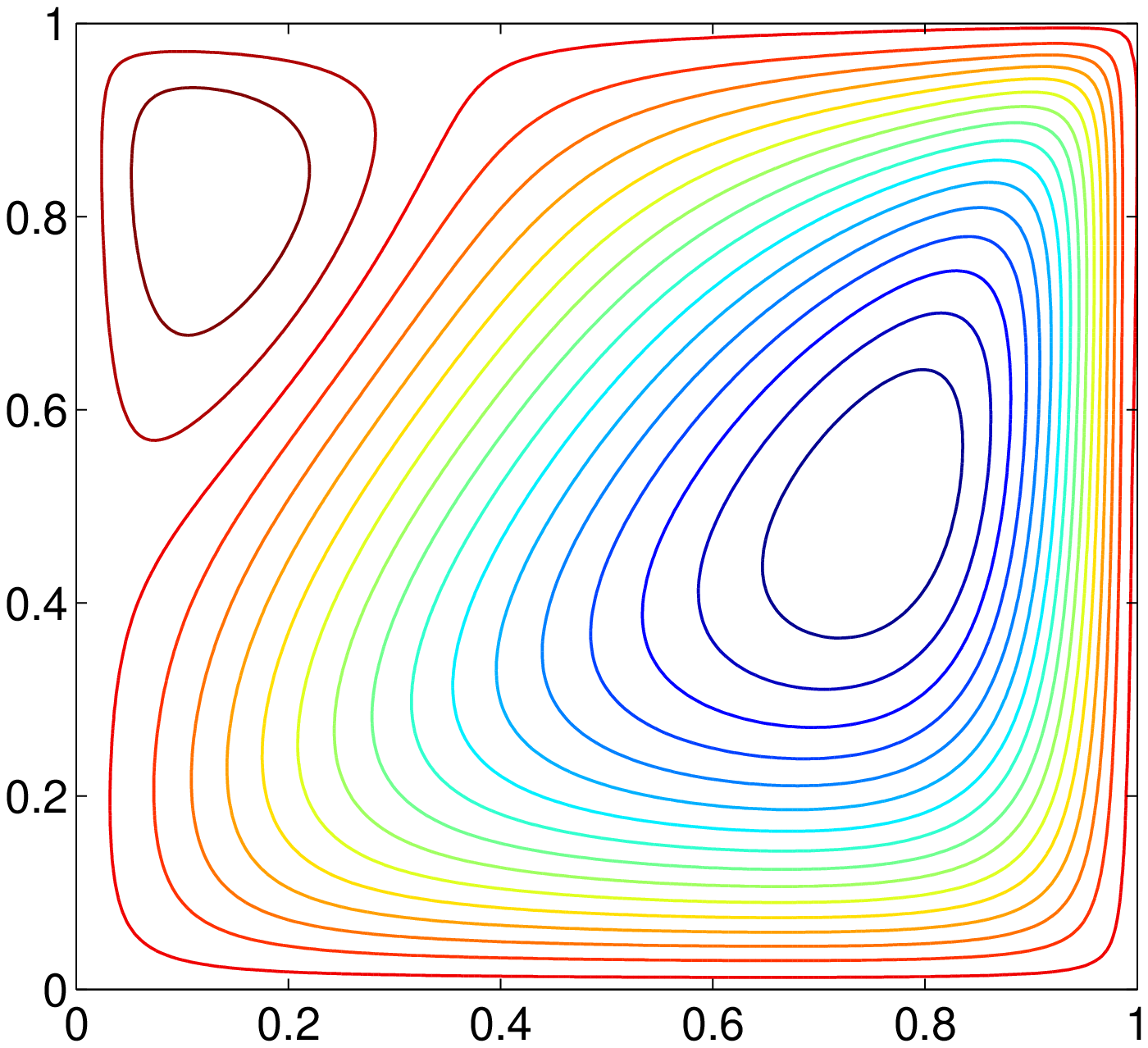}
\includegraphics[width=0.35\textwidth,height=0.25\textheight]{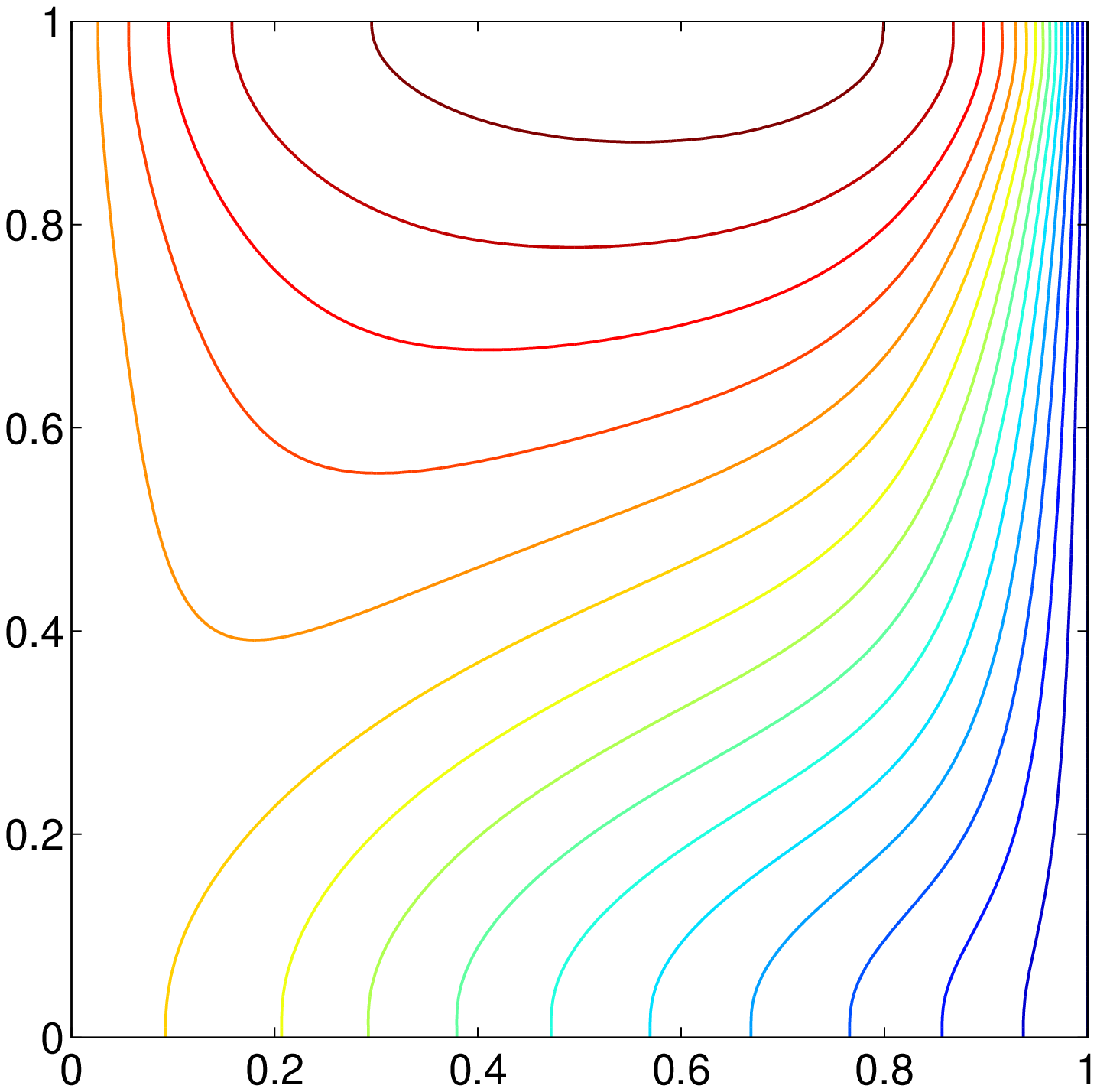}\\
\vspace{-10pt}(b)  $Da=10^{-4}$ \\  \vspace{10pt}
\includegraphics[width=0.35\textwidth,height=0.25\textheight]{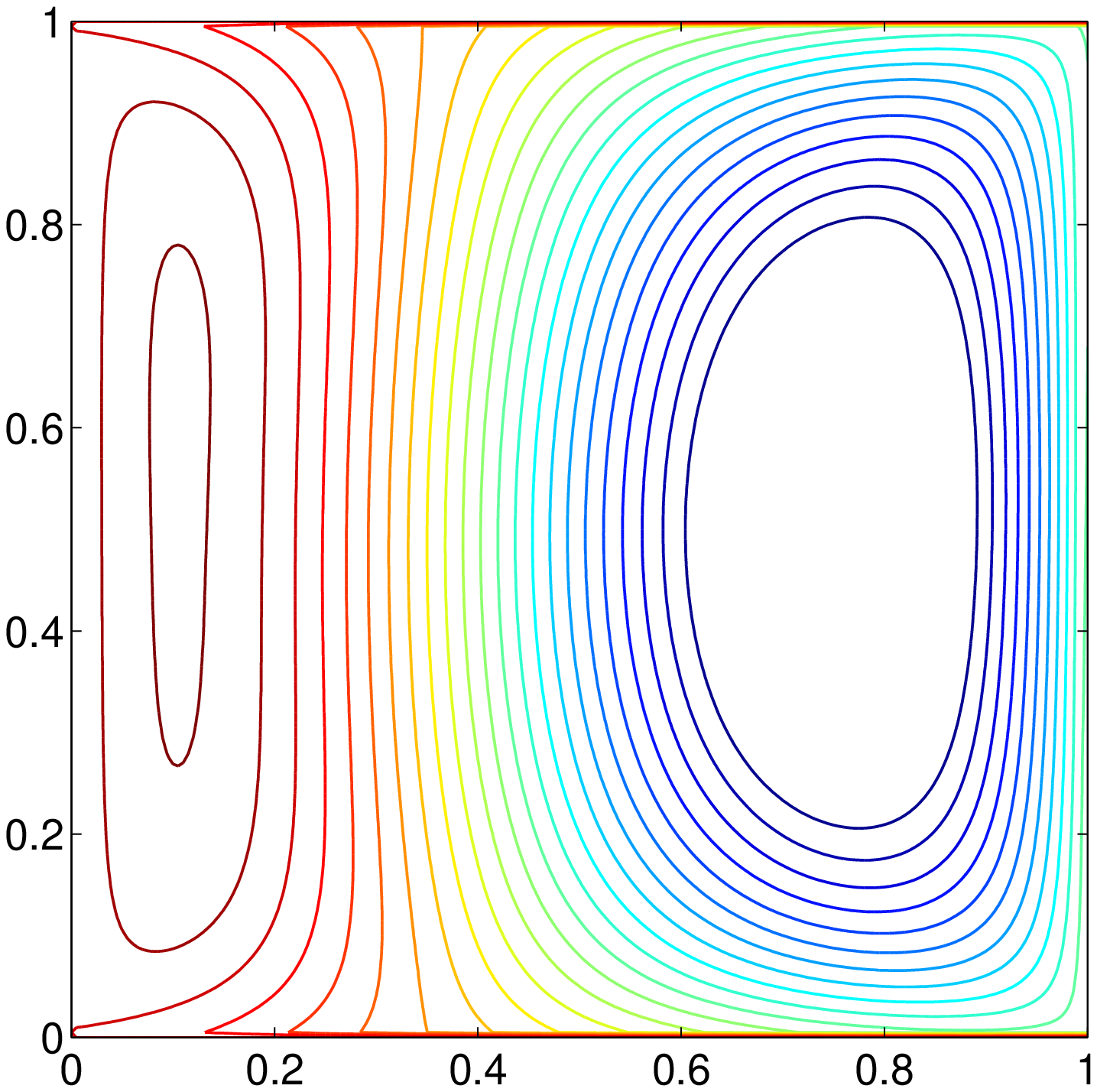}
\includegraphics[width=0.35\textwidth,height=0.25\textheight]{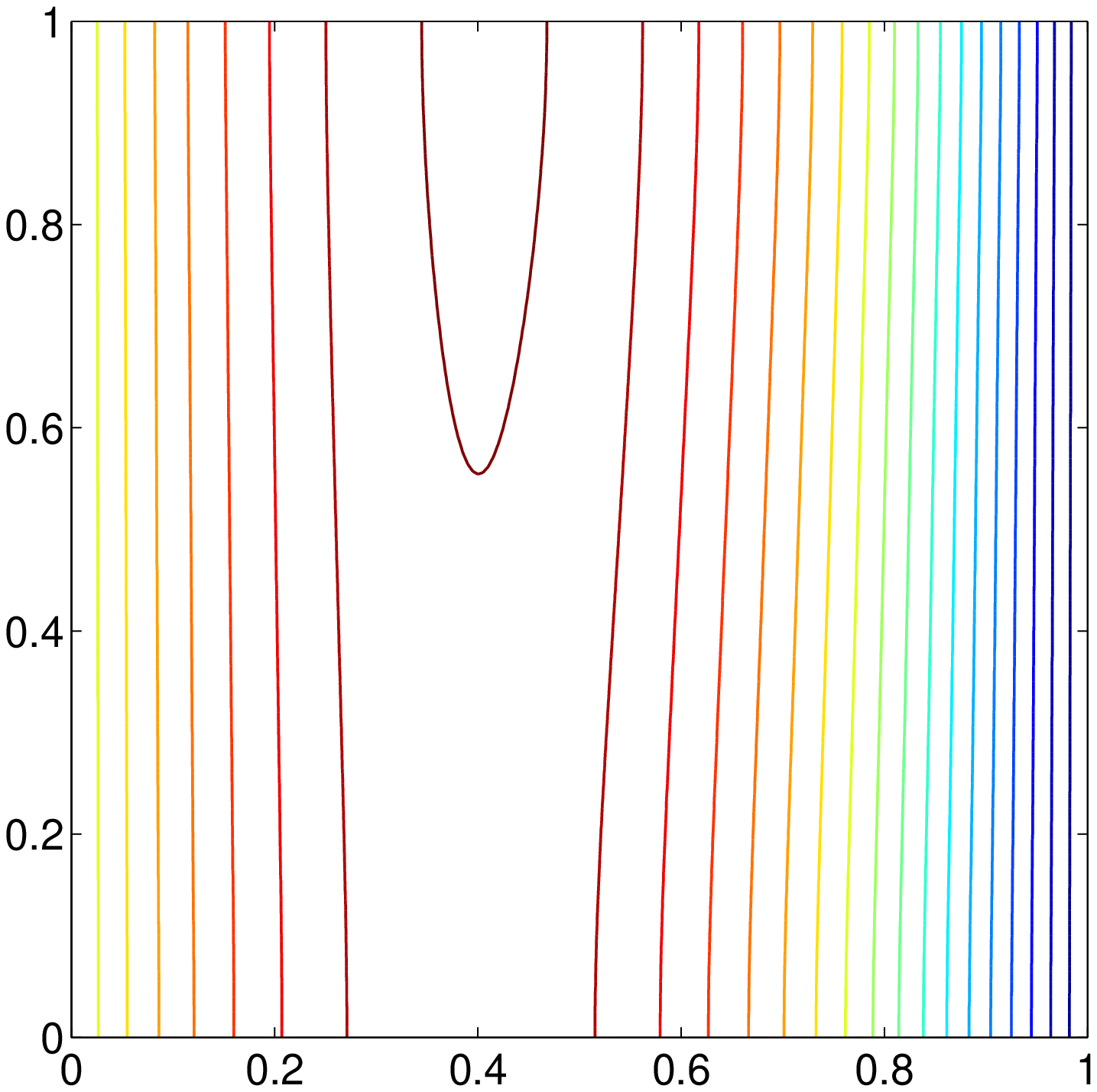}\\
\vspace{-10pt}(c)  $Da=10^{-6}$   \\  \vspace{10pt}
\caption{Streamlines (left) and isotherms (right) for $Ra=10^6$, $\varepsilon=0.4$, $Ra_I=10^7$, and $Pr=0.7$ (gird size: $200\times200$).}
\label{fig:NatualHGPor3}
\end{figure}

The effect of $Da$ on the flow patterns and isotherms at $Ra=10^6, Ra_I=10^7$ and $\varepsilon=0.4$ is illustrated in Fig. \ref{fig:NatualHGPor3}.
It is noted for $Ra_I/Ra\geq 10$ that the heat transfer in the cavity is induced by the internal-heating more intensively than that by the external sidewall-heating. Owing to the combined influence of the internal heat generation and the buoyancy force, a couple of vortices appear respectively in the vicinity of the hot wall and the cold wall. It can be observed from the isotherms that as $Da$ decreases, the convective heat transfer becomes weaker due to the lower strength of flow circulation. In addition, in the internal-heating dominated condition, the maximum dimensionless temperature $\theta_{max}$ increases with the decrease of $Da$. All the observations from Figs. \ref{fig:NatualHGPor}-\ref{fig:NatualHGPor3} are in good agreement with those results in previous studies \cite{Liu14,Jue03}.

To quantify the results, the maximum dimensionless temperature $\theta_{max}$ and the average Nusselt numbers along the hot wall are computed. Table \ref{Tab:NatualHG} lists the present numerical results together with some previous results using the finite element method \cite{Jue03} and the LBM \cite{Liu14} for comparison.
\begin{table}
 \caption{Comparisons of $\overline{Nu}$ and $\theta_{max}$ between the present results and the literature results in Refs. \cite{Liu14,Jue03} at $Pr=0.7$ ($\varepsilon=0.4$, grid sizes: $150\times150$ for $Ra=10^5$, and $200\times200$ for $Ra=10^6$).}
  \vspace{0.4em} \label{Tab:NatualHG}
  \centering
  \begin{tabular*}
   {16cm}[gtb!]{@{\extracolsep{\fill}}cclllllllll}
   \toprule[0.06em]
   \multicolumn{3}{c}{}  &\multicolumn{2}{c}{$Da=10^8, \varepsilon=0.9999$}  &\multicolumn{2}{c}{$Da=10^{-2}$}    &\multicolumn{2}{c}{$Da=10^{-4}$} &\multicolumn{2}{c}{$Da=10^{-6}$}\\
   \cmidrule(lr){4-5} \cmidrule(lr){6-7} \cmidrule(lr){8-9} \cmidrule(lr){10-11}
   $Ra$    &$Ra_I$    &{}     &$\overline{Nu}$  \hspace{2.5em}  &$\theta_{max}$     &$\overline{Nu}$    &$\theta_{max}$   &$\overline{Nu}$    &$\theta_{max}$ &$\overline{Nu}$    &$\sigma_{max}$ \\ \midrule
   $10^5$                 &$10^3$      &Ref. \cite{Jue03}     &4.505     &0.5        &2.884      &0.5      &1.058       &0.5       &0.995     &0.5\\
   {}                     &{}          &Ref. \cite{Liu14}     &4.538     &0.5        &2.903      &0.5      &1.061       &0.5       &0.995     &0.5\\
   {}                     &{}          &Present               &4.536     &0.5        &2.900      &0.5      &1.060       &0.5       &0.995     &0.5\\
   {}                     &$10^5$      &Ref. \cite{Jue03}     &4.021     &0.5        &2.401      &0.5      &0.571       &0.5       &0.506     &0.5\\
   {}                     &{}          &Ref. \cite{Liu14}     &4.057     &0.5        &2.421      &0.5      &0.575       &0.5       &0.506     &0.5\\
   {}                     &{}          &Present               &4.056     &0.5        &2.421      &0.5      &0.574       &0.5       &0.506     &0.5\\
   {}                     &$10^7$      &Ref. \cite{Jue03}     &-42.45    &5.54       &-44.08     &7.30     &-46.37      &10.86     &-48.35    &12.63\\
   {}                     &{}          &Ref. \cite{Liu14}     &-41.35    &5.52       &-43.41     &7.23     &-45.08      &10.76     &-48.31    &13.27\\
   {}                     &{}          &Present               &-40.99    &5.34       &-43.05     &7.11     &-44.79      &10.53     &-48.29    &12.62\\
   $10^6$                 &$10^7$      &Ref. \cite{Jue03}     &4.129     &0.66       &1.242      &0.81     &-1.789      &1.01      &-3.93     &1.32 \\
   {}                     &{}          &Ref. \cite{Liu14}     &4.254     &0.65       &1.296      &0.80     &-1.728      &1.05      &-3.926    &1.32 \\
   {}                     &{}          &Present               &4.244     &0.64       &1.314      &0.80     &-1.729      &1.05      &-3.941    &1.32 \\
   \bottomrule
   \end{tabular*}
 \end{table}
As can be seen from the Table, the present results agree quantitatively well with those reported results in the literature.

\section{Conclusions}
\label{Sec: conclusion}
In this paper, a modified LBGK model is proposed for simulating the incompressible fluid flow and convective heat transfer in porous media at the REV scale. Different from previous LBGK models for thermal flows in porous media, the macroscopic temperature equation can be correctly recovered from the proposed LBGK model by employing a modified EDF and a source term in the evolution equation of temperature field. In addition, following the idea of the LKS, the EDFs for the flow and temperature fields in  the present LBGK model are incorporated with the shear rate and temperature gradient, for which a local computing scheme is proposed instead of the traditionally finite-difference schemes. The key advantage of this treatment is that the fluid viscosity and the thermal diffusivity can be determined by two additional parameters independent from the relaxation times, and thus better numerical stability can be achieved at low viscosities and thermal diffusivities.

The modified LBGK model has been well validated by simulating four two-dimensional convective flow problems, including mixed heat convection in a porous channel, natural convection in a porous cavity, and thermal convection in porous cavity with internal heat generation with isothermally cooled walls/external sidewall-heating. As for the present model for the flow and temperature fields and the local schemes for the gradient operators, they have been demonstrated to be second-order accurate in space by the first test problem where the analytical solution exists. Furthermore, the computational accuracy of the present model and the local scheme for gradient calculations have been investigated with different values of dimensionless relaxation times and the two additional parameters. The results suggest that to obtain more accurate results, the relaxation times should be around unity, and the additional parameter in the flow field should be close to zero, while the other additional parameter in the temperature filed is assigned with smaller value as possible. In addition, the present modified LBGK model is confirmed to be more stable than the non-modified LBGK models at low viscosities and thermal diffusivities. Thus, the present work provides a more useful LBGK model in studying heat (or mass) transfer processes in porous media.

\section*{Acknowledgements}
{
  The financial supports from the National Natural Science Foundation of China under Grant Nos. 51125024 and 51390494 are acknowledged. LW would like to thank Mr. Xuhui Meng for useful discussions.
}

\appendix
\section{Chapman-Enskog analysis on the LBGK model for the flow field}
\label{Appen:MRTMod}
The Chapman-Enskog analysis is provided for the LBE \eqref{LBEeq} to recover the hydrodynamic equations. To this end, the following expansions with a expansion parameter $\xi$ are introduced:
\begin{subequations}\label{ChapmanEn}
\begin{align}
 f_i\;&=\;f_i^{(0)}+\xi f_i^{(1)}+\xi^2 f_i^{(2)}+\cdots,\label{eqm1} \\
 \partial_t&=\;\xi\partial_{t_1}+ \xi^2\partial_{t_2},\label{eqm2} \\
  \nabla\;&=\;\xi\nabla_1,\qquad \bm{F}=\xi \bm{F}^{(1)}\label{eqm3}.
\end{align}
\end{subequations}
Note that the shear rate $\bm{S}$ resided in the EDF \eqref{EQFIN} is related with the spatial gradient. Thus, the EDF is expanded as the following multiscale form:
\begin{equation}\label{EqExEDF}
  f_i^{(eq)}=f_i^{e(0)}+\xi f_i^{e(1)},
\end{equation}
where $f_i^{e(0)}$ and $f_i^{e(1)}$ are given by
\begin{equation}\label{EQFINApp}
  f_i^{e(0)}=\begin{cases}
     \rho_0-(1-\omega_0)\frac{\varepsilon p}{c_s^2}+\rho_0s_0(\bm{u}),\quad i=0\\
     \omega_i\frac{\varepsilon p}{c_s^2}+\rho_0s_i(\bm{u}),\qquad\qquad\qquad\; i\neq0
  \end{cases},
  \qquad f_i^{e(1)}=\omega_i\rho_0\frac{A\delta_t\bm{S}_1:(\bm{c}_i\bm{c}_i-c_s^2\bm{I})}{2c_s^2},
\end{equation}
where $\bm{S}_1$ is defined as $\bm{S}=\xi \bm{S}_1$. With these definitions, one can easily obtain that:
\begin{align}
    \sum_if_i^{e(0)}=\rho_0,\hspace{10pt} \sum_i\bm{c}_if_i^{e(0)}=\rho_0\bm{u},\hspace{10pt}
     \sum_i\bm{c}_i\bm{c}_if_i^{e(0)}= \varepsilon p\bm{I}+\frac{\rho_0\bm{u}\bm{u}}{\varepsilon},\hspace{10pt}\sum_i\bm{c}_i\bm{c}_i\bm{c}_if_i^{e(0)}= c_s^2\rho_0\Delta\cdot\bm{u},\label{EqZEqu}\\
     \sum_if_i^{e(1)}=0,\hspace{35pt} \sum_i\bm{c}_if_i^{e(1)}=0,\hspace{35pt} \sum_i\bm{c}_i\bm{c}_if_i^{e(1)}=c_s^2\rho_0 A\delta_t\bm{S}_1,  \label{EqFEqu}
\end{align}
where $\Delta\cdot\bm{u}=u_\alpha\delta_{\beta\gamma}+u_\beta\delta_{\alpha\gamma}+u_\gamma\delta_{\alpha\beta}$, and $\delta_{\alpha\beta}$ denotes the Kronecker tensor.

Expanding $f_i(\bm{x}+\bm{c}_i\delta_t, t+\delta_t)$ in Eq. \eqref{LBEeq} with the Taylor theorem about space $\bm{x}$ and time $t$, and applying the above multiscale expansions to the resulting equations, one can obtain the following set of successive equations in the order of $\xi$:
\begin{subequations}\label{MulSepEq}
\begin{align}
  \xi^0\;&:\quad f_i^{(0)}=f_i^{e(0)}, \label{Seeqm1} \\
  \xi^1\;&:\quad D_{1i}f_i^{(0)}=-\frac{1}{\tau_f\delta_t}\bigl[f_i^{(1)}-f_i^{e(1)}\bigr]+F_i^{(1)},  \label{Seeqm2} \\
  \xi^2\;&:\quad \partial_{t_2}f_i^{(0)}+ D_{1i}f_i^{(1)}+ \frac{1}{2}\delta_tD_{1i}^2f_i^{(0)}= -\frac{1}{\tau_f\delta_t}f_i^{(2)} .\label{Seeqm3}
\end{align}
\end{subequations}
where $D_{1i}=\partial_{t_1}+\bm{c}_i\cdot\nabla_1$. Substituting $f_i^{(1)}$ from Eq. \eqref{Seeqm2} into Eq. \eqref{Seeqm3} yields
\begin{equation}\label{Seeqm3new}
   \xi^2\;:\quad \partial_{t_2}f_i^{(0)}+\left(1-\frac{1}{2\tau_f}\right)D_{1i}f_i^{(1)}+\frac{1}{2\tau_f}D_{1i}f_i^{e(1)}
   +\frac{\delta_t}{2}D_{1i}F_i^{(1)}=-\frac{1}{\tau_f\delta_t}f_i^{(2)}.
\end{equation}

Based on the mass and momentum conservation laws in combination with the definitions of fluid density and velocity, the following equations can be established:
\begin{subequations}
  \begin{equation}\label{EqfeqM}
    \rho_0=\sum_if_i^{(eq)}=\sum_if_i,
  \end{equation}
   \begin{equation}
     \rho_0\bm{u}=\sum_i\bm{c}_if_i^{(eq)}=\sum_i\bm{c}_if_i+\frac{\delta_t}{2}\rho_0\bm{F}.
  \end{equation}
\end{subequations}
Along with these results, we can get:
\begin{subequations}\label{ConstrEq}
\begin{equation}
   \sum_if_i^{(0)}=\rho_0, \qquad\quad  \sum_if_i^{(k)}=0\quad (k>0),
\end{equation}
\begin{equation}
    \sum_i\bm{c}_if_i^{(0)}=\rho_0\bm{u}, \qquad\quad \sum_i\bm{c}_if_i^{(1)}=-\frac{\delta_t}{2}\rho_0\bm{F}^{(1)},\qquad\quad  \sum_i\bm{c}_if_i^{(k)}=0\quad (k>1).
\end{equation}
\end{subequations}
From Eq. \eqref{FtermF} for the forcing term $F_i$, the following velocity moments of $F_i$ can be obtained
\begin{equation}\label{FmomEq}
  \sum_iF^{(1)}=0,\qquad \sum_i\bm{c}_iF^{(1)}=\left(1-\frac{1}{2\tau_f}\right)\rho_0\bm{F}^{(1)},
  \qquad \sum_i\bm{c}_i\bm{c}_iF^{(1)}=\left(1-\frac{1}{2\tau_f}\right)\left(\frac{\rho_0\bm{u}\bm{F}^{(1)}}{\varepsilon}
 +\frac{\rho_0\bm{F}^{(1)}\bm{u}}{\varepsilon}\right).
\end{equation}

By taking the velocity moments of Eq. \eqref{Seeqm2} and combining Eqs. \eqref{Seeqm1}, \eqref{EqZEqu}, \eqref{EqFEqu}, \eqref{ConstrEq} and \eqref{FmomEq}, the first-order macroscopic equations can be obtained
\begin{subequations}\label{Eqt1Flu}
  \begin{equation}
    \partial_{t_1}\rho_0+\nabla_1\cdot(\rho_0\bm{u})=0,
  \end{equation}
  \begin{equation}
    \partial_{t_1}(\rho_0\bm{u}) +\nabla_1\cdot\left(\varepsilon p\bm{I}+\frac{\rho_0\bm{u}\bm{u}}{\varepsilon}\right)=\rho_0\bm{F}^{(1)}.
  \end{equation}
\end{subequations}
Similarly, taking the moments of Eq. \eqref{Seeqm3new} leads to the following equations at $O(\xi^2)$:
\begin{subequations}
  \begin{equation}
    \partial_{t_2}\rho_0=0,
  \end{equation}
  \begin{equation}\label{Eqt2Flow}
   \begin{split}
    \partial_{t_2}(\rho_0\bm{u})+\left(1-\frac{1}{2\tau_f}\right)\nabla_1\cdot\left(\sum_i\bm{c}_i\bm{c}_if_i^{(1)}\right)
    +\frac{1}{2\tau_f}\nabla_1\cdot\left(c_s^2\rho_0 A\delta_t\bm{S}_1\right)\\
    +\frac{\delta_t}{2}\left(1-\frac{1}{2\tau_f}\right)\nabla_1\cdot\left(\frac{\rho_0\bm{u}\bm{F}^{(1)}}{\varepsilon}
    +\frac{\rho_0\bm{F}^{(1)}\bm{u}}{\varepsilon}\right)=0.
    \end{split}
  \end{equation}
\end{subequations}
To proceed further, the momentum flux $\sum_i\bm{c}_i\bm{c}_if_i^{(1)}$ needs to be evaluated. By making use of Eqs. \eqref{Seeqm2} and \eqref{Eqt1Flu}, and with some standard algebraic manipulations, we obtain that
\begin{align}
  \sum_i\bm{c}_i\bm{c}_if_i^{(1)}&=-\tau_f\delta_t\sum_i\bm{c}_i\bm{c}_i\left(D_{1i}f_i^{(0)}-F_i^{(1)}\right)+\sum_i\bm{c}_i\bm{c}_if_i^{e(1)}\notag\\
  &=-\tau_f\delta_t\left[\partial_{t_1}\left(\varepsilon p \bm{I}+\frac{\rho_0\bm{uu}}{\varepsilon}\right)+c_s^2\nabla_1\cdot(\Delta\cdot\bm{u})
  -\left(1-\frac{1}{2\tau_f}\right)\left(\frac{\rho_0\bm{u}\bm{F}^{(1)}}{\varepsilon}+\frac{\rho_0\bm{F}^{(1)}\bm{u}}{\varepsilon}\right)\right]
  +c_s^2\rho_0A\delta_t\bm{S}_1\notag\\
  &=c_s^2\rho_0(A-\tau_f)\delta_t\bm{S}_1-\frac{\delta_t}{2}\left(\frac{\rho_0\bm{u}\bm{F}^{(1)}}{\varepsilon}+\frac{\rho_0\bm{F}^{(1)}\bm{u}}{\varepsilon}\right). \label{Eq:Shear}
\end{align}
In the above derivations, as used in Ref. \cite{GuoS00}, we have neglected the terms of $O(Ma^2)$ from the fact that
\begin{equation}\label{EqSeMa}
 \frac{\partial p}{\partial t}=O(Ma^2), \qquad O(\delta p)=O(Ma^2), \qquad O(\bm{u})=O(Ma).
\end{equation}
Here $Ma$ is the Mach number of the flow. Therefore, the final form of Eq. \eqref{Eqt2Flow} can be written as
\begin{equation}\label{Eqt2FlowTer}
  \partial_{t_2}(\rho_0\bm{u})+\nabla_1\cdot\left[\rho_0c_s^2\left(A-\tau_f+\frac{1}{2}\right)\delta_t\bm{S}_1\right]=0.
\end{equation}

Combining the first- and second-order equations and noting that $\rho_0$ is a constant, the final hydrodynamic equations for the flows in porous media can be derived:
\begin{subequations}
  \begin{equation}\label{MsReceq1}
 \nabla\cdot\bm{u}=0,
 \end{equation}
 \begin{equation}\label{MMReceq2}
   \frac{\partial\bm{u}}{\partial t}+(\bm{u}\cdot \nabla)\left(\frac{\bm{u}}{\varepsilon}\right)=-\frac{1}{\rho_0}\nabla(\varepsilon p)+\nu_e\nabla^2\bm{u}+\bm{F},
 \end{equation}
\end{subequations}
where $\nu_e$ is the effect viscosity given by
\begin{equation}
  \nu_e=c_s^2\left(\tau_f-A-\frac{1}{2}\right)\delta_t.
\end{equation}

Now, we derive the formula to compute the pressure $p$ via the distribution function. Note that $r_0(\bm{u})=0$ due to $\nabla\cdot\bm{u}=0$. From the expression of $f_0^{(eq)}$ and the Taylor expansion for the LBE \eqref{LBEeq}, we then have
\begin{align*}
  \frac{\varepsilon(1-\omega_0)}{c_s^2}p(\bm{x},t)&=\rho_0-f_0^{(eq)}(\bm{x},t)+\rho_0s_0(\bm{u}(\bm{x},t))\\
   &=\rho_0-\left[f_0(\bm{x},t)+\tau_f\delta_t(D_0f_0(\bm{x},t)-F_0(\bm{x},t))\right]+\rho_0s_0(\bm{u}(\bm{x},t))+O(\delta_t^2)\\
   &=\rho_0-f_0(\bm{x},t)-\tau_f\delta_t\partial_tf_0(\bm{x},t)+\tau_f\delta_tF_0(\bm{x},t)+\rho_0s_0(\bm{u}(\bm{x},t))+O(\delta_t^2).
\end{align*}
It can be demonstrated that the term of $\partial_tf_0(\bm{x},t)$ in the above equation can be neglected. As for this result, we come back again to Eqs. \eqref{EqExEDF} and \eqref{EQFINApp} for $f_0(\bm{x},t)$, and recur to Eq. \eqref{EqSeMa} to derive
\begin{align*}
   \partial_tf_0(\bm{x},t)&=\partial_tf_0^{e(0)}(\bm{x},t)+O(\xi)\\
    &=-\frac{\varepsilon(1-\omega_0)}{c_s^2}\frac{\partial p}{\partial t} -\frac{\omega_0\rho_0}{2\varepsilon c_s^2}\frac{\partial \mid\bm{u}\mid^2}{\partial t}+O(\xi)\\
    &=O(\xi+Ma^2),
\end{align*}
Therefore, we can obtain
\begin{equation}
  \begin{split}
    \frac{\varepsilon(1-\omega_0)}{c_s^2}p(\bm{x},t)&=\rho_0-f_0(\bm{x},t)+\tau_f\delta_tF_0(\bm{x},t)
    +\rho_0s_0(\bm{u}(\bm{x},t))+O(\delta_t^2+\xi\delta_t+Ma^2\delta_t)\\
    &=\sum_{i\neq0}f_i(\bm{x},t)+\tau_f\delta_tF_0(\bm{x},t)
    +\rho_0s_0(\bm{u}(\bm{x},t))+O(\delta_t^2+\xi\delta_t+Ma^2\delta_t),
   \end{split}
\end{equation}
where Eq. \eqref{EqfeqM} has been used. As a consequence, the pressure $p$ can be calculated with accuracy of order $O(\delta_t^2+\xi\delta_t+Ma^2\delta_t)$ as
\begin{equation}
  p(\bm{x},t)=\frac{c_s^2}{\varepsilon(1-\omega_0)}\left[\sum_{i\neq0}f_i(\bm{x},t)+\rho_0s_0(\bm{u}(\bm{x},t))
  +\tau_f\delta_tF_0(\bm{x},t)\right],
\end{equation}
where $s_0(\bm{u})$ and $F_0$ can be written in a definite form as
\begin{equation}
 s_0(\bm{u})=-\frac{\omega_0}{2\varepsilon c_s^2}\mid\bm{u}\mid^2, \qquad F_0=-\omega_0\left(1-\frac{1}{2\tau_f}\right)\frac{\bm{u}\cdot\bm{F}}{\varepsilon c_s^2}.
\end{equation}




\begin{thebibliography}{99}
\bibitem{ChengP78}P. Cheng, Heat transfer in geothermal systems, Adv. Heat Transfer 14 (1978) 1-105.
\bibitem{Bejan06}D. A. Nield, A. Bejan, Convection in Porous Media, 3rd ed., Springer, New York, 2006.
\bibitem{Vafai05}K. Vafai, Handbook of porous media, 2nd ed, Taylor \& Francis, New York, 2005.
\bibitem{ChenS98}S. Chen, G. D. Doolen, Lattice Boltzmann method for fluid flows, Annu. Rev. Fluid Mech. 30 (1998) 329-364.
\bibitem{Succi01}S. Succi, The Lattice Boltzmann Method for Fluid Dynamics and Beyond, Oxford University Press, New York, 2001.
\bibitem{GuoS13}Z. L. Guo, C. Shu, Lattice Boltzmann method and its applications in engineering, World Scientific, Singapore, 2013.
\bibitem{Succi08}S. Succi, Lattice Boltzmann across scales: from turbulence to DNA translocation, Eur. Phys. J. B 64 (2008) 471-479.
\bibitem{Mohammad11}A. A. Mohammad, Lattice Boltzmann method fundamentals and engineering applications with computer codes, Springer-Verlag, 2011.
\bibitem{Guo02}Z. Guo, T. S. Zhao, Lattice Boltzmann model for incompressible flows through porous media, Phys. Rev. E 66 (2002) 036304.
\bibitem{Guo05}Z. Guo, T. S. Zhao, A lattice Boltzmann model for convection heat transfer in porous media, Numer. Heat Transfer B 47 (2005) 157-177.
\bibitem{Seta06}T. Seta, E. Takegoshi, K. Okui, Lattice Boltzmann simulation of natural convection in porous media, Math. Comput. Simul. 72 (2006) 195-200.
\bibitem{Succi89}S. Succi, E. Foti, F. Higuera, Three-dimensional flows in complex geometries with the lattice Boltzmann method, Europhys. Lett. 10 (1989) 433.
\bibitem{Martys96}N. S. Martys, H. Chen, Simulation of multicomponent fluids in complex three-dimensional geometries by the lattice Boltzmann method, Phys. Rev. E 53 (1996) 743-751.
\bibitem{Pan06}C. Pan, L. S. Luo, C. T. Miller, An evaluation of lattice Boltzmann schemes for porous medium flow simulation, Comput. Fluids 35 (2006) 898-909.
\bibitem{Kang07}Q. Kang, P. C. Lichtner, D. Zhang, An improved lattice Boltzmann model for multicomponent reactive transport in porous media at the pore scale, Water Resour. Res. 43 (12) (2007).
\bibitem{Parmigiani11}A. Parmigiani, C. Huber, O. Bachmann, B. Chopard, Pore-scale mass and reactant transport in multi-phase porous media flows, J. Fluid Mech. 686 (2011) 40-76.
\bibitem{Landry14}C. J. Landry, Z. T. Karpyn, O. Ayala, Relative permeability of homogenous-wet and mixed-wet porous media as determined by pore-scale lattice Boltzmann modeling, Water Resour. Res. 50 (2014) 3672-3689.
\bibitem{Spaid97}M. A. A. Spaid, F. R. Phelan, Lattice Boltzmann methods for modeling microscale flow in fibrous porous media, Phys. Fluids E 9 (1997) 2468-2474.
\bibitem{Dardis98}O. Dardis, J. McCloskey, Lattice Boltzmann scheme with real numbered solid density for the simulation of flow in porous media, Phys. Rev. E 57 (1998) 4834.
\bibitem{Kang02}Q. Kang, D. Zhang, S. Chen, Unified lattice Boltzmann method for flow in multi-scale porous media, Phys. Rev. E 66 (2002) 056307.
\bibitem{Zarghami14}A. Zarghami, C. Biscarini, S. Succi, S. Ubertini, Hydrodynamics in Porous Media: A Finite Volume Lattice Boltzmann Study, J. Sci. Comput 59 (2014) 80-103.
\bibitem{Shokouhmand09}H. Shokouhmand, F. Jam, M.R. Salimpour, Simulation of laminar flow and convective heat transfer in conduits filled with porous media using lattice Boltzmann method, Int. Commun. Heat Mass Transfer 36 (2009) 378-384.
\bibitem{Rong10}F. Rong, Z. Guo, Z. Chai, B. Shi, A lattice Boltzmann model for axisymmetric thermal flows through porous media, Int. J. Heat Mass Transfer 53 (2010) 5519-5527.
\bibitem{Abrach13}H. El Abrach, H. Dhahri, A. Mhimid, Lattice Boltzmann method for modeling heat and mass transfers during drying of deformable porous medium, J. Porous Media, 16 (2013) 837-855.
\bibitem{Gaoetal14}D. Y. Gao, Z. Q. Chen, L. H. Chen, A thermal lattice Boltzmann model for natural convection in porous media under local thermal non-equilibrium conditions, Int. J. Heat Mass Transfer 70 (2014) 979-989.
\bibitem{BGK54}P. L. Bhatnagar, E. P. Gross, M. Krook, A model for collision processes in gases. I. Small amplitude processes in charged and neutral one-component systems, Phys. Rev. 94 (1954) 511-525.
\bibitem{Higuera89}F. J. Higuera, S. Succi, R. Benzi, Lattice gas dynamics with enhanced collisions, Europhys. Lett. 9 (1989) 345-349.
\bibitem{Hume92}D. d'Humi$\grave{e}$res, Generalized lattice-Boltzmann equations, in: B. D. Shizgal, D. P. Weaver (Eds.), Rarefied Gas Dynamics: Theory and Simulations, Prog. Astronaut. Aeronaut., vol. 159, AIAA, Washington, DC, 1992, pp. 450-458.
\bibitem{Lalleme&Luo00}P. Lallemand, L.-S. Luo, Theory of the lattice Boltzmann method: dispersion, dissipation, isotropy, and stability, Phys. Rev. E 61 (2000) 6546-6562.
\bibitem{Liu14}Q. Liu, Y.-L. He, Q. Li, W.-Q. Tao, A multiple-relaxation-time lattice Boltzmann model for convection heat
  transfer in porous media, Int. J. Heat Mass Transfer 73 (2014) 761-775.
\bibitem{Inamuro02}T. Inamuro, A lattice kinetic scheme for incompressible viscous flows with heat transfer, Phil. Trans. R. Soc. Lond. A, 360 (2002) 477-484.
\bibitem{Inamuro06}T. Inamuro, Lattice Boltzmann methods for viscous fluid flows and for two-phase fluid flows, Fluid Dyn. Res. 38 (2006) 641-659.
\bibitem{Nishiyama13}T. Nishiyama, T. Inamuro, S. Yasuda, Numerical simulation of the dispersion of aggregated Brownian particles under shear flows, Comput. Fluids, 86 (2013) 395-404.
\bibitem{Yoshino07}M. Yoshino, Y. Hotta, T. Hirozane, M. Endo, A numerical method for incompressible non-Newtonian fluid flows based on the lattice Boltzmann method, J. Non-Newton. Fluid Mech. 147 (2007) 69-78.
\bibitem{WangL14}L. Wang, J. C. Mi, X. H. Meng, Z. L. Guo, A localized mass-conserving lattice Boltzmann approach for non-Newtonian fluid flows, Commun. Comput. Phys, 17(4) (2015) 908-924.
\bibitem{PengY04}Y. Peng, C. Shu, Y. T. Chew, Lattice kinetic scheme for the incompressible viscous thermal flows on arbitrary meshes, Phys. Rev. E 69 (2004) 016703.
\bibitem{ChengP90}C. T. Hsu, P. Cheng, Thermal dispersion in a porous medium, Int. J. Heat Mass Transfer 33 (1990) 1587-1597.
\bibitem{GuoS00}Z. Guo, B. Shi, N. Wang, Lattice BGK model for incompressible Navier-Stokes equation, J. Comput. Phys. 165 (2000) 288-306.
\bibitem{Chai13}Z. H. Chai, T. S. Zhao, Lattice Boltzmann model for the convection-diffusion equation, Phys. Rev. E 87 (2013) 063309.
\bibitem{Gao11}D. Y. Gao, Z. Q. Chen, Lattice Boltzmann simulation of natural convection dominated melting in a rectangular cavity filled with porous media, Int. J. Therm. Sci 50 (2011) 493-501.
\bibitem{Chopard09}B. Chopard, J. L. Falcone, J. Latt, The lattice Boltzmann advection-diffusion model revisited, Eur. Phys. J. Special Topics 171 (2009) 245-249.
\bibitem{Dixit06}H.N. Dixit, V. Babu, Simulation of high Rayleigh number natural convection in a square cavity using the lattice Boltzmann method, Int. J. Heat Mass Transfer 49 (2006) 727-739.
\bibitem{Boyd06}J. Boyd, J. Buick, S. Green, A second-order accurate lattice Boltzmann non-Newtonian flow model, J. Phys. A: Math. Gen, 39 (2006) 14241-14247.
\bibitem{YongL12}W. A. Yong, L. S. Luo, Accuracy of the viscous stress in the lattice Boltzmann equation with simple boundary conditions, Phys. Rev. E. 86 (2012) 065701(R).
\bibitem{GuoNEES}Z. L. Guo, C. G. Zheng, B. C. Shi, An extrapolation method for boundary conditions in lattice Boltzmann method, Phys Fluids 14 (2007) 2007-2010.
\bibitem{Beckermann88}C. Beckermann, R. Viskanta, S. Ramadhyani, Natural convection in vertical enclosures containing simultaneously fluid and porous layers, J. Fluid Mech. 186 (1988) 257-284.
\bibitem{Lauriat89}G. Lauriat, V. Prasad, Non-Darcian effects on natural convection in a vertical porous enclosure, Int. J. Heat Mass Transfer 32 (1989) 2135-2148.
\bibitem{Nithiarasu97}P. Nithiarasu, K.N. Seetharamu, T. Sundararajan, Natural convective heat transfer in a fluid saturated variable porosity medium, Int. J. Heat Mass Transfer 40 (1997) 3955-3967.
\bibitem{Davis83}G. de Vahl Davis, Natural convection of air in a square cavity: a bench mark numerical solution, Int. J. Numer. Methods Fluids 3 (1983) 249-264.
\bibitem{Hortmann90}M. Hortmann, M. Peri\'{c} , G. Scheuerer, Finite volume multigrid prediction of laminar natural convection: bench-mark solutions, Int. J. Numer. Methods Fluids 11 (1990) 189-207.
\bibitem{GuoB02}Z. L. Guo, B. C. Shi, C. G. Zheng, A coupled lattice BGK model for the Boussinesq equations, Int. J. Numer. Meth. Fluids 39 (2002) 325-342.
\bibitem{Khanafer98}K. M. Khanafer, A. J. Chamkha, Hydromagnetic natural convection from an inclined porous square enclosure with heat generation, Numer. Heat Transfer A 33 (1998) 891-910.
\bibitem{Jue03}T. C. Jue, Analysis of thermal convection in a fluid-saturated porous cavity with internal heat generation, Heat Mass Transfer 40 (2003) 83-89.

\end{thebibliography}



\end{document}